\documentclass{emulateapj}

\usepackage{natbib} 
\usepackage{multirow} 
\usepackage[caption=false]{subfig} 
\usepackage{enumerate} 
\usepackage[T1]{fontenc} 
\usepackage{amsmath} 

\def\reff@jnl#1{{\rm#1\/}}

\def\aj{\reff@jnl{AJ}}                  
\def\araa{\reff@jnl{ARA\&A}}            
\def\apj{\reff@jnl{ApJ}}                
\def\apjl{\reff@jnl{ApJ}}               
\def\apjs{\reff@jnl{ApJS}}              
\def\ao{\reff@jnl{Appl.Optics}}         
\def\apss{\reff@jnl{Ap\&SS}}            
\def\aap{\reff@jnl{A\&A}}               
\def\aapr{\reff@jnl{A\&A~Rev.}}         
\def\aaps{\reff@jnl{A\&AS}}             
\def\azh{\reff@jnl{AZh}}                        
\def\baas{\reff@jnl{BAAS}}              
\def\jrasc{\reff@jnl{JRASC}}            
\def\memras{\reff@jnl{MmRAS}}           
\def\mnras{\reff@jnl{MNRAS}}            
\def\pra{\reff@jnl{Phys.Rev.A}}         
\def\prb{\reff@jnl{Phys.Rev.B}}         
\def\prc{\reff@jnl{Phys.Rev.C}}         
\def\prd{\reff@jnl{Phys.Rev.D}}         
\def\prl{\reff@jnl{Phys.Rev.Lett}}      
\def\pasp{\reff@jnl{PASP}}              
\def\pasj{\reff@jnl{PASJ}}              
\def\qjras{\reff@jnl{QJRAS}}            
\def\skytel{\reff@jnl{S\&T}}            
\def\solphys{\reff@jnl{Solar~Phys.}}    
\def\sovast{\reff@jnl{Soviet~Ast.}}     
 \def\ssr{\reff@jnl{Space~Sci.Rev.}}     
\def\zap{\reff@jnl{ZAp}}                        
\def\nat{\reff@jnl{Nature}}             

\shorttitle{A multi-wavelength investigation of RCW175}
\shortauthors{Tibbs et al.}

\begin{document}


\title{A multi-wavelength investigation of RCW175: an H\textsc{ii} region harboring spinning dust emission}

\author{C.~T.~Tibbs\altaffilmark{1}, R.~Paladini\altaffilmark{2}, M.~Compi{\`e}gne\altaffilmark{1,3}, C.~Dickinson\altaffilmark{4}, M.~I.~R.~Alves\altaffilmark{5}, N.~Flagey\altaffilmark{6}, S.~Shenoy\altaffilmark{7}, A.~Noriega-Crespo\altaffilmark{8}, S.~Carey\altaffilmark{1}, S.~Casassus\altaffilmark{9}, R.~D.~Davies\altaffilmark{4}, R.~J.~Davis\altaffilmark{4}}
\email{ctibbs@ipac.caltech.edu}

\altaffiltext{1}{\textit{Spitzer} Science Center, California Institute of Technology, Pasadena, CA 91125, USA}
\altaffiltext{2}{NASA \textit{Herschel} Science Center, California Institute of Technology, Pasadena, CA 91125, USA}
\altaffiltext{3}{Laboratoire d'Optique Atmosph{\'e}rique, UMR8518, CNRS-INSU, Universit{\'e} Lille 1, Villeneuve d'Ascq, France}
\altaffiltext{4}{Jodrell Bank Centre for Astrophysics, School of Physics and Astronomy, The University of Manchester, M13 9PL, UK}
\altaffiltext{5}{Institut d'Astrophysique Spatiale, Universit{\'e} Paris Sud XI, B{\^a}timent 121, 91405 Orsay, France}
\altaffiltext{6}{Jet Propulsion Laboratory, California Institute of Technology, Pasadena, CA 91125, USA}
\altaffiltext{7}{Space Science Division, NASA Ames Research Center, M/S 245-6, Moffett Field, CA 94035, USA}
\altaffiltext{8}{Infrared Processing and Analysis Center, California Institute of Technology, Pasadena, CA 91125, USA}
\altaffiltext{9}{Departamento de Astronom{\'{\i}}a, Universidad de Chile, Casilla 36-D, Santiago, Chile}


\begin{abstract}

Using infrared, radio continuum and spectral observations, we performed a detailed investigation of the H\textsc{ii} region RCW175. We determined that RCW175, which actually consists of two separate H\textsc{ii} regions, G29.1-0.7 and G29.0-0.6, is located at a distance of 3.2~$\pm$~0.2~kpc. Based on the observations we infer that the more compact G29.0-0.6 is less evolved than G29.1-0.7 and was possibly produced as a result of the expansion of G29.1-0.7 into the surrounding interstellar medium. We compute a star formation rate for RCW175 of (12.6~$\pm$~1.9)~$\times$~10$^{-5}$~M$_{\sun}$~yr$^{-1}$, and identified 6 possible young stellar object candidates within its vicinity. Additionally, we estimate that RCW175 contains a total dust mass of~215~$\pm$~53~M$_{\sun}$. 

RCW175 has previously been identified as a source of anomalous microwave emission~(AME), an excess of emission at cm wavelengths often attributed to electric dipole radiation from the smallest dust grains. We find that the AME previously detected in RCW175 is not correlated with the smallest dust grains (polycyclic aromatic hydrocarbons or small carbonaceous dust grains), but rather with the exciting radiation field within the region. This is a similar result to that found in the Perseus molecular cloud, another region which harbors AME, suggesting that the radiation field may play a pivotal role in the production of this new Galactic emission mechanism. Finally, we suggest that these observations may hint at the importance of understanding the role played by the major gas ions in spinning dust models.

\end{abstract}




\section{Introduction}
\label{Sec:Introduction}

Dust is known to be ubiquitous in our Galaxy, playing a crucial role in the heating, cooling and reprocessing of the interstellar medium~(ISM). It is therefore no surprise that dust exists within H\textsc{ii} regions, and several studies have been aimed at revealing the distribution of dust within these ionized regions~\citep[e.g.][]{Inoue:02, Draine:11, Paladini:12}. H\textsc{ii} regions are the result of the formation of early type OB stars, with their exact shape and size dependent on the surrounding environment and the balance between photoionization and recombination. They provide us with an opportunity to investigate the star formation rate~(SFR) within our own Galaxy, and there has also been substantial work suggesting that H\textsc{ii} regions play an important role in triggering star formation~\citep[e.g.][]{Deharveng:05, Zavagno:06}. Given the importance of H\textsc{ii} regions, it is essential to fully understand the physics and dynamics of these sources.

\begin{figure*}
\begin{center}$
\begin{array}{cc}
\includegraphics[scale=0.65, viewport=50 130 550 680]{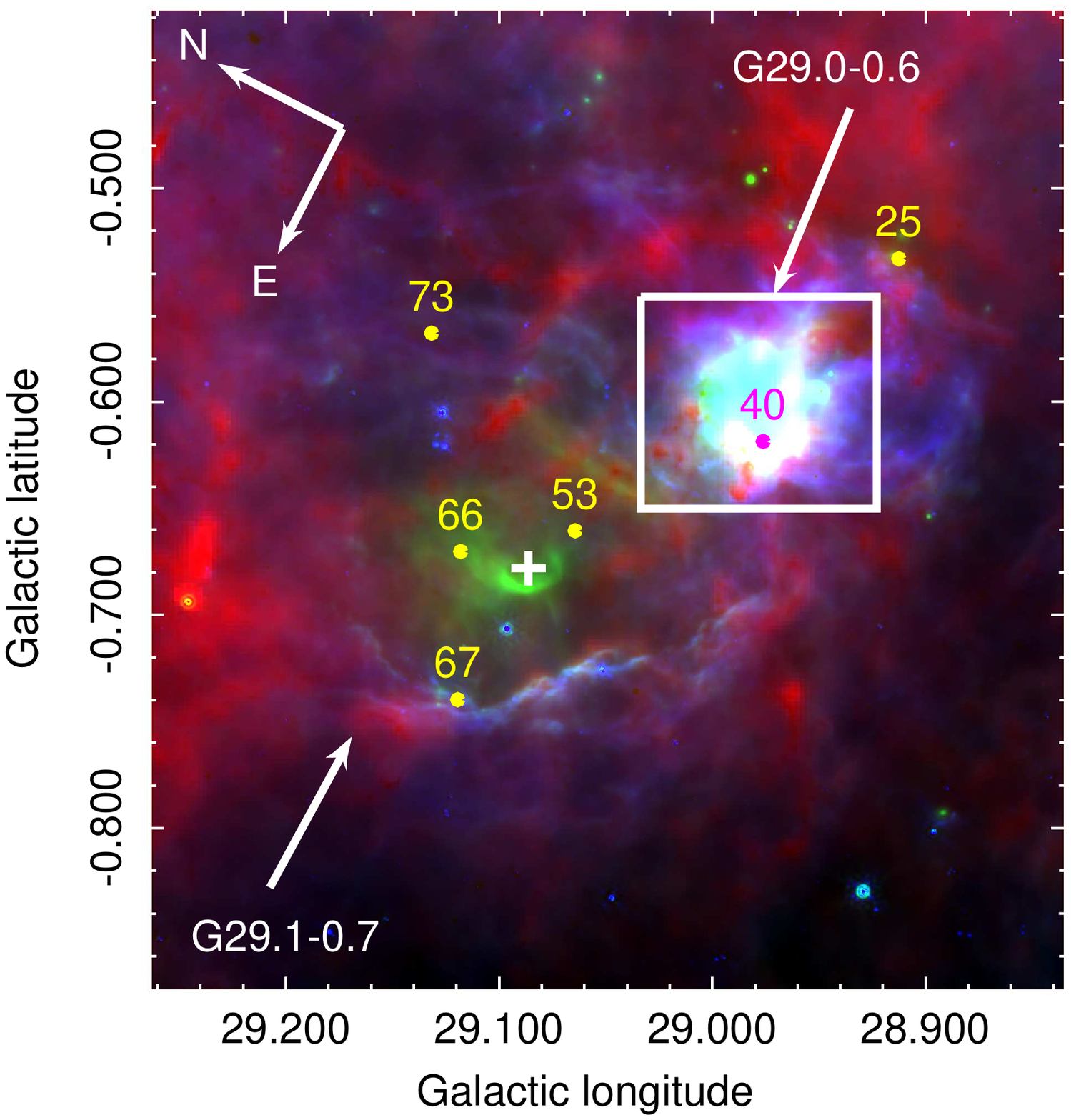} &
\includegraphics[scale=0.2, viewport=20 -400 550 50]{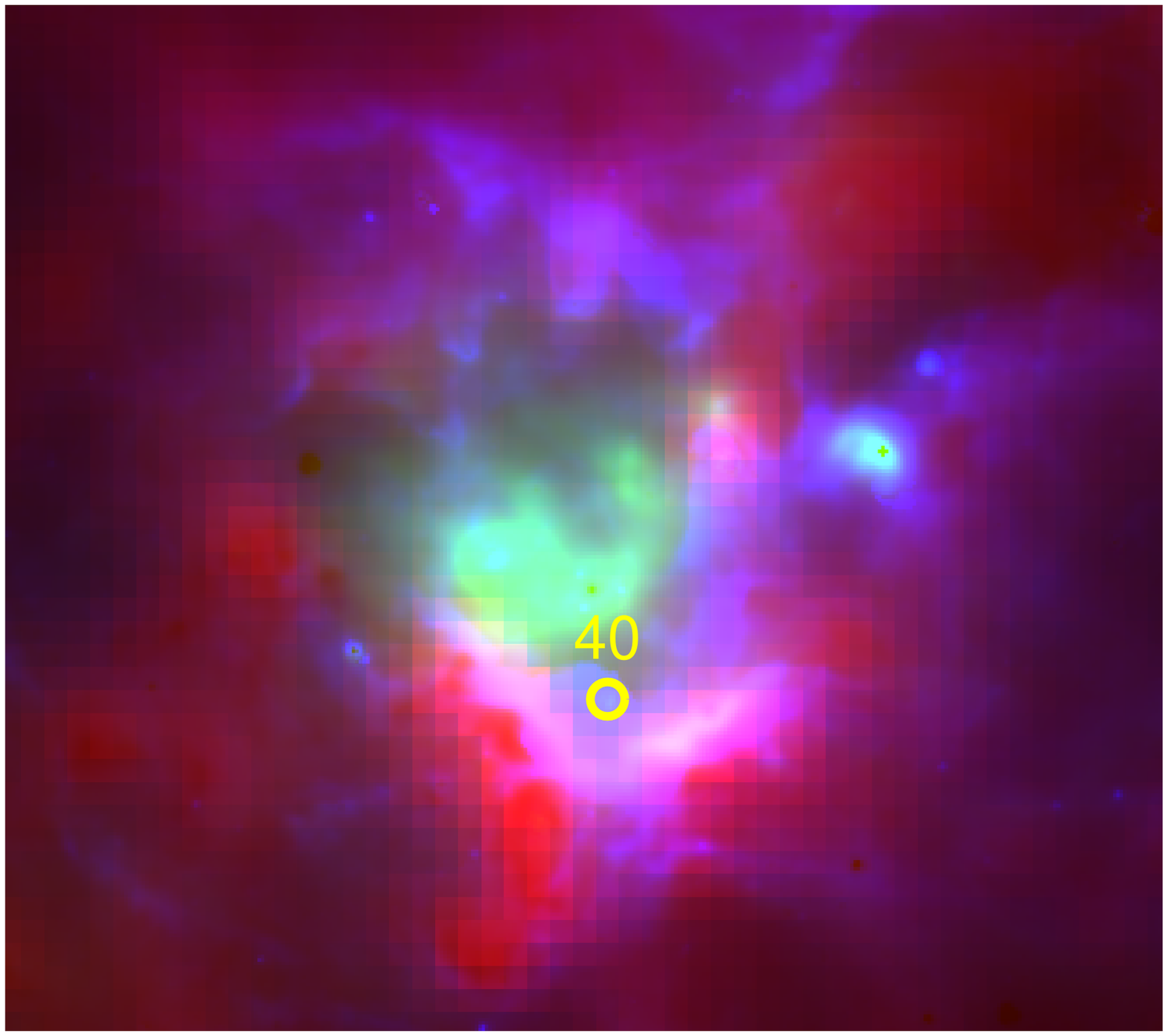} \\
\end{array}$
\end{center}
\caption{Three color image of RCW175 comprising of the IRAC 8~$\mu$m~(blue), MIPS 24~$\mu$m~(green) and the SPIRE 350~$\mu$m~(red) maps highlighting the existing two-shell structure. The region bounded by the box is displayed separately to allow the spatial detail in this region to be seen. The 24~$\mu$m emission traces the warmer dust originating from within the shells while the 8~$\mu$m emission, which traces the PAH emission, is originating from the swept up dust in the shell itself. The location of the B1~II star S65-4 is marked with a cross. G29.1-0.7, the larger, diffuse component to the east, is not a complete shell, and is only bounded on one side by the swept up material. We believe that this is the result of a rupture on the north side of the H\textsc{ii} region where the ionized gas has erupted into the surrounding ISM. G29.0-0.6, shown in the cut-out, appears as more of a complete shell with pillars pointing towards the centre. Also displayed is the location of the six identified YSO candidates~(see Section~\ref{Subsec:YSOs}).}
\label{Fig:3_color_RCW175}
\end{figure*}

In this analysis we focus solely on the H\textsc{ii} region RCW175. RCW175, located on the Galactic plane at $l$~=~29.$\!^{\circ}$07 and $b$~=~$-$0.$\!^{\circ}$68, was originally identified and catalogued by~\citet{Sharpless:59} and was observed in emission with an H$\alpha$ filter by~\citet{Rodgers:60}. RCW175 is actually just one component of the entire complex that will be discussed in this work. As shown in Figure~\ref{Fig:3_color_RCW175}, the complex consists of two separate components: G29.1-0.7, a diffuse component to the east that is visible in H$\alpha$~(this is the component detected by~\citealt{Sharpless:59} and~\citealt{Rodgers:60}) and G29.0-0.6, a more compact component to the west that is heavily obscured by dust, and as a consequence, was not detectable in H$\alpha$ emission. Throughout this paper we will use RCW175 to refer to the entire complex. 

RCW175 is a particularly interesting H\textsc{ii} region, as it has been observed to be a source of anomalous microwave emission~(AME;~\citealp{Dickinson:09}). AME is observed as an excess of emission with respect to the other Galactic emission mechanisms at cm wavelengths, namely: thermal bremsstrahlung or free-free emission produced as a result of the interaction between the free electrons and ions in the ionized gas; synchrotron emission originating from the acceleration of relativistic electrons in the Galactic magnetic fields; and thermal dust emission arising from the vibration of large dust grains. AME is often attributed to electric dipole radiation arising from rapidly spinning small dust grains, each containing a dipole moment~\citep{DaL:98}, and has been observed in other Galactic H\textsc{ii} regions~\citep{Dickinson:07, Todorovic:10} along with a number of additional Galactic environments such as molecular clouds~\citep{Watson:05, Casassus:08, Tibbs:10, Planck_AME:11}, dark clouds~\citep{AMI:09, Dickinson:10, Vidal:11} and reflection nebulae~\citep{Castellanos:11, Genova:11}.

In this work we attempt to provide a complete analysis of RCW175 in order to improve our understanding of both H\textsc{ii} regions in general and also the AME. The layout of the paper is as follows: Section~\ref{Sec:Data} presents the data used in this work. In Section~\ref{Sec:Results_Discussion} we describe the morphology of the region and compute the kinematic distance, the SFR and the dust mass. We also produce a complete spectral energy distribution~(SED) of the source and discuss the presence of AME, including its possible origin and excitation mechanism within the specific environment of RCW175. Finally, in Section~\ref{Sec:Summary} we summarize our results and discuss our conclusions.


\section{Data}
\label{Sec:Data}

Performing a thorough investigation of RCW175 requires data spanning a wide range of wavelengths. However, given the angular size of RCW175 ($\sim$~25~arcmin~--~see Figure~\ref{Fig:3_color_RCW175}), we decided to only select data with an angular resolution $\lesssim$~15~arcmin to ensure the region was at least partially resolved. All the data used in this analysis are described in Sections~\ref{Subsec:IR_Data}, \ref{Subsec:Radio_Data} and \ref{Subsec:Spectral_Data} corresponding to infrared~(IR) data, radio continuum data and radio spectral data, respectively.

\subsection{IR Data}
\label{Subsec:IR_Data}

\subsubsection{IRIS Data}

The \textit{Infrared Astronomical Satellite}~(\textit{IRAS}) data provide almost complete all-sky coverage at four bands centered at 12, 25, 60 and 100~$\mu$m. The Improved Reprocessing of the \textit{IRAS} Survey~(IRIS;~\citealp{Miville-Deschenes:05}) data represent a significant improvement over the original \textit{IRAS} data in regards to absolute calibration, destriping and the removal of the zodiacal light. The IRIS data have a typical gain uncertainty of~$\approx$~15~$\%$ and an angular resolution of~$\approx$~4~arcmin. To mitigate the effect of the limited resolving power of the IRIS data, we also include higher angular resolution data from the more recent \textit{Spitzer Space Telescope} and \textit{Herschel Space Observatory} satellites.

\subsubsection{\textit{Spitzer} Data}

To trace the warm dust in RCW175 we used mid-IR observations from the \textit{Spitzer} GLIMPSE \citep{Churchwell:09} and MIPSGAL~\citep{Carey:09} Legacy Programmes. Both of these progammes involved mapping the inner Galactic plane~($-$60$^{\circ}$~$<$~$l$~$<$~60$^{\circ}$) and when combined produce a wavelength coverage of 3.6, 4.5, 5.8, 8.0, 24 and 70~$\mu$m. In this analysis, we decided to use only the 8~$\mu$m emission from the GLIMPSE survey and the 24~$\mu$m from the MIPSGAL survey. The reasons for neglecting the three shortest wavelength IRAC bands is that the point source density increases significantly with the decrease in wavelength, and at these shorter wavelengths the spectrum can be contaminated by the presence of ionic and molecular lines, therefore making it substantially more difficult to accurately investigate extended emission. We also ignored the MIPS 70~$\mu$m observations, which suffer from responsivity variations, as there are now superior quality data obtained with \textit{Herschel} at the same wavelength~(see Section~\ref{Subsubsec:Herschel_Data}).

Both the IRAC 8~$\mu$m data, with an angular resolution of 2~arcsec, and the MIPS 24~$\mu$m data, with an angular resolution of 6~arcsec, were reprocessed independently from the~\textit{Spitzer Science Centre} pipeline. First, an extended emission correction was applied to the 8~$\mu$m data to ensure that the IRAC calibration, which is based on point sources, was effectively applied to the extended emission. Secondly, both the 8 and 24~$\mu$m data were corrected for the contribution of zodiacal light, which if not subtracted can result in an overestimate of the emission at these wavelengths by up to a factor of two. Both data sets were then subjected to source extraction, as the removal of the sources allows us to better investigate the diffuse emission. The final step in the reprocessing was to perform an overlap correction to ensure a consistent background at each wavelength. Given the data reprocessing, we assume a typical uncertainty of~10~$\%$ for the extended emission in both the IRAC 8~$\mu$m and the MIPS 24~$\mu$m data.

\subsubsection{\textit{Herschel} Data}
\label{Subsubsec:Herschel_Data}

To complement the mid-IR \textit{Spitzer} observations, data from the \textit{Herschel Space Observatory}~\citep{Pilbratt:10}, observed as part of the Hi-Gal Key Programme~\citep{Molinari:10}, were also incorporated into this analysis. Hi-Gal, like both GLIMPSE and MIPSGAL, is a photometric survey of the inner Galactic plane ($-$60$^{\circ}$~$<$~$l$~$<$~60$^{\circ}$, |$b$|~$<$~1$^{\circ}$) using both the PACS~\citep{Poglitsch:10} and SPIRE~\citep{Griffin:10} photometers at 70, 160, 250, 350 and 500~$\mu$m. These data, with an angular resolution between 6~arcsec at 70~$\mu$m and 35~arcsec at 500~$\mu$m, were reduced with the \textsc{romagal} pipeline as described by~\citet{Traficante:11}. The absolute zero level was determined through the IRIS 60~$\mu$m data for the PACS 70~$\mu$m observations and the IRIS 100~$\mu$m data for the other four bands, in a similar manner to that described in~\citet{Bernard:10}. We assume a 20~$\%$ uncertainty for the extended emission in the PACS and SPIRE data.

\begin{figure*}
\begin{center}
\includegraphics[scale=0.95, viewport=40 170 600 650]{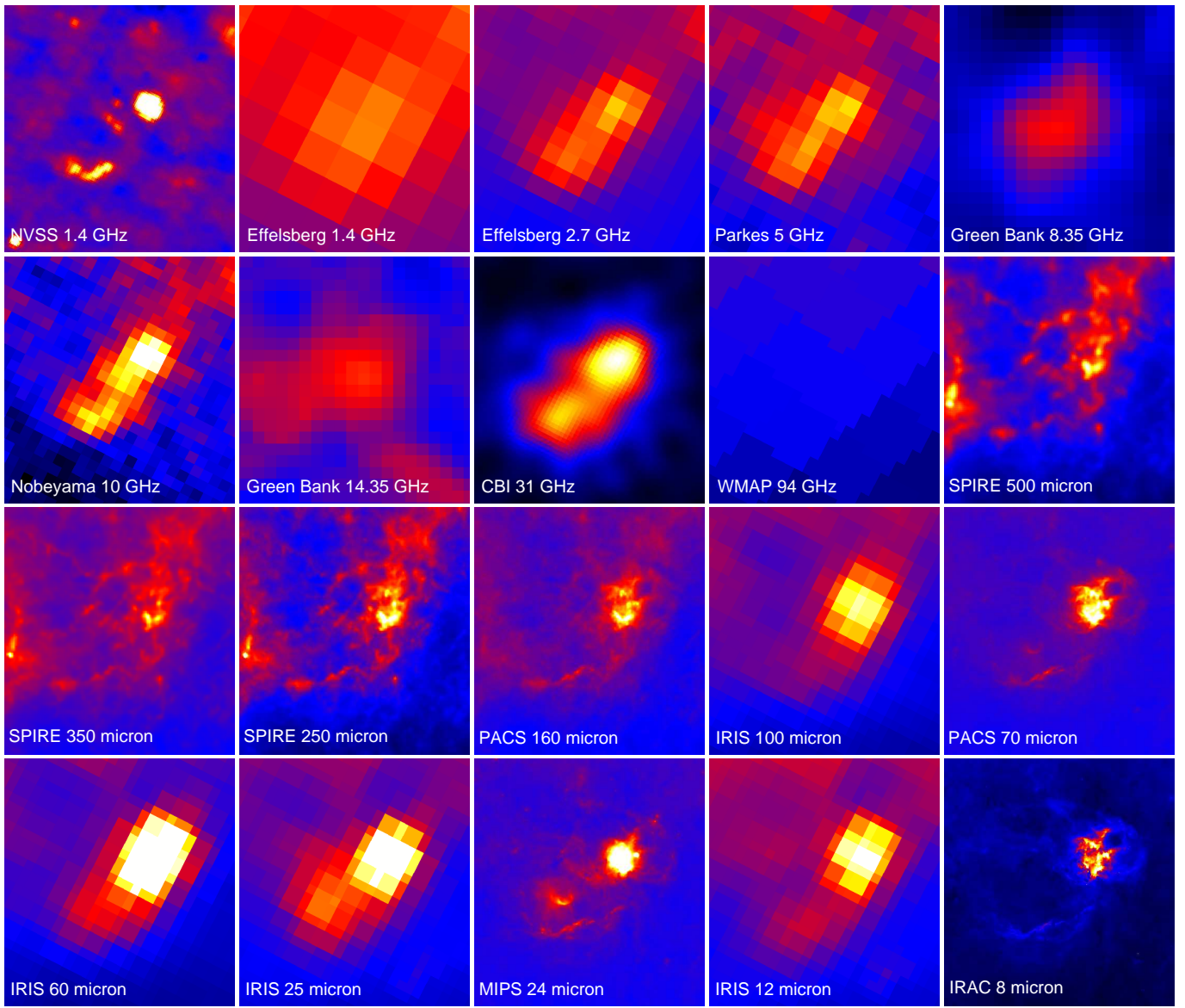} 
\end{center}
\caption{Maps of RCW175 all the way from the radio to the mid-IR. There are two maps at 1.4~GHz, as the NVSS map provides a high resolution image but, due to flux loss, it cannot be used to estimate the total flux density, and so we also have the Effelsberg 1.4~GHz map. Where the angular resolution is $\lesssim$~5~arcmin, it is possible to identify both G29.0-0.6 and G29.1-0.7. All maps are on the same co-ordinate system as Figure~\ref{Fig:3_color_RCW175}.}
\label{Fig:Data}
\end{figure*}

\subsection{Radio Continuum Data}
\label{Subsec:Radio_Data}

\subsubsection{NVSS Data}

The National Radio Astronomy Observatory (NRAO) Very Large Array~(VLA) Sky Survey~(NVSS) data at 1.4~GHz~\citep{Condon:98} represent the highest angular resolution~(0.75~arcmin) radio map of RCW175. However, the NVSS data were observed with the VLA interferometer in D and DnC configuration, and due to a lack of short spacings they are characterized by a lack of sensitivity on scales greater than~$\sim$~15~arcmin. This makes the NVSS map useful to identify high resolution radio structures present in RCW175, but it cannot be used to determine the total flux density within the region.

\subsubsection{Effelsberg Data}

Given the limitation of the NVSS data discussed above, data from both the 21~cm~\citep{Reich:90a} and 11~cm~\citep{Reich:90b} Galactic plane surveys performed with the Effelsberg 100~m telescope were used in this analysis. The total power level of the 21~cm~(1.4~GHz) survey data ($-$3$^{\circ}$~$<$~$l$~$<$~240$^{\circ}$, |$b$|~$<$~4$^{\circ}$) was obtained based on the Stockert 21~cm survey~\citep{Reich:82, Reich:86} and they are characterized by an angular resolution of 9.4~arcmin.

Due to the data reduction techniques used to process the 11~cm~(2.7~GHz) survey data ($-$2$^{\circ}$~$<$~$l$~$<$~240$^{\circ}$, |$b$|~$<$~5$^{\circ}$), the large-scale structure had to be added separately using low resolution~($\approx$~19~arcmin) Stockert 11~cm data~\citep{Reif:87}. The final map has an angular resolution of 4.3~arcmin. 

Based on observations of radio sources, the temperature calibration accuracy is believed to be of the order of~10~$\%$ for both the 21 and 11~cm data.

\subsubsection{Parkes Data}

To complement the Effelsberg data, we used the Parkes 6cm~(5~GHz) data that were obtained as part of the~\citet{Haynes:78} southern hemisphere Galactic plane survey ($-$170$^{\circ}$~$<$~$l$~$<$~40$^{\circ}$, |$b$|~$<$~2$^{\circ}$) using the Parkes 64~m telescope. These data have an angular resolution of 4.1~arcmin and typical uncertainties are estimated to be~10~$\%$.

\subsubsection{Nobeyama Data}

\citet{Handa:87} observed the Galactic plane ($-$5$^{\circ}$~$<$~$l$~$<$~56$^{\circ}$, |$b$|~$<$~1.5$^{\circ}$) at 3~cm~(10~GHz) using the 45~m telescope at the Nobeyama Radio Observatory. These data have an angular resolution of 3~arcmin and we assume a typical uncertainty of~$\approx$~10~$\%$.

\subsubsection{Green Bank Data}

Data observed simultaneously at 3.6 and 2.1~cm (8.35 and 14.35~GHz) with the Green Bank 13.7~m telescope~\citep{Langston:00} as part of the Green Bank Galactic plane survey ($-$15$^{\circ}$~$<$~$l$~$<$~255$^{\circ}$, |$b$|~$<$~5$^{\circ}$) were also included in this analysis. These data have an angular resolution of 9.7 and 6.6~arcmin at 3.6 and 2.1~cm, respectively, and we assume a typical uncertainty of 10~$\%$.

\subsubsection{Cosmic Background Imager Data}

The Cosmic Background Imager~(CBI;~\citealp{Taylor:11}) was a 13-element interferometer that operated in the wavelength range 1.2~--~0.8~cm~(26~--~36~GHz). It was with the CBI that~\citet{Dickinson:09} first detected the presence of an AME component in RCW175. Although the CBI was an interferometer, the entire RCW175 complex was observed within a single CBI primary beam~(28.2~arcmin) and hence any flux loss on these scales should be negligible. The CBI synthesized beam is 4.3~arcmin and we assume a conservative uncertainty of 5~$\%$ for the CBI data.

\subsubsection{\textit{WMAP} Data}

\textit{Wilkinson Microwave Anisotropy Probe} (\textit{WMAP}) 7~yr total intensity sky maps~\citep{Jarosik:11} represent all-sky observations at 1.3, 0.9, 0.7, 0.5 and 0.3~cm~(23, 33, 41, 61 and 94~GHz) with an angular resolution of 52.8, 39.6, 30.6, 21.0 and 13.2~arcmin, respectively. However, given the angular resolution restrictions discussed previously, we use only the \textit{WMAP} 94~GHz data in this analysis. Although the \textit{WMAP} data are calibrated using the CMB dipole, and hence are very accurately calibrated, due to beam errors and other systematics we assume a conservative 5~$\%$ uncertainty.

\subsection{Radio Spectral Data}
\label{Subsec:Spectral_Data}

To complement all of the photometric data described in Sections~\ref{Subsec:IR_Data} and~\ref{Subsec:Radio_Data} we also utilize spectral data to help investigate the kinematics of RCW175.

\begin{figure*}
\begin{center}$
\begin{array}{ccccc}
\includegraphics[scale=0.175, viewport=20 10 500 650]{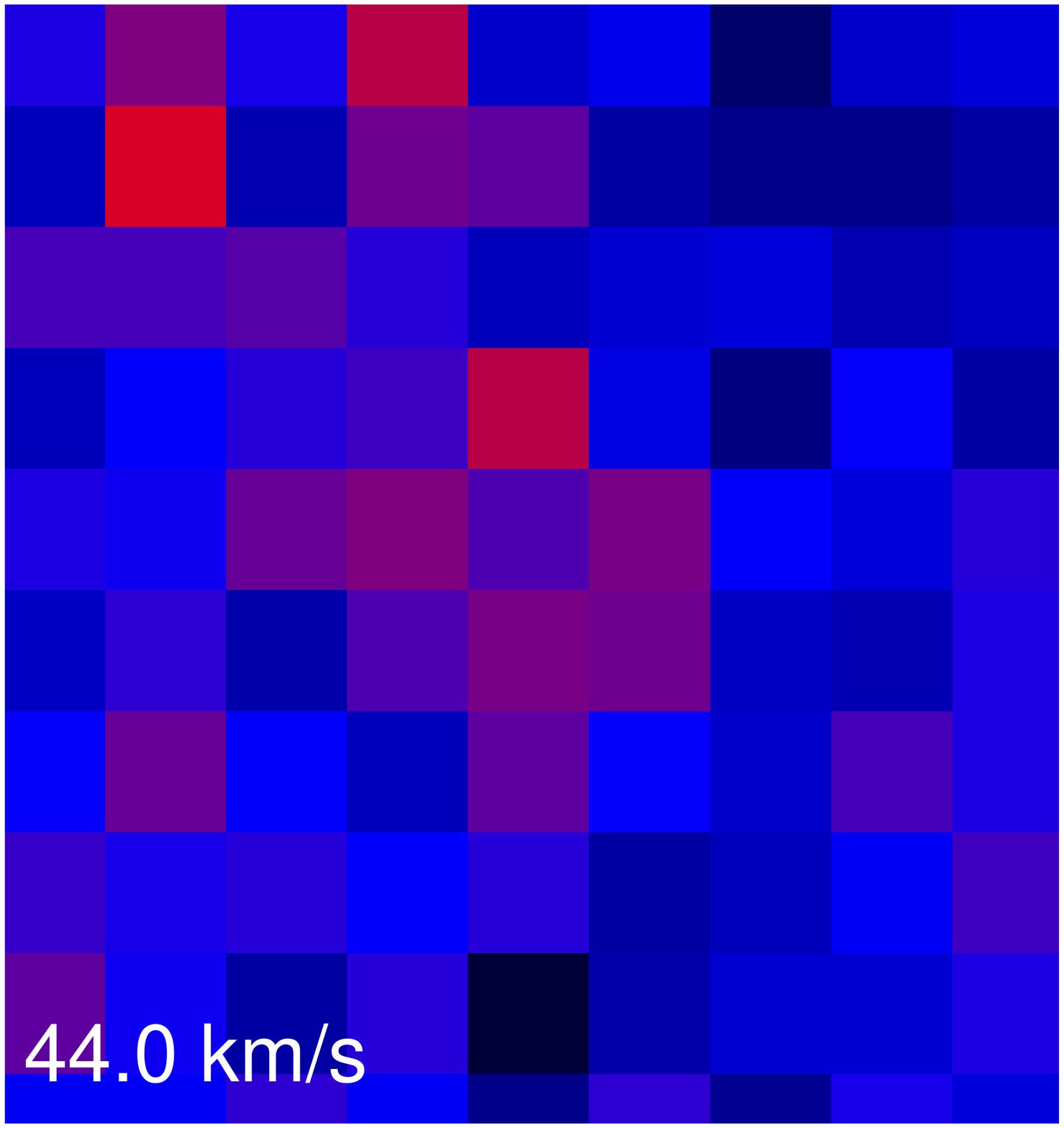} &
\hspace{2.50mm}
\includegraphics[scale=0.175, viewport=20 10 500 650]{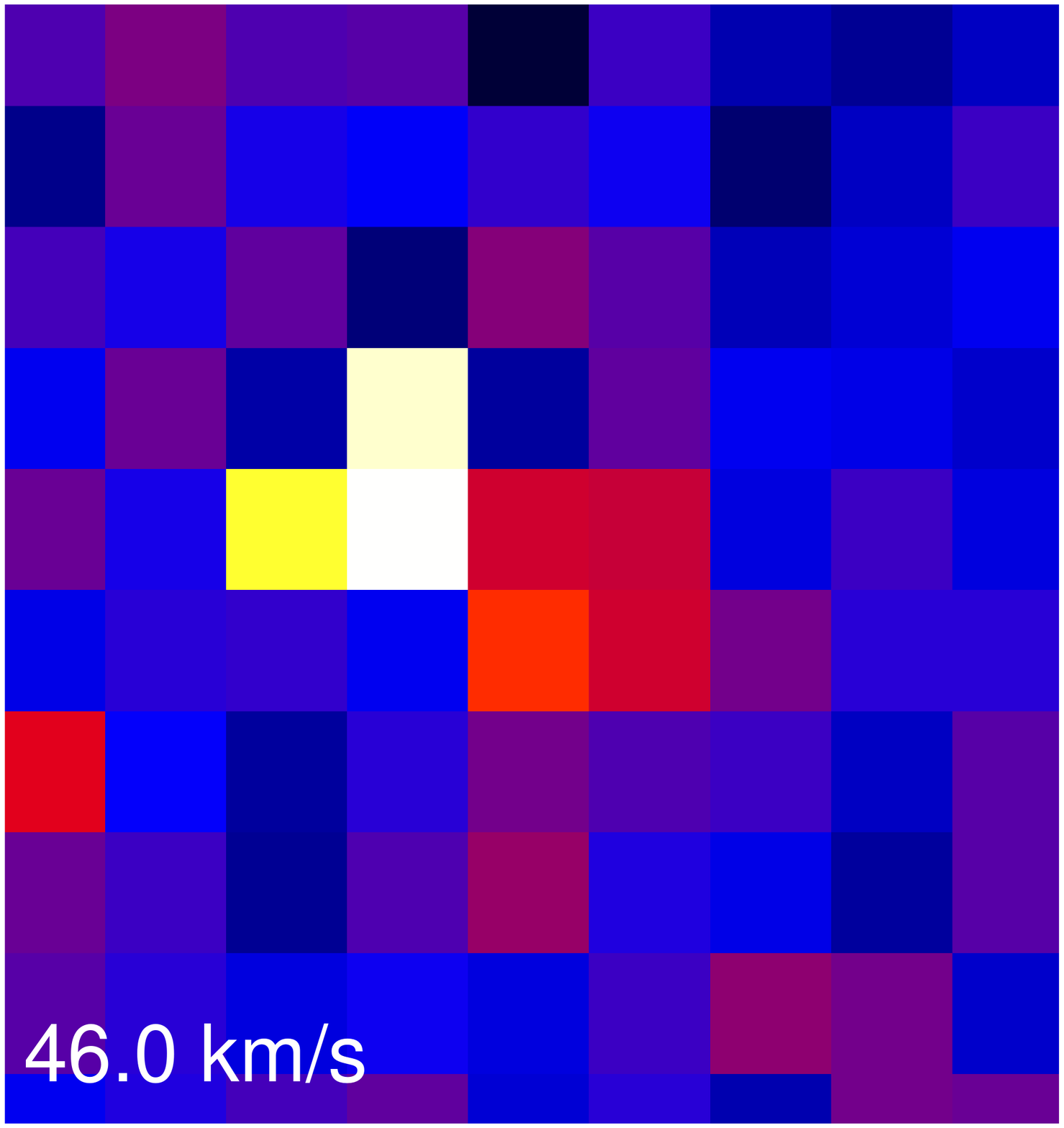} &
\hspace{2.50mm}
\includegraphics[scale=0.175, viewport=20 10 500 650]{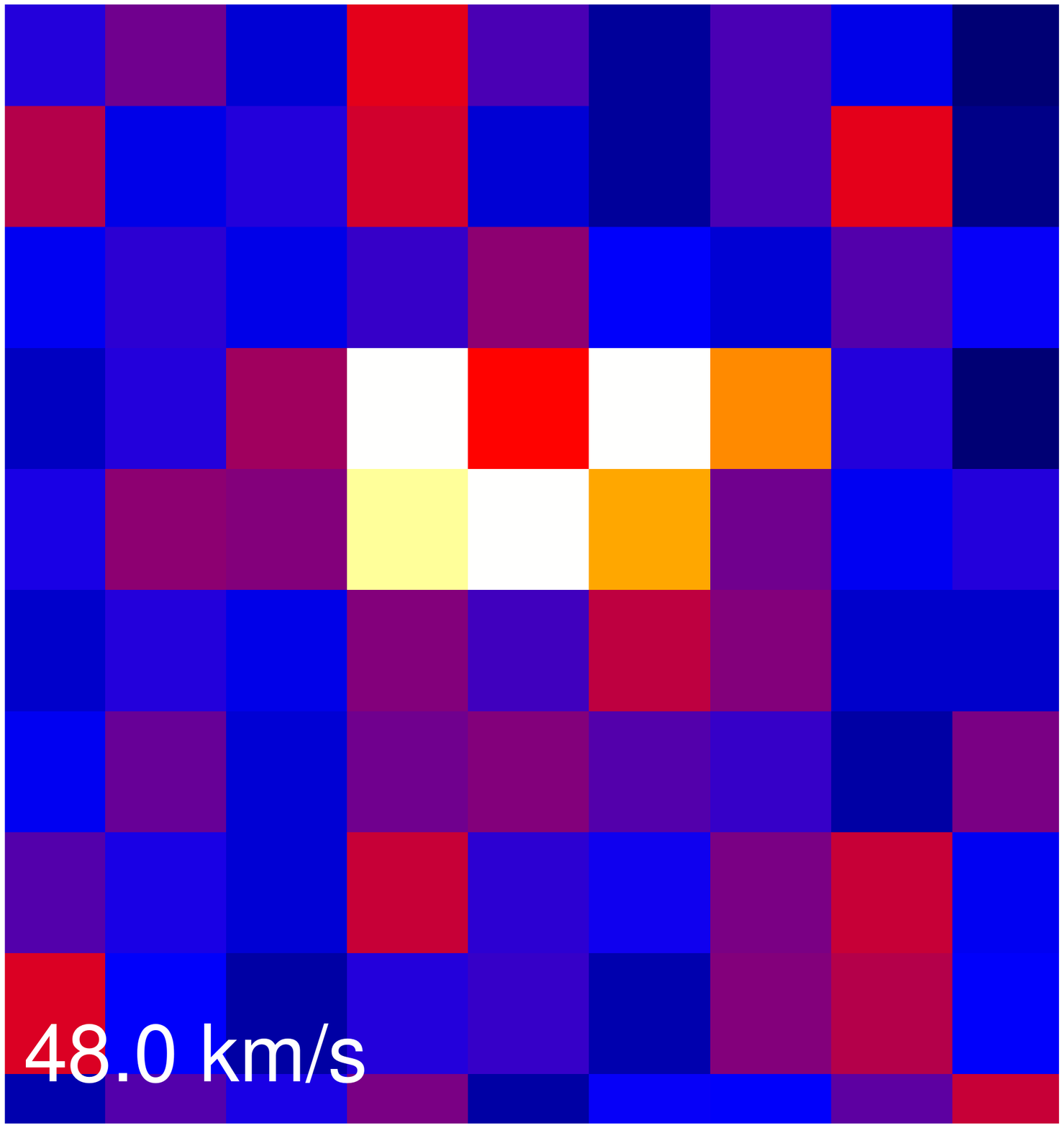} &
\hspace{2.50mm}
\includegraphics[scale=0.175, viewport=20 10 500 650]{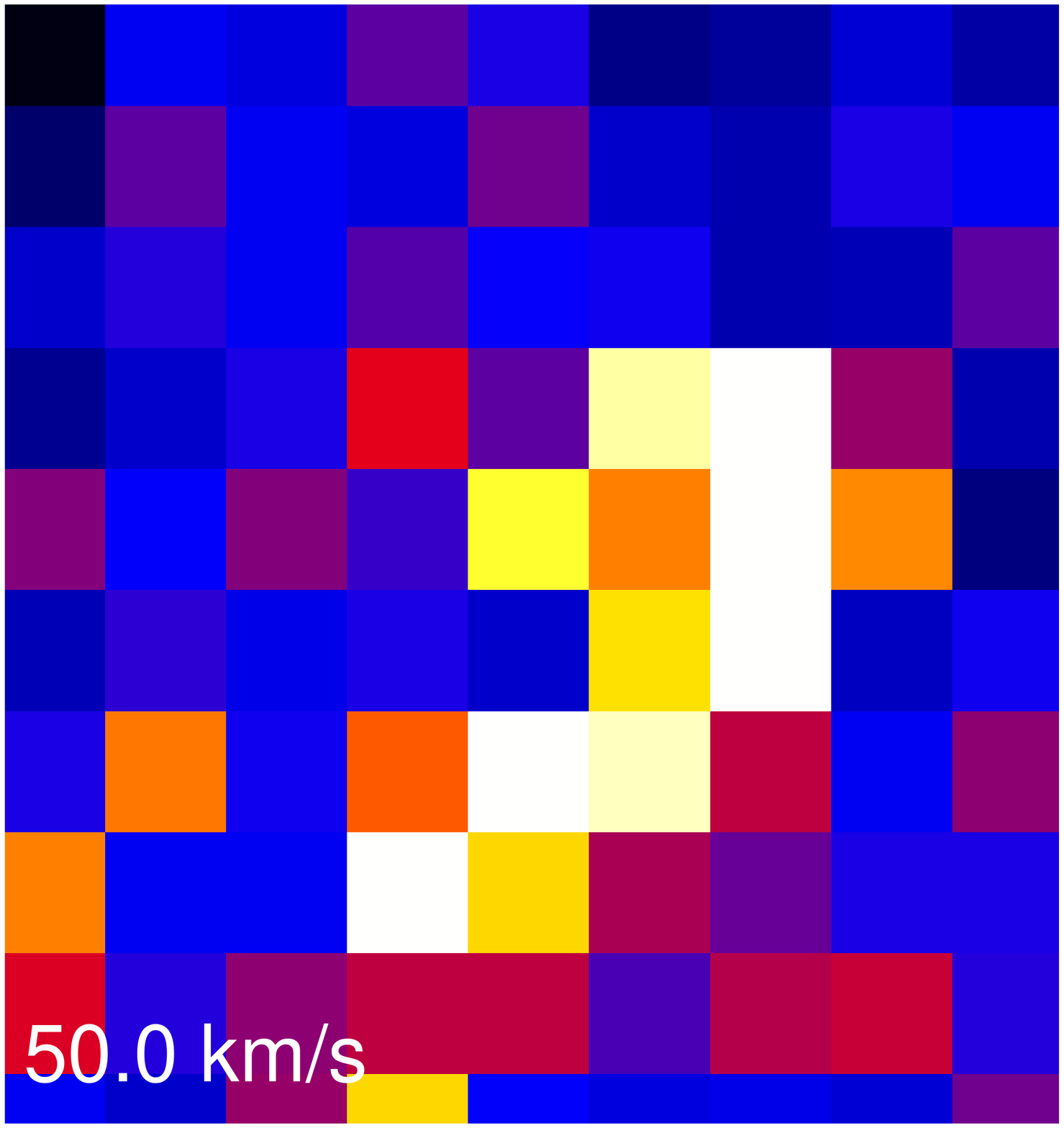} &
\hspace{2.50mm}
\includegraphics[scale=0.175, viewport=20 10 500 650]{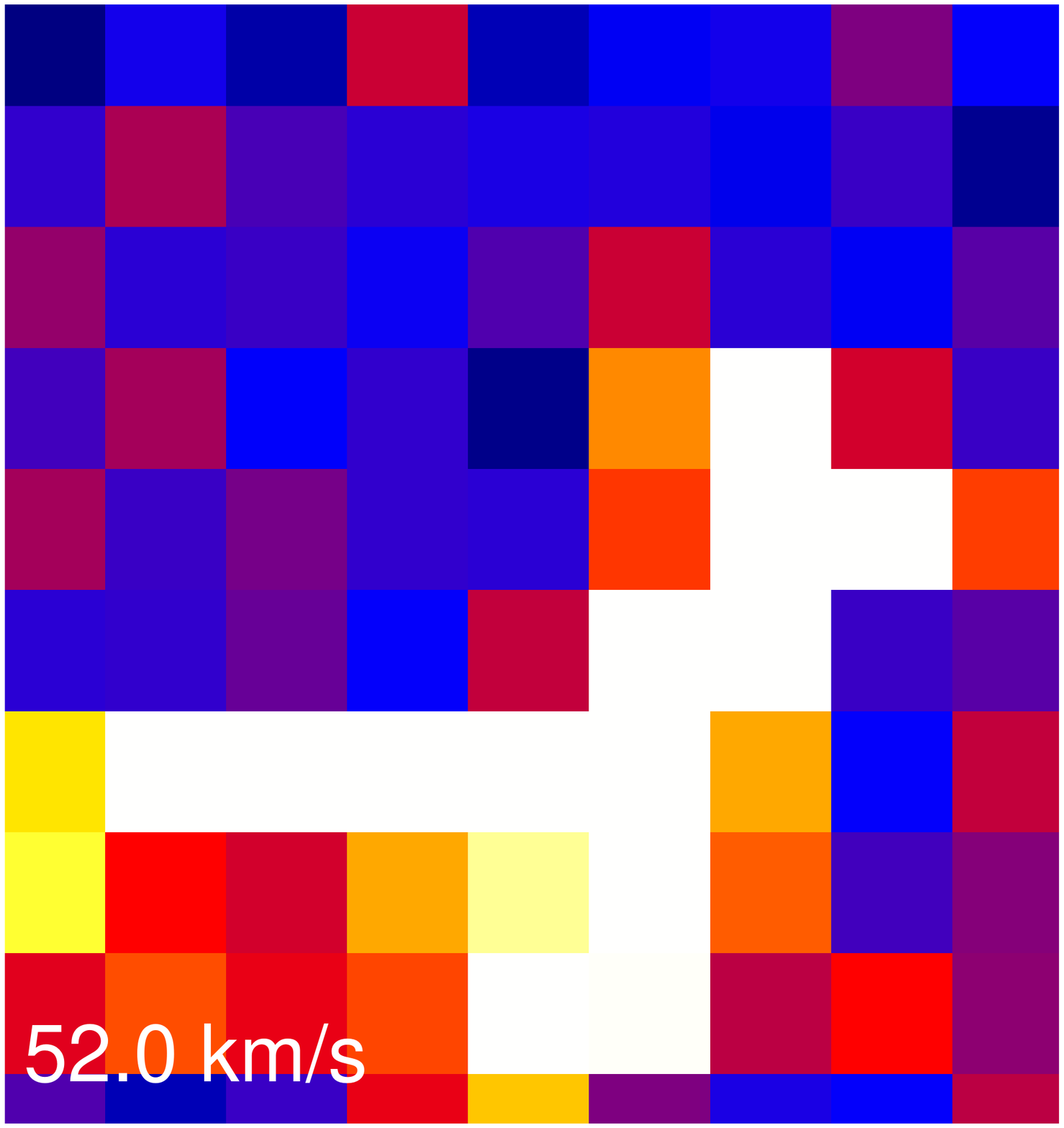} \\
\end{array}$
$
\begin{array}{cc}
\includegraphics[scale=0.175, viewport=20 100 500 600]{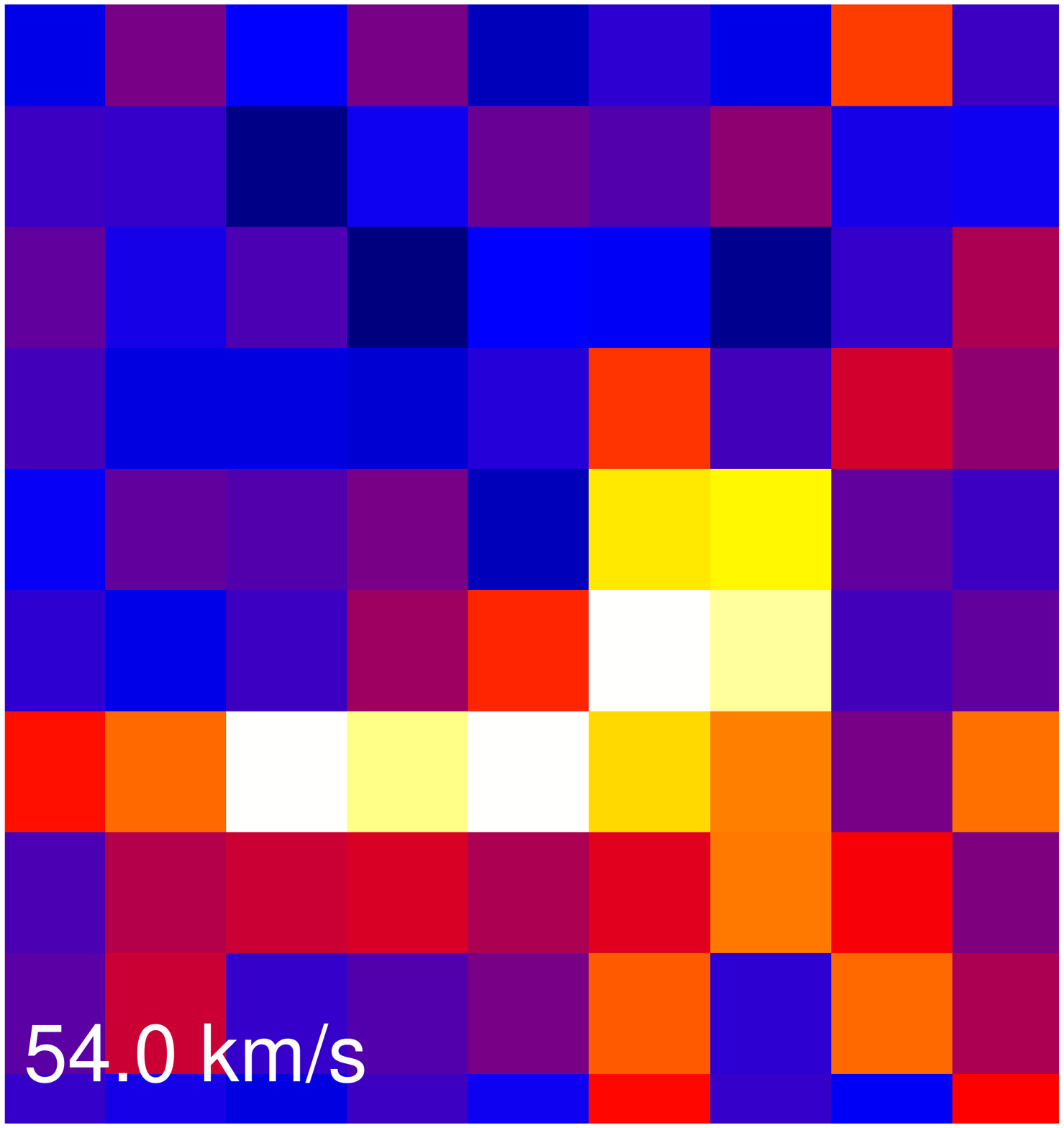} &
\hspace{2.50mm}
\includegraphics[scale=0.175, viewport=20 100 500 600]{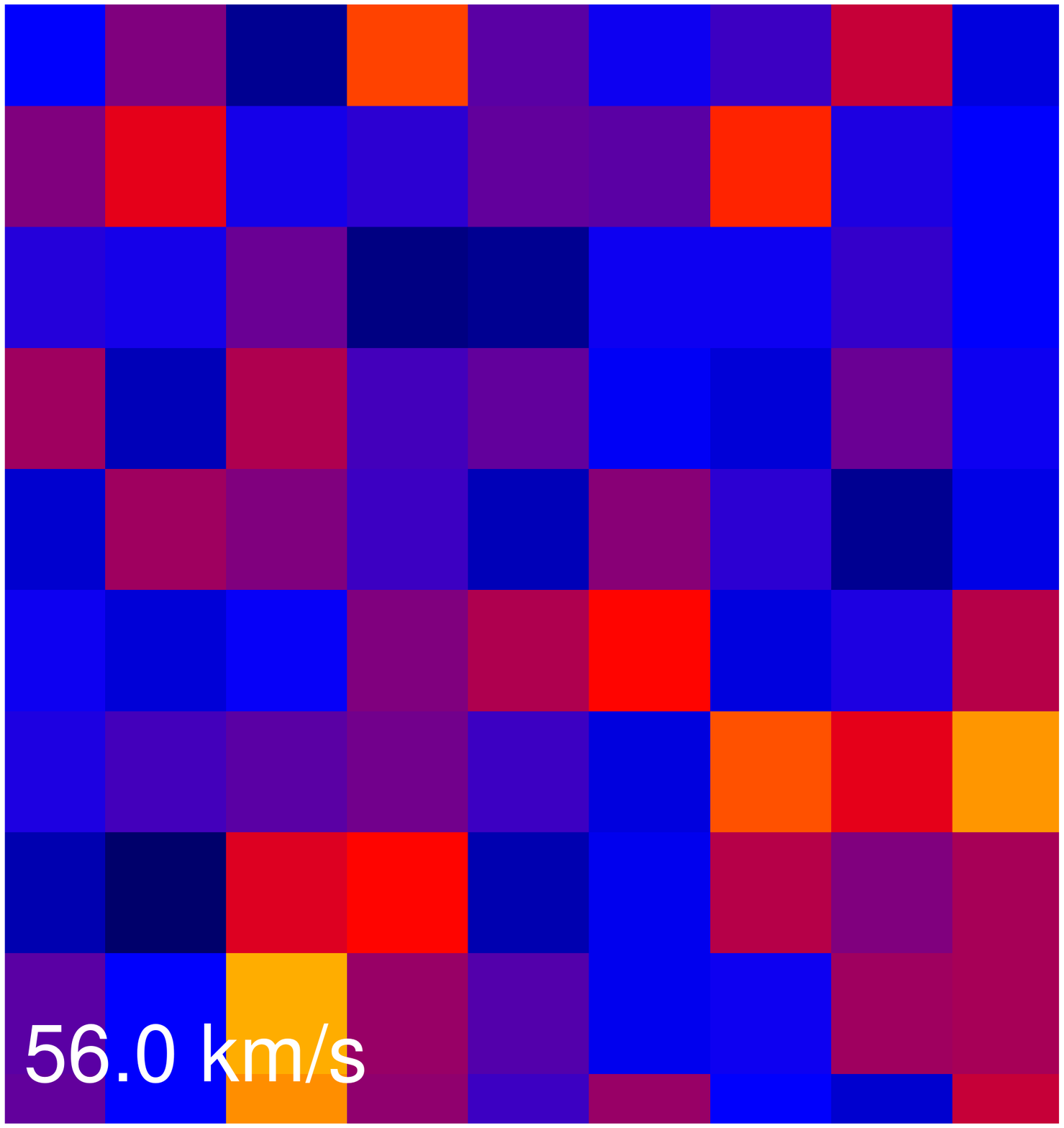} \\
\end{array}$
\end{center}
\caption{$^{12}$CO data at seven channels across RCW175. All maps are on the same co-ordinate system as Figure~\ref{Fig:3_color_RCW175}.}
\label{Fig:12CO_Data}
\end{figure*}

\begin{figure*}
\begin{center}$
\begin{array}{ccccc}
\includegraphics[scale=0.175, viewport=20 10 500 600]{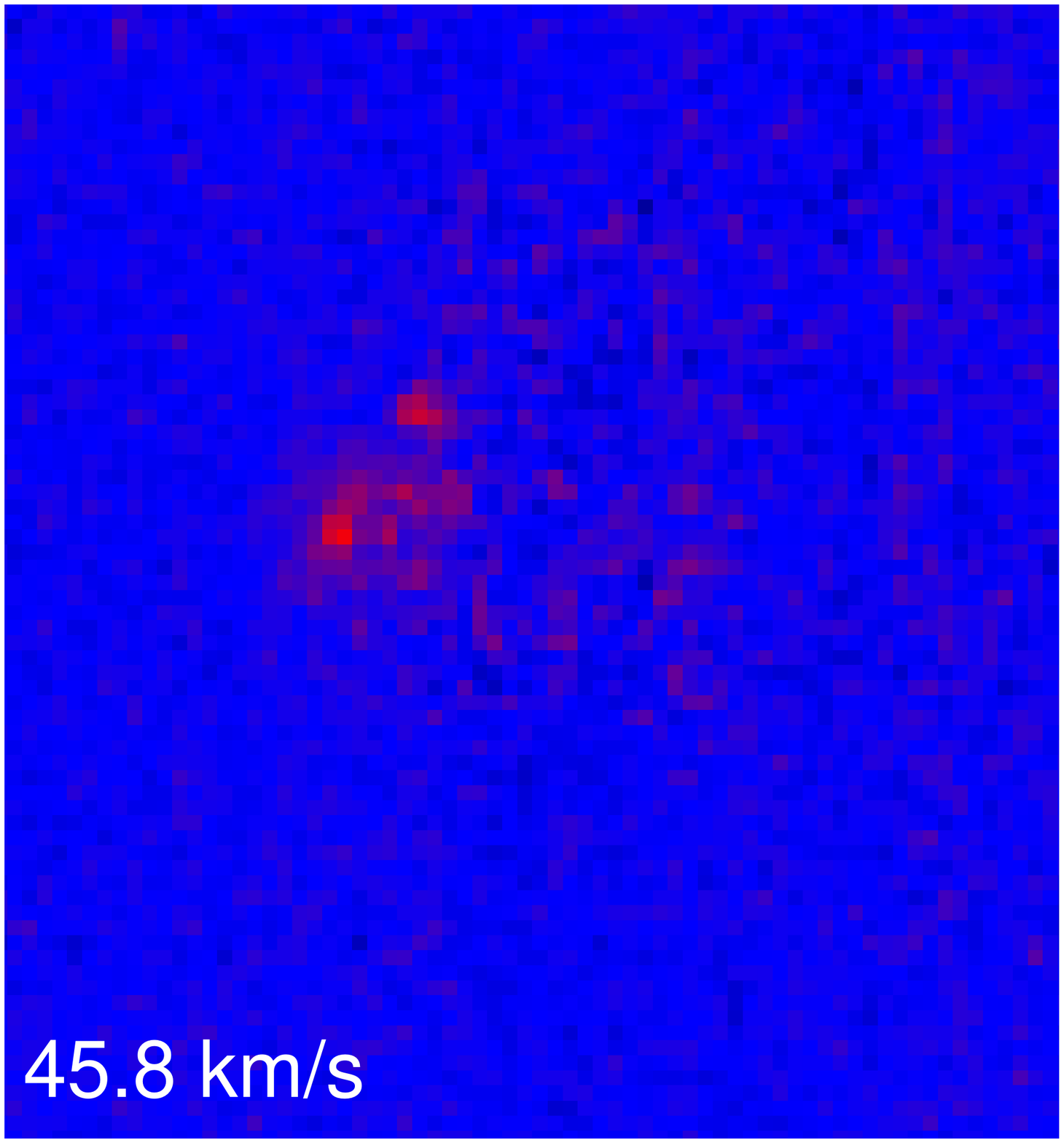} &
\hspace{2.50mm}
\includegraphics[scale=0.175, viewport=20 10 500 600]{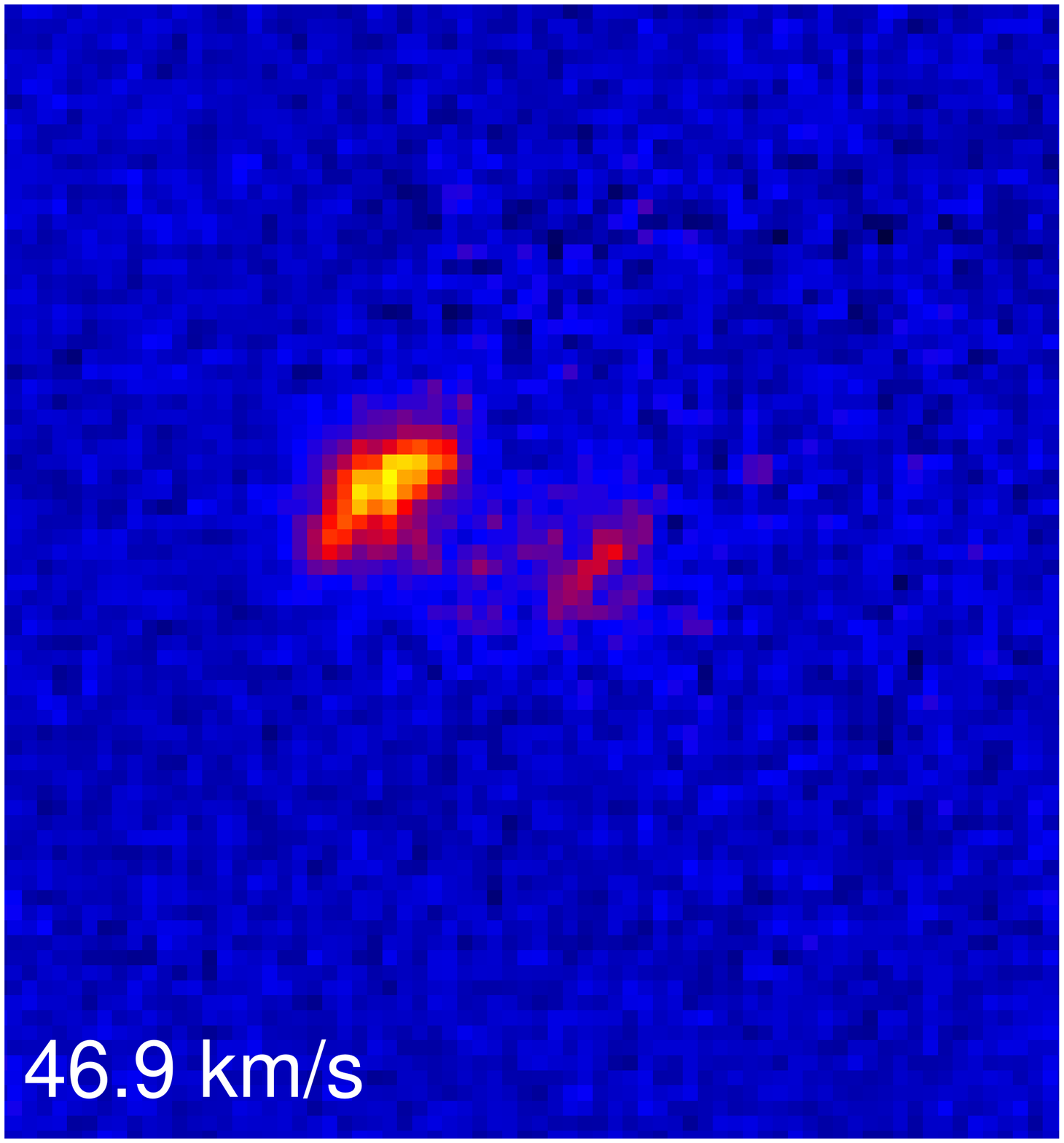} &
\hspace{2.50mm}
\includegraphics[scale=0.175, viewport=20 10 500 600]{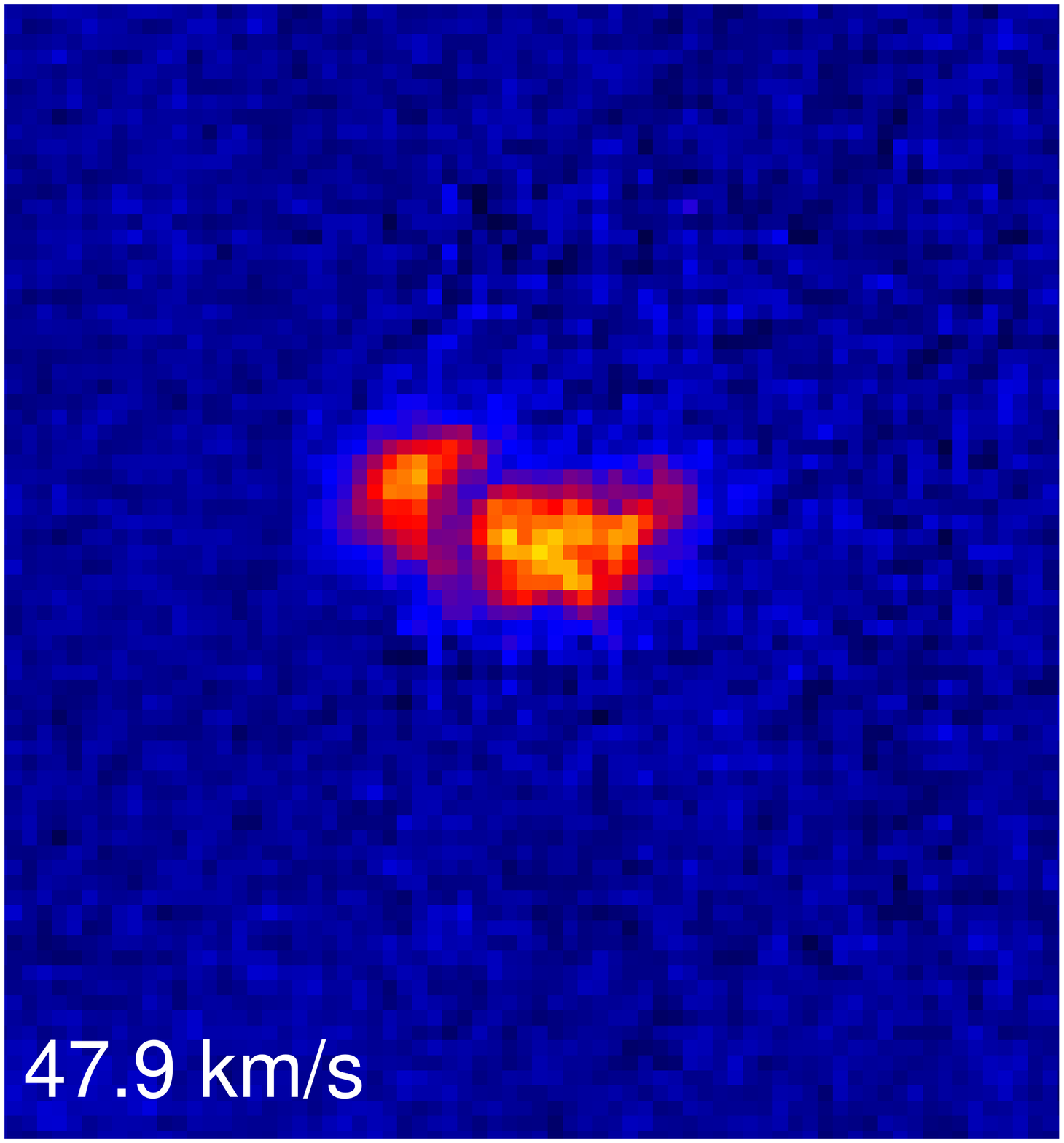} &
\hspace{2.50mm}
\includegraphics[scale=0.175, viewport=20 10 500 600]{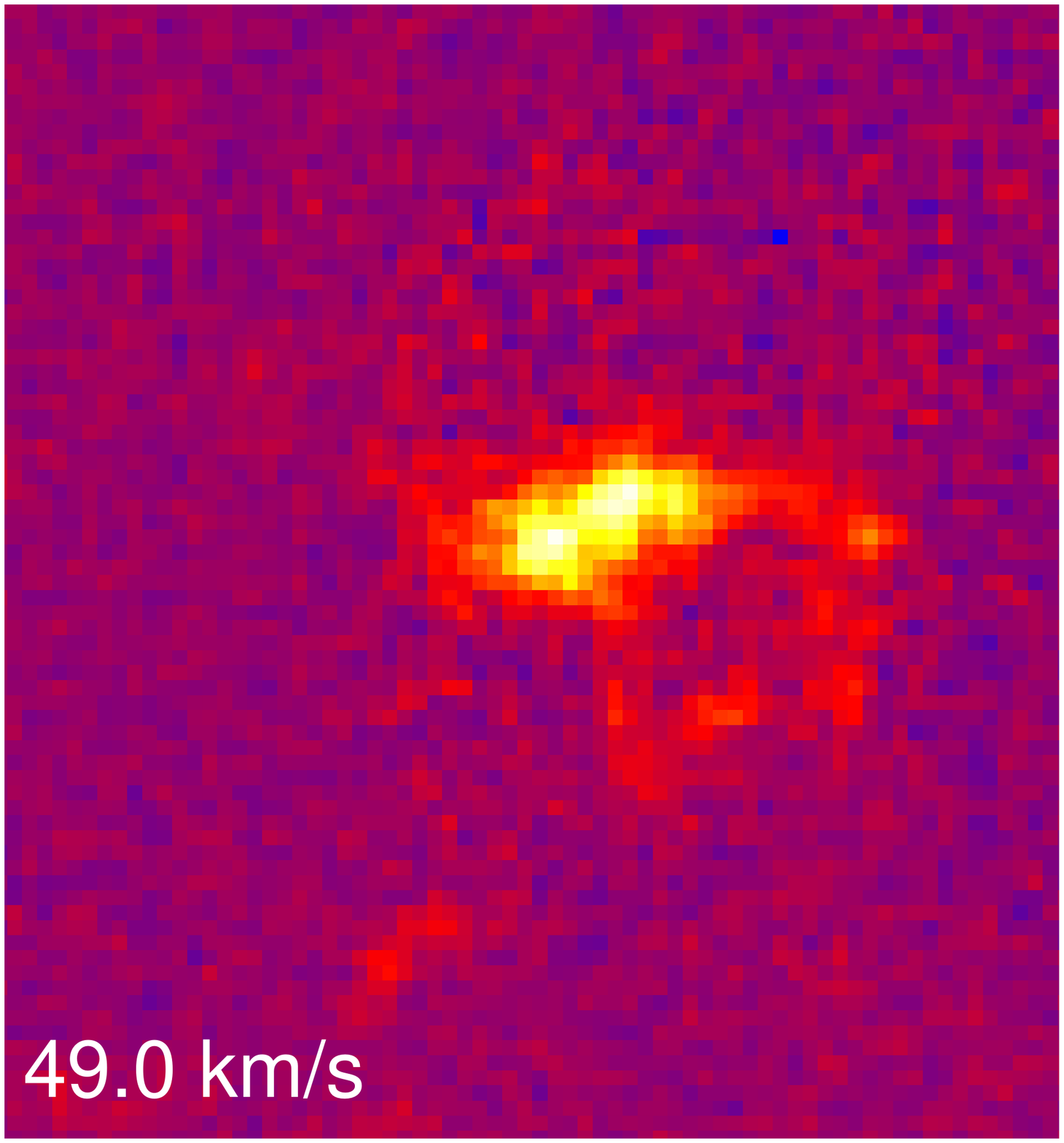} &
\hspace{2.50mm}
\includegraphics[scale=0.175, viewport=20 10 500 600]{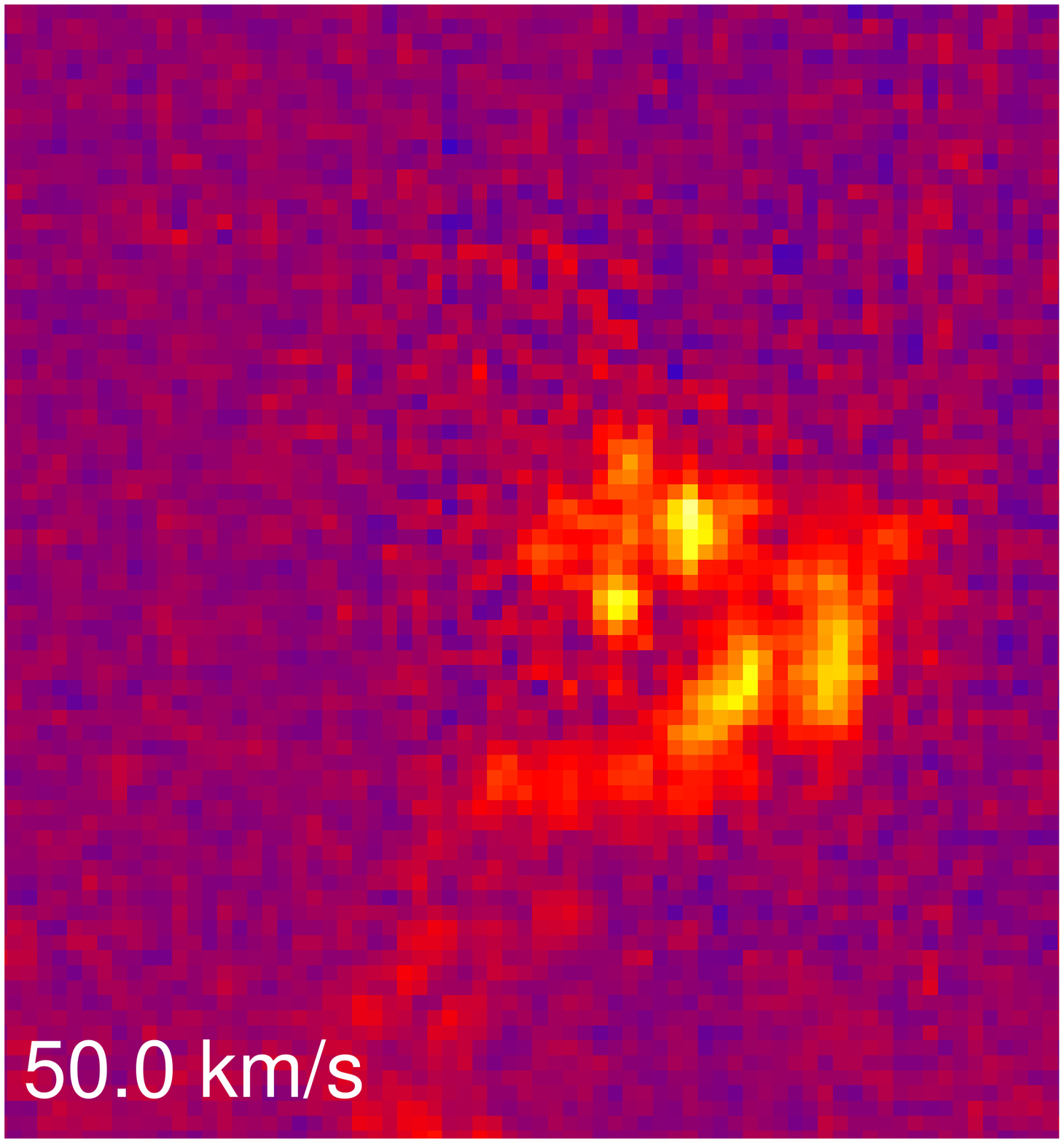} \\

\includegraphics[scale=0.175, viewport=20 100 500 600]{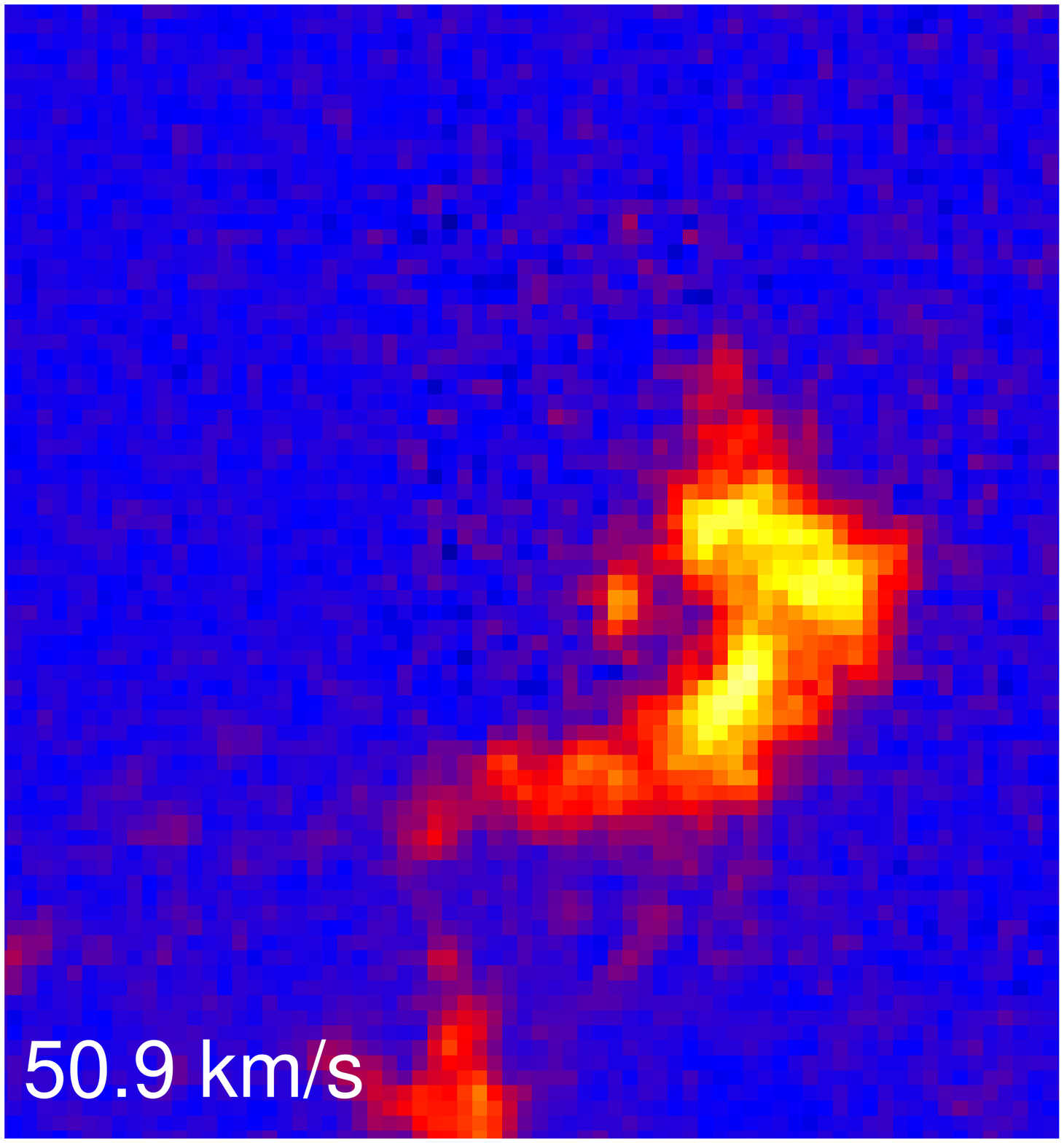} &
\hspace{2.50mm}
\includegraphics[scale=0.175, viewport=20 100 500 600]{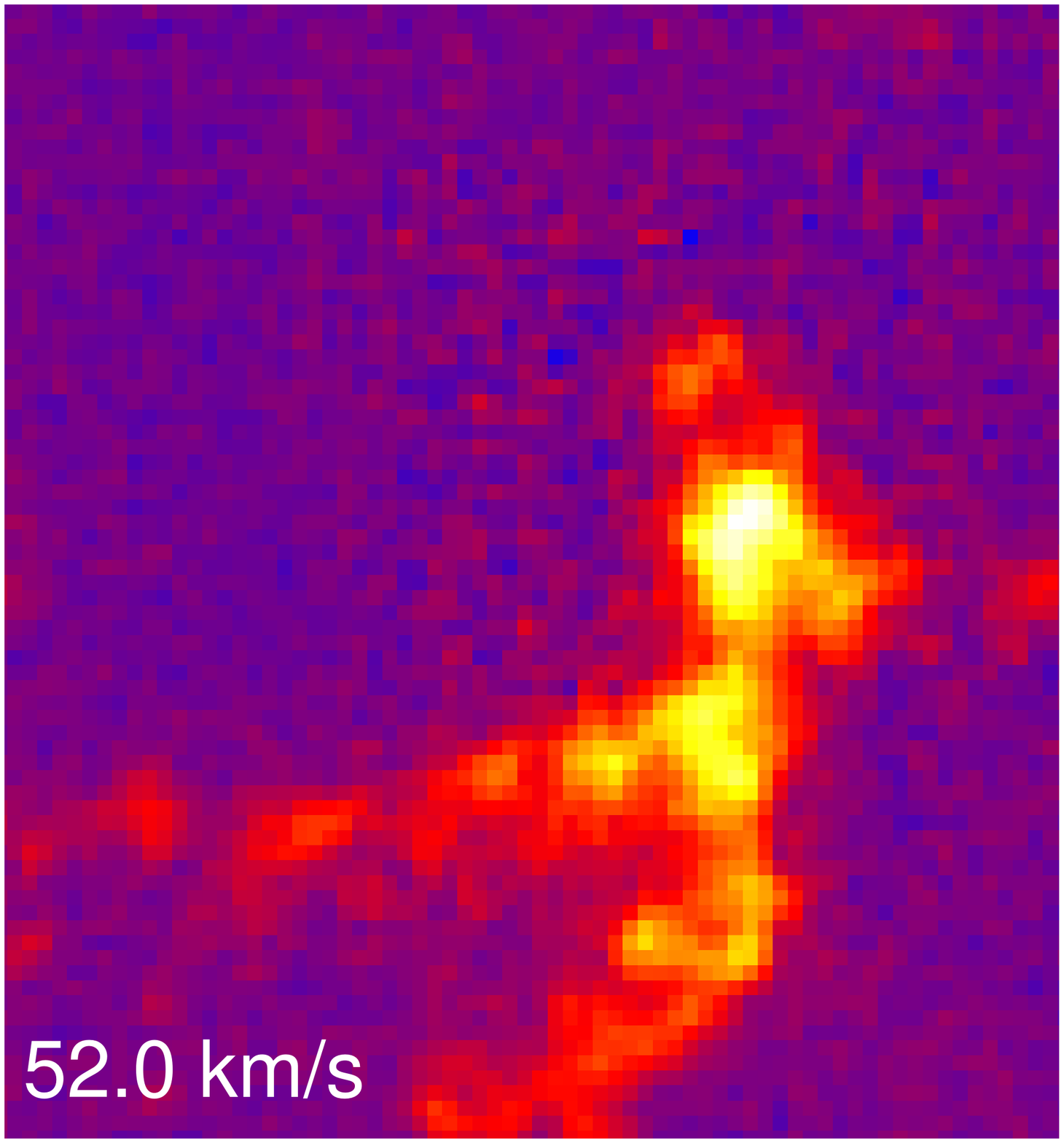} &
\hspace{2.50mm}
\includegraphics[scale=0.175, viewport=20 100 500 600]{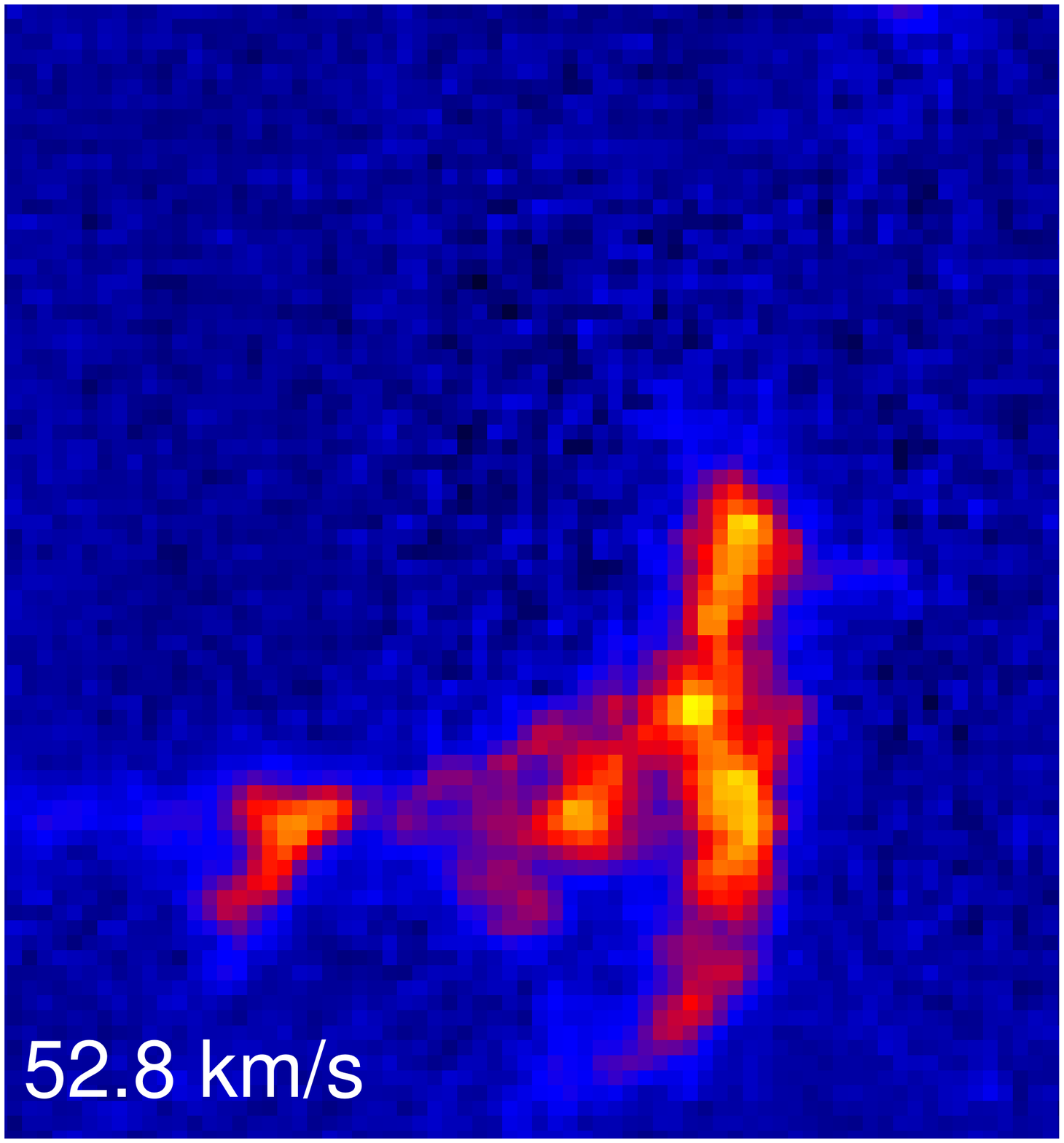} &
\hspace{2.50mm}
\includegraphics[scale=0.175, viewport=20 100 500 600]{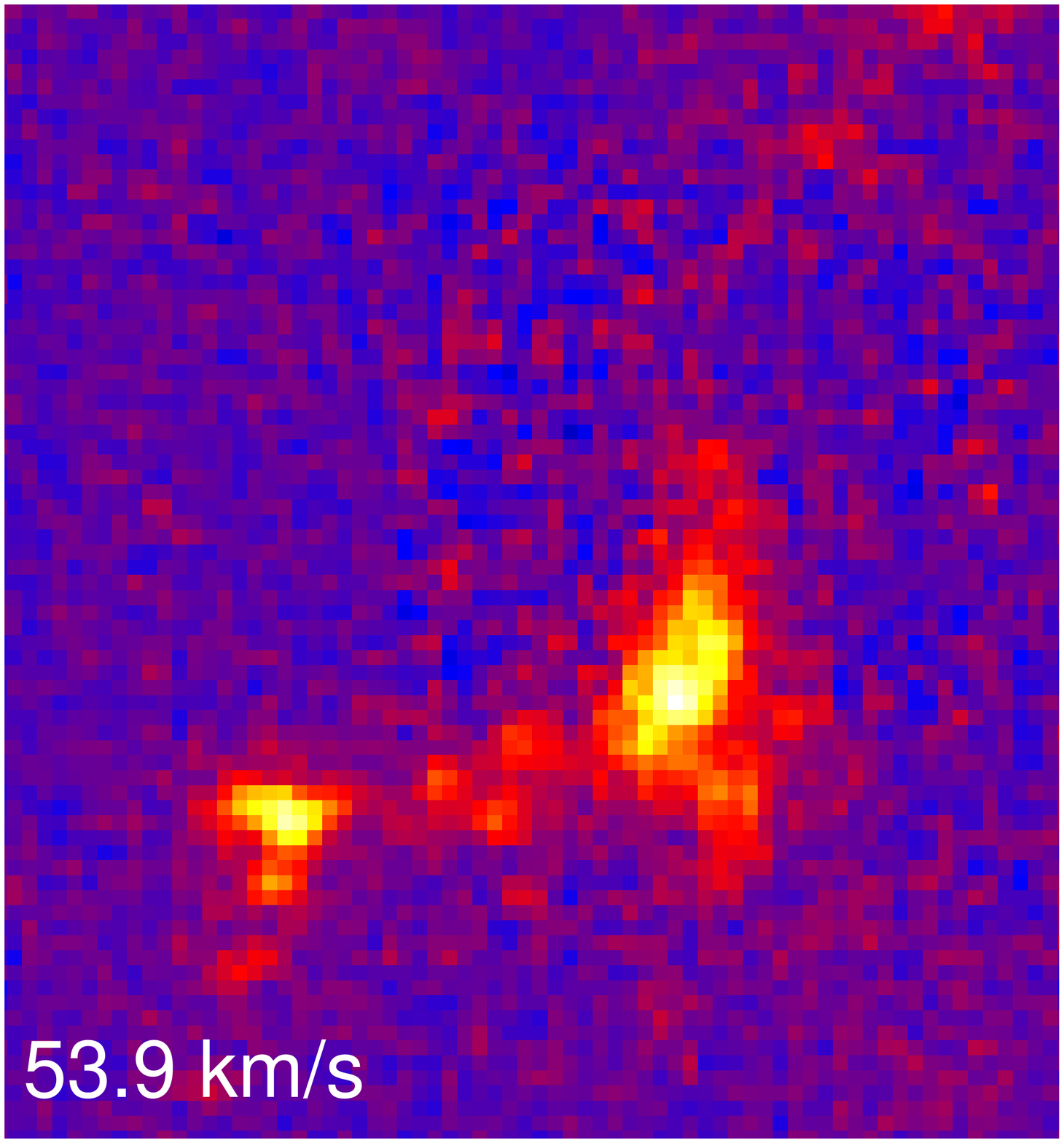} &
\hspace{2.50mm}
\includegraphics[scale=0.175, viewport=20 100 500 600]{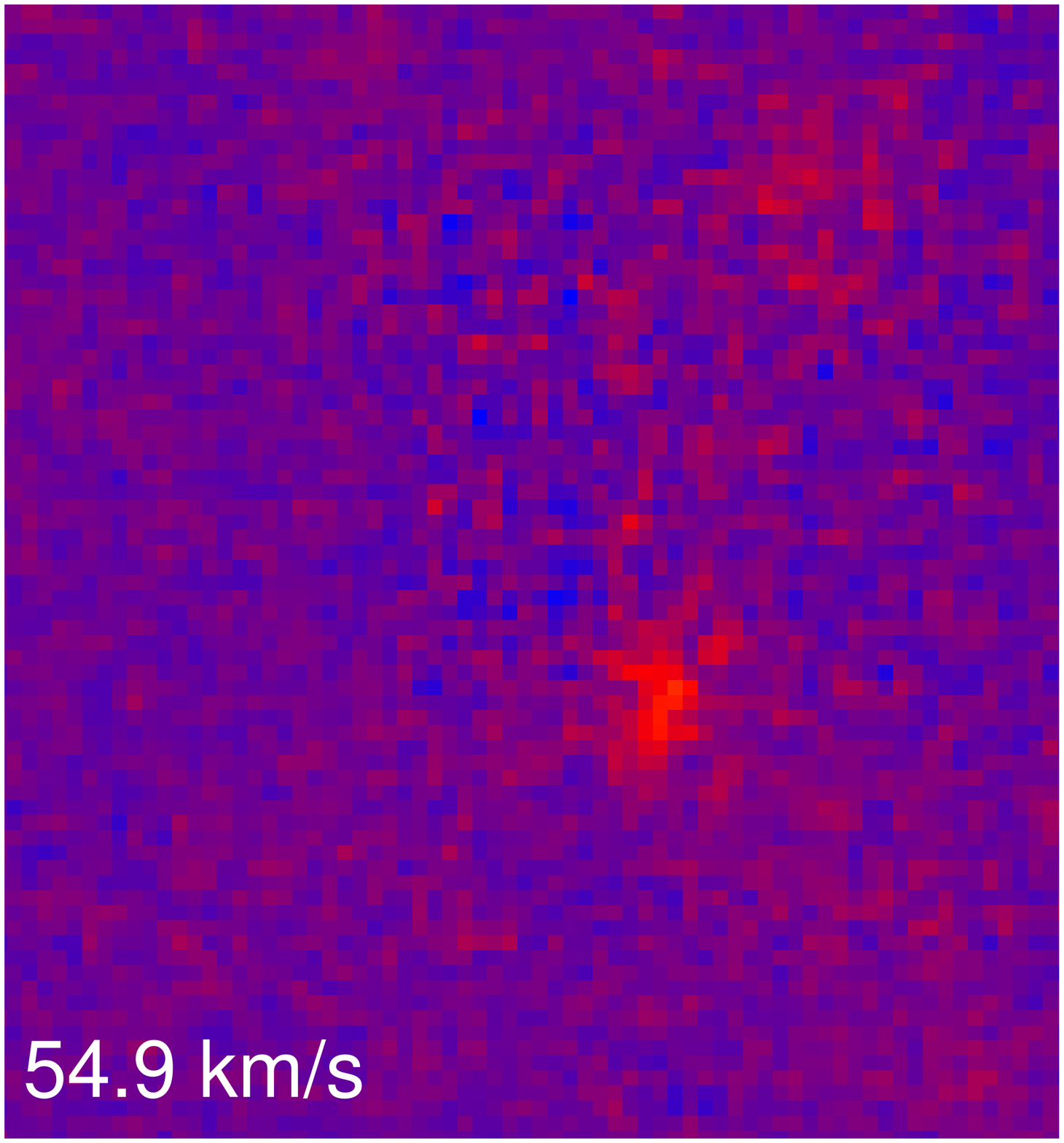} \\
\end{array}$
\end{center}
\caption{$^{13}$CO data at ten channels across RCW175. All maps are on the same co-ordinate system as Figure~\ref{Fig:3_color_RCW175}.}
\label{Fig:13CO_Data}
\end{figure*}

\begin{figure*}
\begin{center}$
\begin{array}{ccccc}
\includegraphics[scale=0.175, viewport=20 100 500 580]{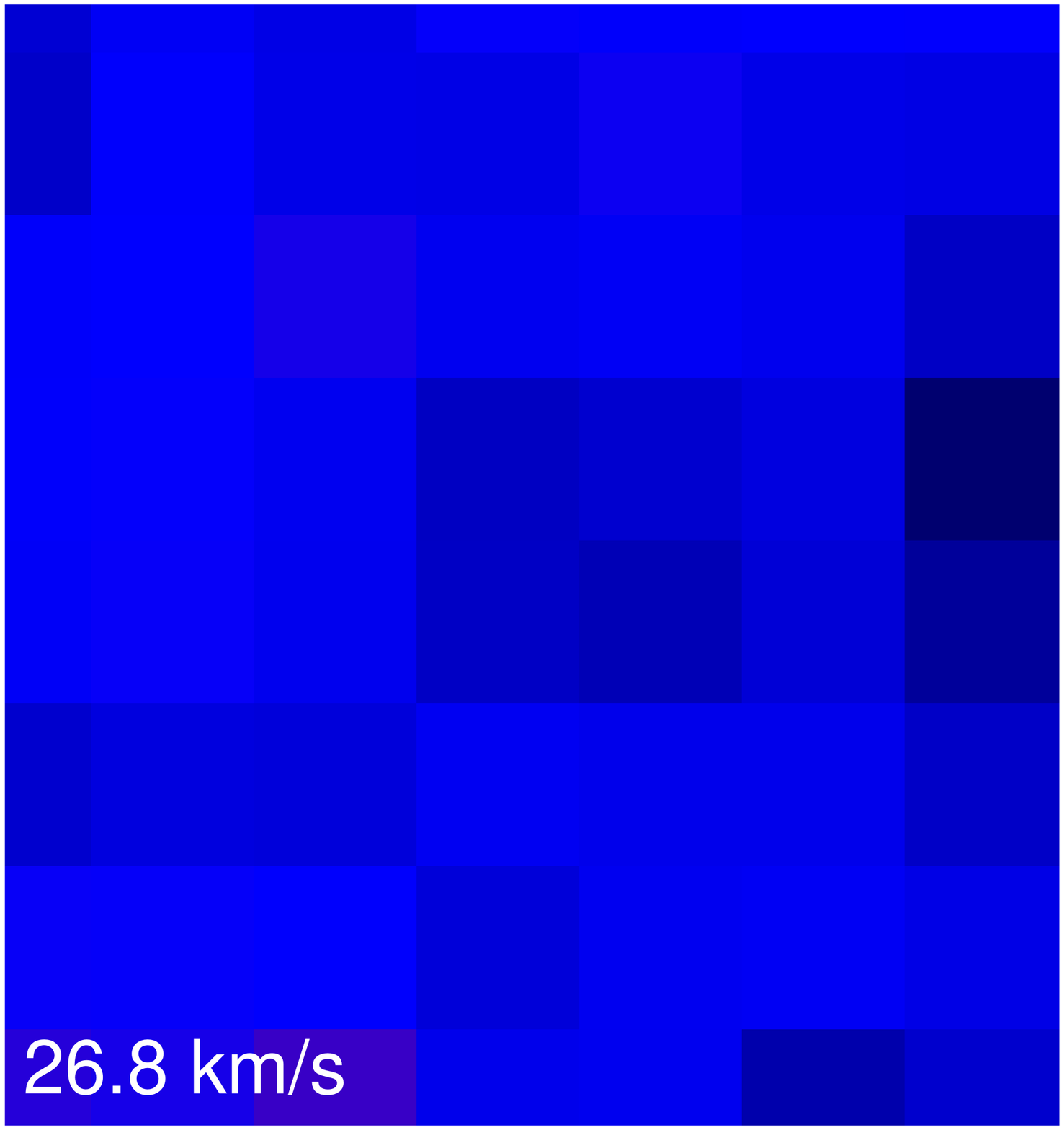} &
\hspace{2.50mm}
\includegraphics[scale=0.175, viewport=20 100 500 580]{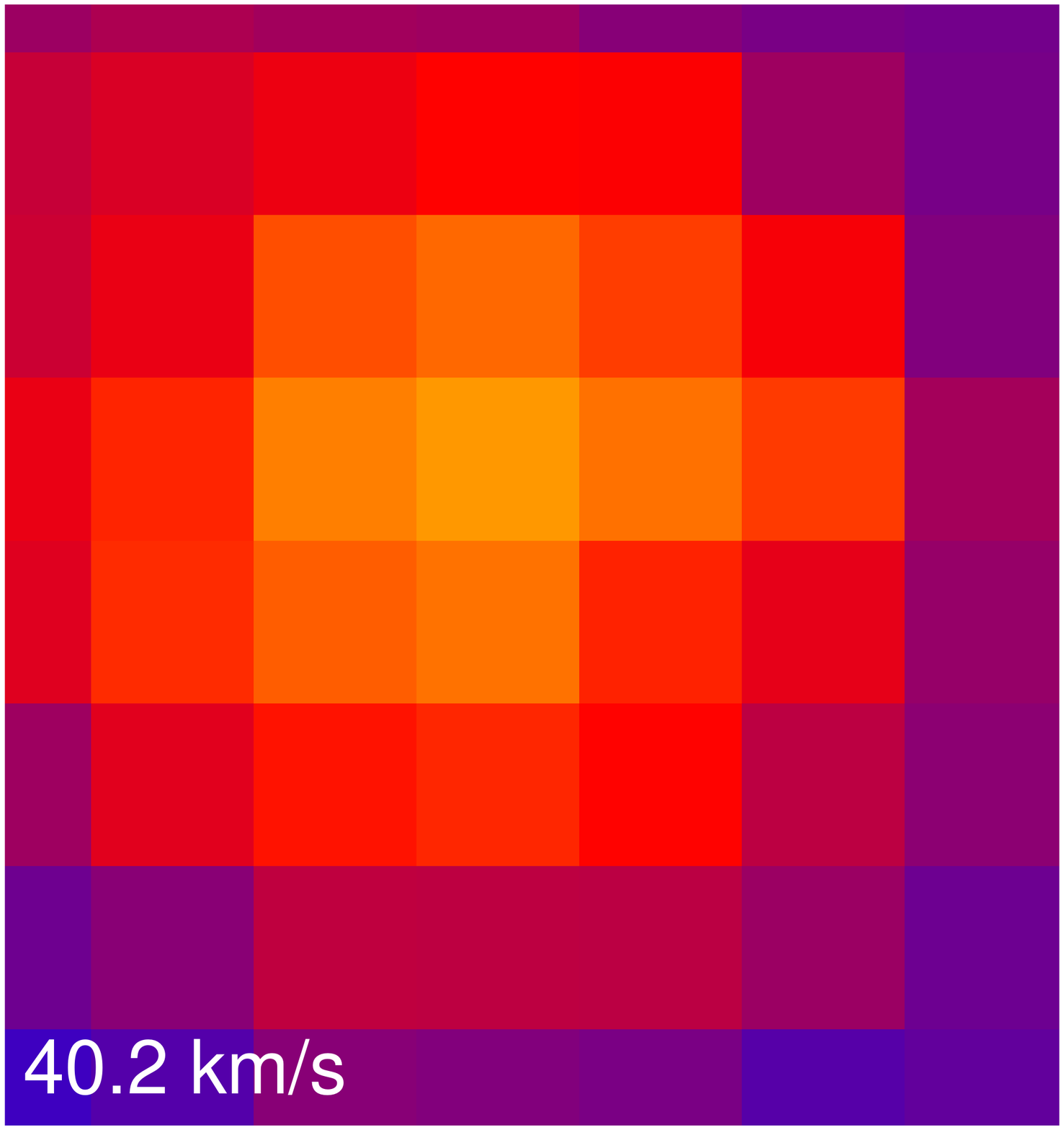} &
\hspace{2.50mm}
\includegraphics[scale=0.175, viewport=20 100 500 580]{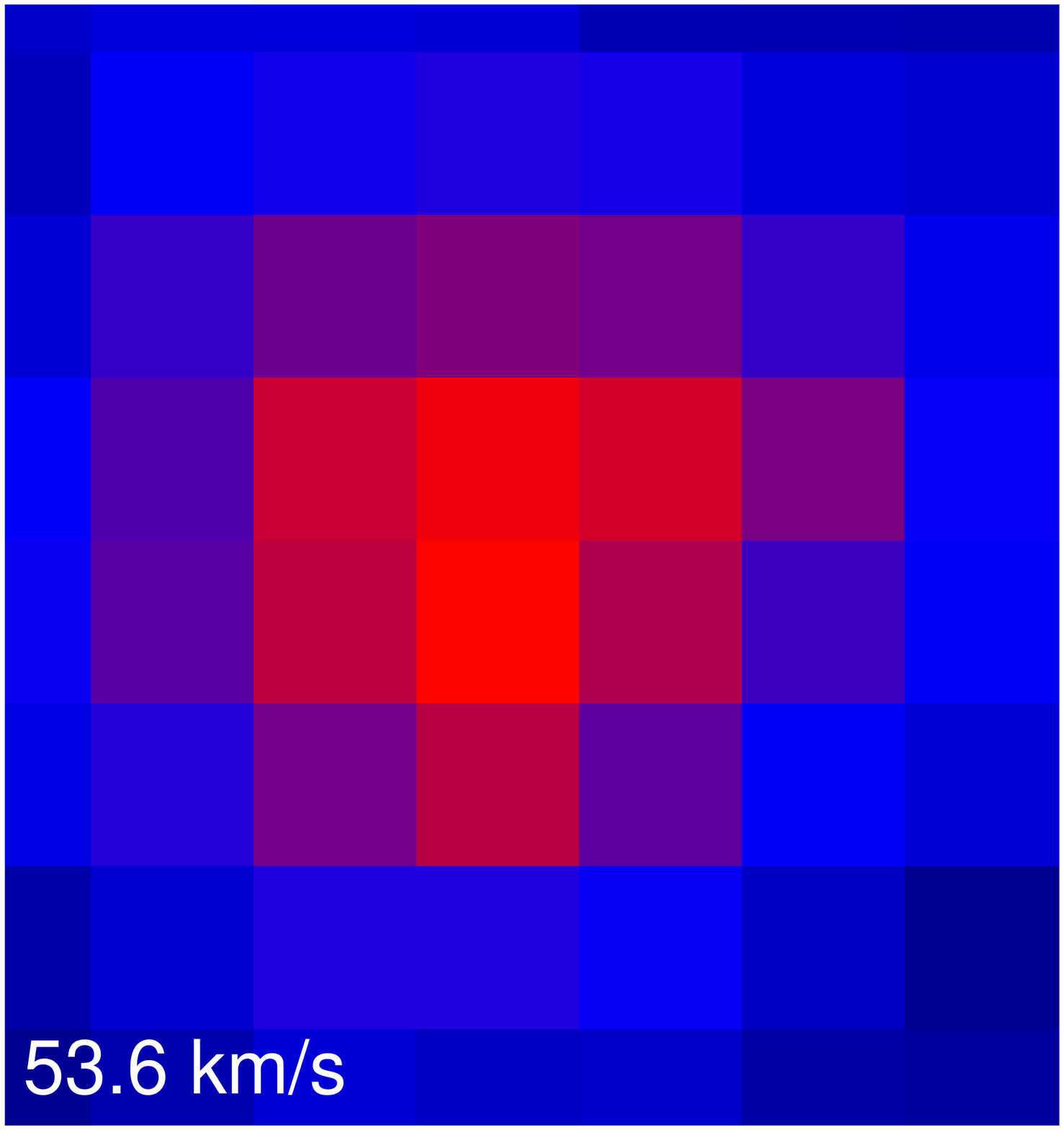} &
\hspace{2.50mm}
\includegraphics[scale=0.175, viewport=20 100 500 580]{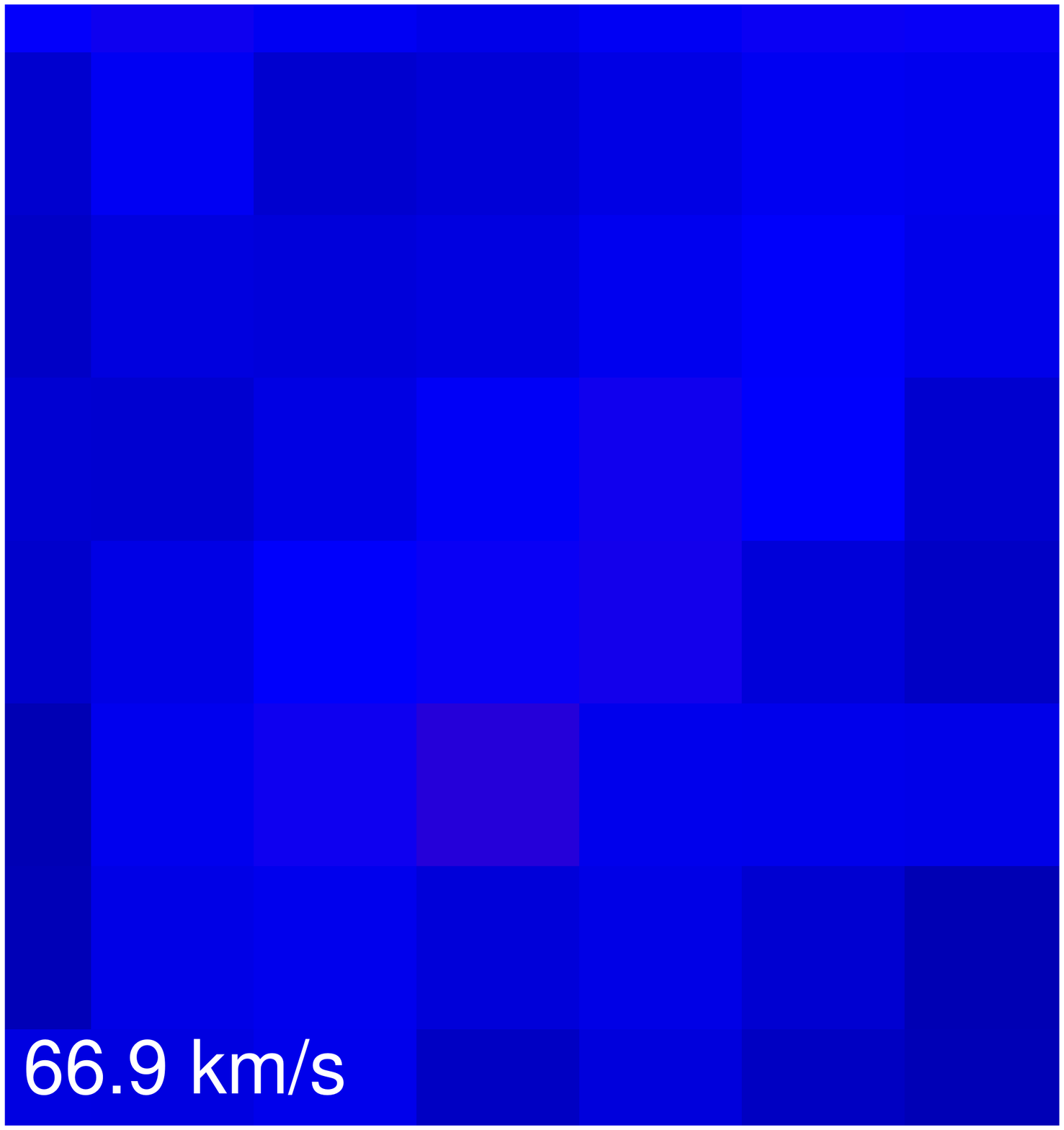} &
\hspace{2.50mm}
\includegraphics[scale=0.175, viewport=20 100 500 580]{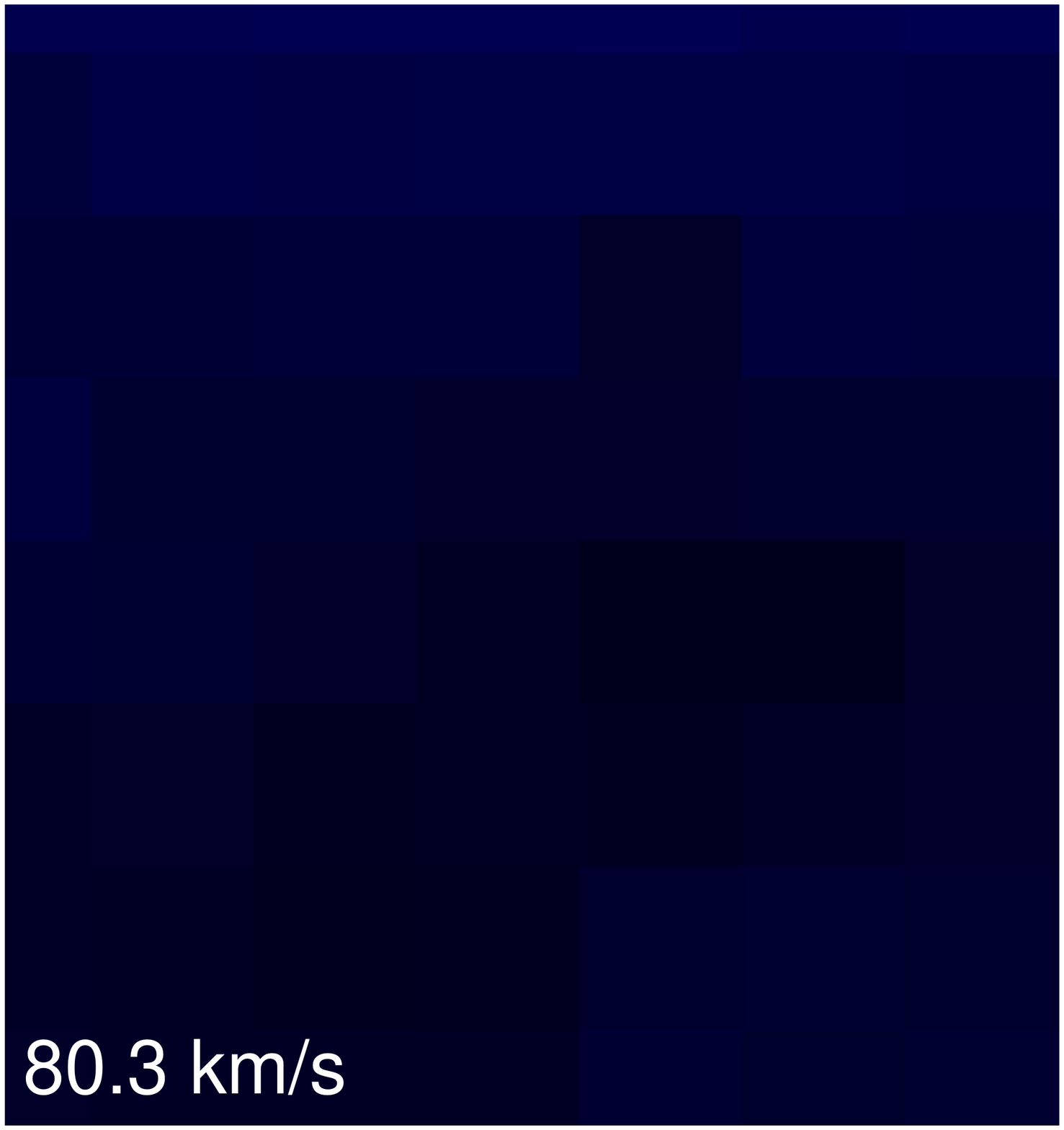} \\
\end{array}$
\end{center}
\caption{RRL data at five channels across RCW175. All maps are on the same co-ordinate system as Figure~\ref{Fig:3_color_RCW175}.}
\label{Fig:RRL_Data}
\end{figure*}

\subsubsection{H\textsc{i} Data}

The atomic hydrogen component of the ISM is known to be widely distributed throughout the Galaxy, and therefore, even though RCW175 is an H\textsc{ii} region, we still wish to investigate the atomic hydrogen in the vicinity of the source. H\textsc{i} data of RCW175 were obtained from the VLA Galactic Plane Survey~\citep{Stil:06}. These observations were performed as part of a H\textsc{i} and 21~cm continuum emission survey covering the longitude range 18$^{\circ}$~$<$~$l$~$<$~67$^{\circ}$, with a latitude coverage from |$b$|~$<$~1.3$^{\circ}$ to |$b$|~$<$~2.3$^{\circ}$. The short spacing information that is missing from the VLA interferometric observations is obtained by additional observations with the Green Bank Telescope. The H\textsc{i} spectral line data have an angular resolution of 1~arcmin, a spectral resolution of 1.56~km~s$^{-1}$ and an r.m.s. noise of 2~K.

\subsubsection{CO Data}

CO data is widely regarded as the most effective tracer of molecular hydrogen. With this in mind, we incorporate both $^{12}$CO and $^{13}$CO observations of RCW175 in this analysis.

The $^{12}$CO observations were performed as part of the Massachusetts~--~Stony Brook Galactic plane CO survey~\citep{Sanders:86}. These observations were performed at 2.6~mm~(115.3~GHz) and mapped the J=1--0 transition in the longitude range~8$^{\circ}$~$<$~$l$~$<$~90$^{\circ}$ and the latitude range |$b$|~$<$~1.05$^{\circ}$. These data have an angular resolution of~$\approx$~3~arcmin, a spectral resolution of 1~km~s$^{-1}$ and an r.m.s. sensitivity of~$\approx$~0.4~K.

The $^{13}$CO data were obtained from the Galactic Ring Survey~\citep{Jackson:06}. Mapping the J=1--0 transition~(2.7~mm or 110.2~GHz), these observations cover the longitude range~18$^{\circ}$~$<$~$l$~$<$~55.7$^{\circ}$, and the latitude range |$b$|~$<$~1$^{\circ}$. These data have an angular resolution of 46~arcsec, a spectral resolution of 0.21~km~s$^{-1}$ and an r.m.s. sensitivity of 0.13~K.

\subsubsection{Radio Recombination Line Data}

The Radio Recombination Line~(RRL) data were produced by~\citet{Alves:11} by combining the three RRLs found within the 64~MHz bandwidth of the Parkes Zone of Avoidance H\textsc{i} survey~\citep{Staveley-Smith:98} data. The three RRLs, H166$\alpha$, H167$\alpha$ and H168$\alpha$, occur at a wavelength of 21.0, 21.4 and 21.8~cm~(1.42, 1.40 and 1.37~GHz), respectively. The RRL data have an angular resolution of 14.8~arcmin, a spectral resolution of 20~km~s$^{-1}$ and an r.m.s. sensitivity of~$\approx$~0.003~K.


\section{Results and Discussion}
\label{Sec:Results_Discussion}


\subsection{Morphology}
\label{Subsec:Morphology}

In Figure~\ref{Fig:3_color_RCW175} we have identified the two separate components that make up RCW175. Both of these components, G29.0-0.6 and G29.1-0.7, appear shell-like: G29.1-0.7 is a larger, more diffuse component, $\approx$~8~arcmin across, while G29.0-0.6 is more compact and only~$\approx$~4~arcmin across.

Based on its shape, and using 20 and 90~cm VLA continuum data in conjunction with \textit{Midcourse Space Experiment}~(\textit{MSX}) 21~$\mu$m data, \citet{Helfand:06} tentatively classified G29.1-0.7 as a supernova remnant~(SNR) candidate. However, using all the available data presented in this analysis, we identify G29.1-0.7 as an H\textsc{ii} region, which is consistent with the recent analysis performed by~\citet{Alves:11}.

Figure~\ref{Fig:Data} displays maps of RCW175 from the radio all the way to the mid-IR. Although some of the radio maps lack the resolution to completely resolve both components, G29.0-0.6 is clearly the brighter component. This is true at all wavelengths, including the IR data. The \textit{Herschel} PACS and SPIRE IR emission originating from G29.1-0.7 is restricted mainly to a dust filament located along the edge of the shell visible in the IRAC 8~$\mu$m map, while G29.0-0.6 appears strongly in all the \textit{Herschel} bands. Although G29.0-0.6 is the dominant component, both components are visible at all the radio frequencies, indicating the presence of ionized gas in both of them. 

From Figure~\ref{Fig:3_color_RCW175} it is possible to see that the 24~$\mu$m emission appears to originate within both the diffuse shell of G29.1-0.7 and the compact shell of G29.0-0.6, while conversely the 8~$\mu$m emission largely avoids these regions, with the bulk of the 8~$\mu$m emission being concentrated around the edge of the shells in filamentary structures. This concept has previously been observed in other H\textsc{ii} regions~\citep[e.g.][]{Povich:07, Deharveng:10, Paladini:12}, and is generally explained in terms of the classical scenario of the formation of H\textsc{ii} regions: the ionized gas, produced by the ionizing OB star, expands into the surrounding environment due to pressure gradients. This expansion sweeps up the gas and dust creating an interface between the ionized gas and the surrounding material where photons of lower energy dominate, known as a photodissociation region~(PDR; \citealp[see][]{Hollenbach:97} for a complete review). In this hypothesis, the dust within the shell is warm, and thought to consist of small grains, while the smaller and less durable dust grains and large molecules, such as the Polycyclic Aromatic Hydrocarbons~(PAHs), may be destroyed by the harsh radiation field or swept up by the radiation pressure from the ionizing source. Since this scenario is consistent with what we observe in RCW175, we infer that RCW175 actually consists of two separate H\textsc{ii} regions.

In addition to the two-shell structure, also visible in Figure~\ref{Fig:3_color_RCW175} is the location of the B1~II type star, S65-4, which appears to be towards the very centre of G29.1-0.7. S65-4 is part of a five star cluster identified by~\citet{Forbes:89}, and is the only OB star in the~\citet{Reed:03} catalogue within the vicinity of RCW175. We, therefore, believe that S65-4 is the ionization source that produced the diffuse shell, G29.1-0.7. The lack of any known OB stars associated with G29.0-0.6 does not suggest that there are none present, but instead can probably be explained by the fact that this more compact component is much more heavily obscured by dust~--~as can be seen by looking at the longer wavelength IR maps in Figure~\ref{Fig:Data}.

Having identified what we believe to be two individual H\textsc{ii} regions within RCW175, we use the available spectral data to determine the relationship between them. By looking at the $^{12}$CO and $^{13}$CO spectral data displayed in Figures~\ref{Fig:12CO_Data} and~\ref{Fig:13CO_Data}, we determined that both G29.0-0.6 and G29.1-0.7 appear at a similar velocity of~$\sim$~50~km~s$^{-1}$. In addition to both components appearing at the same velocity, they also appear to be spatially associated with each other, with the CO emission in G29.1-0.7 restricted to the dust filament along the shell boundary and G29.0-0.6 being located at one end of this filament. This association implies that both components are at a similar distance and are actually interacting with each other.

In general, we find that the CO emission appears to spatially replicate the \textit{Herschel} PACS and SPIRE data, which is not entirely unexpected as both are tracing the denser environments within RCW175. As has previously been mentioned, although the shell structure of G29.1-0.7 can be seen at 8~$\mu$m, the longer wavelength PACS and SPIRE emission is concentrated in a dust filament along the edge of the shell, but does not entirely enclose the shell. Assuming the H\textsc{ii} region formation hypothesis, this implies that the ISM is much more dense in the direction of the filament, while in the other direction, the surrounding ISM provided less resistance to the radiation pressure and hence the swept up material formed a much lower density shell.

From Figure~\ref{Fig:RRL_Data} it is possible to identify an RRL component associated with RCW175. However, due to the coarse angular resolution of data, it is difficult to determine exactly where the emission is originating from within RCW175. The RRL data trace the ionized gas, therefore confirming its presence within the region. In fact,~\citet{Lockman:89} detected RRL emission from both G29.1-0.7 and G29.0-0.6, confirming that ionized gas is present in both components of RCW175, hence reinforcing the idea that both G29.1-0.7 and G29.0-0.6 are H\textsc{ii} regions.

To further understand the morphology of RCW175, we model the IR emission using a dust emission model.


\subsection{Dust Modeling}
\label{Subsec:Dust_Modeling}

In the past few years, there have been substantial improvements of dust models, which have significantly increased our understanding of the ISM dust properties. One such model has been developed by~\citet{Compiegne:11} and is called \textsc{DustEm}.\footnote{http://www.ias.u-psud.fr/DUSTEM/} \textsc{DustEm} is a sophisticated model that allows the user to parameterize various physical dust quantities to compute the corresponding emission and extinction properties. It has previously been used to characterize the dust properties in a variety of environments, such as regions of diffuse emission on the Galactic plane~\citep{Compiegne:10}, the star forming region the Eagle nebula~\citep{Flagey:11} and the Perseus molecular cloud~\citep{Tibbs:11}. 

In this analysis we use \textsc{DustEm} in a similar manner as was used in the previous works, modeling the dust in RCW175 in terms of five dust populations: small carbonaceous dust grains, large carbonaceous dust grains, large silicate dust grains and neutral and ionized PAHs. We merge the neutral and ionized PAHs into a single component of PAHs, use the small carbonaceous dust grains as the Very Small Grain~(VSG) dust component and combine the large carbonaceous and silicate dust grains to form a Big Grain~(BG) dust component. Since \textsc{DustEm} itself does not contain a fitting routine, we used an updated version of the SED-fitting routine developed for the analysis performed on the Perseus molecular cloud by~\citet{Tibbs:11}. This fitting routine is based on the \textsc{idl} routine MPFIT~\citep{Markwardt:09} and combines the dust emissivities computed by~\textsc{DustEm} with the instrumental bandpass filters, deriving an estimate of the flux that would be effectively observed at a given wavelength.

\begin{figure*}
\begin{center}$
\begin{array}{cc}
\includegraphics[scale=0.45, viewport=50 20 500 350]{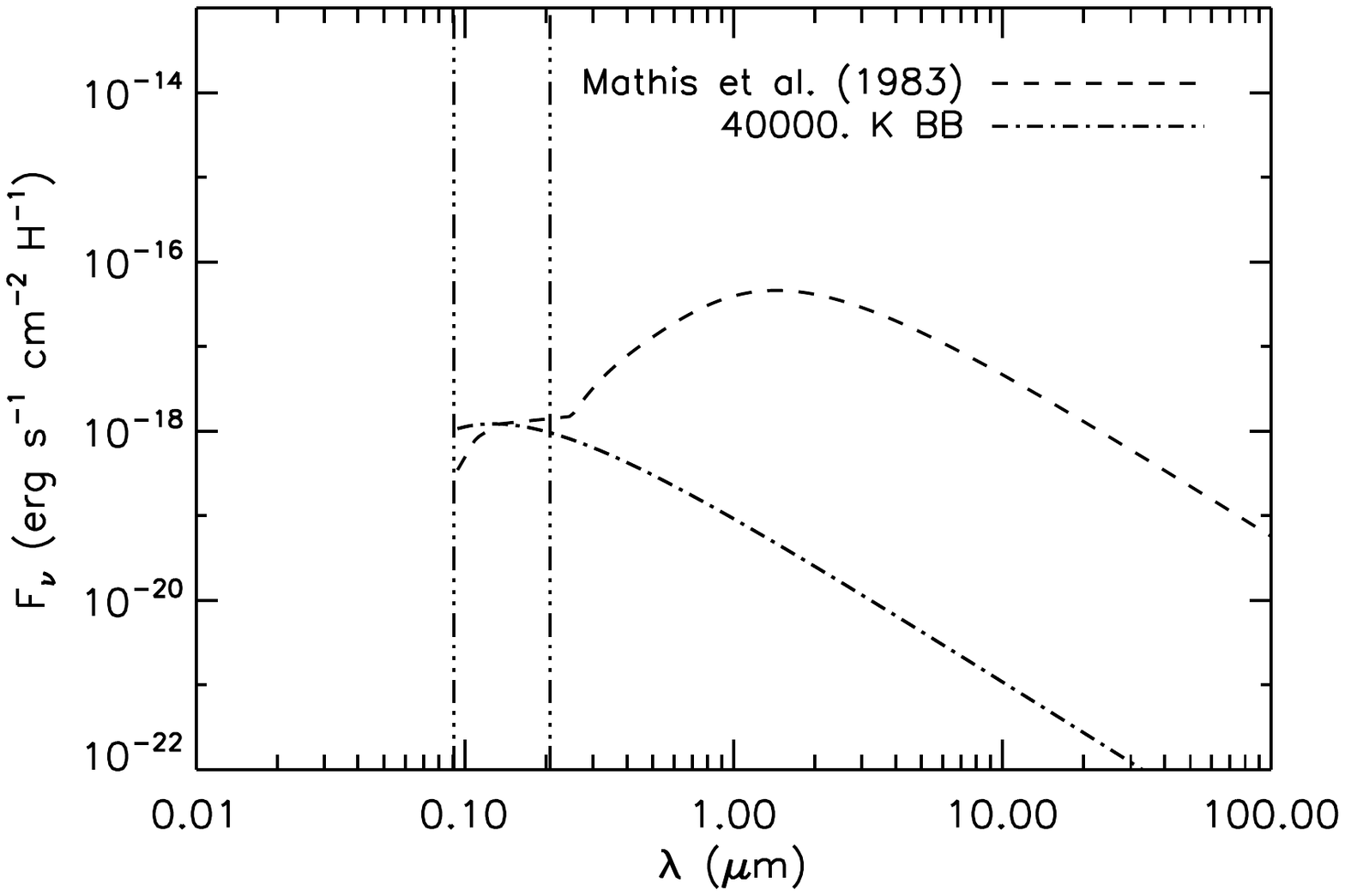} &
\includegraphics[scale=0.45, viewport=50 20 500 350]{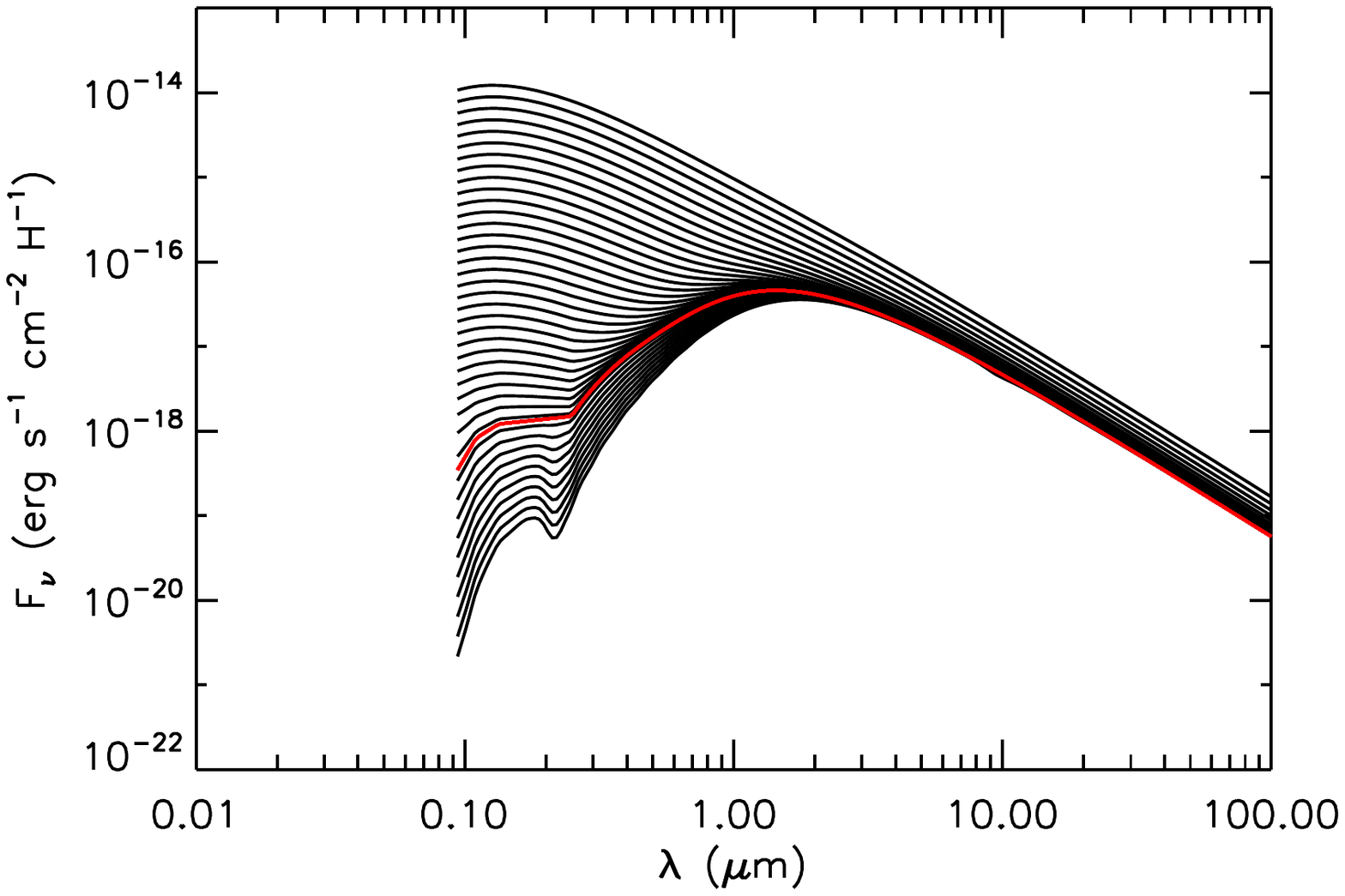} \\
\end{array}$
\end{center}
\caption{Plots showing the parameterization of the ERF. \textit{Left}: The two components used to parameterize the ERF: the standard \citet{Mathis:83} radiation field; and a 40,000~K blackbody representing the ionization source for the region. The two components are normalized to have the same flux between 6~--~13.6~eV. \textit{Right}: The full parameterization of the ERF for values from 0.05 to 10,000, illustrating the combination of the two components and the extinction applied to the \citet{Mathis:83} radiation field for values of $\chi_{ERF}$~$<$~1. The \citet{Mathis:83} radiation field~(i.e. $\chi_{ERF}$~=~1) is plotted in red.}
\label{Fig:ISRF}
\end{figure*}

For this analysis we take advantage of the new \textit{Herschel} data with their superior angular resolution, and ignore the IRIS and \textit{WMAP} data. Combining the \textit{Herschel} data with the IRAC 8~$\mu$m data and the MIPS 24~$\mu$m data, results in seven bands which provide a measure of the thermal dust emission across the entire IR spectrum. Therefore, combining \textsc{DustEm} with these seven IR bands allows us to characterize the three dust grain populations of interest: PAHs, VSGs and BGs. To do this, all seven IR maps were convolved to the common angular resolution of the SPIRE 500~$\mu$m map~(35~arcsec) and regridded onto a unified grid of 11.5~arcsec pixels. Running \textsc{DustEm} with our SED-fitting routine on a pixel-by-pixel basis, we fit the seven IR data points by adjusting the abundance of the PAHs and VSGs relative to the BGs~($Y_{PAH}$ and $Y_{VSG}$), the strength of the exciting radiation field~(ERF;~$\chi_{ERF}$), and the column density of hydrogen~($N_{H}$). In the previous work on the Perseus molecular cloud,~\citet{Tibbs:11} parameterized the radiation field by simply scaling the~\citet{Mathis:83} solar neighborhood radiation field, which is realistic given the lack of high-mass star formation in the cloud. However, since there is clearly a source of ionization present in RCW175, scaling the standard~\citet{Mathis:83} solar neighborhood field alone would not be a good approximation. Therefore, we decided to parameterize the ERF by combining the~\citet{Mathis:83} radiation field with a 40,000~K blackbody, which is the typical temperature of an OB star, representing the source of ionization. To combine these two separate components of the ERF, we normalized the 40,000~K blackbody to the~\citet{Mathis:83} radiation field between 6 and 13.6~eV~(see Figure~\ref{Fig:ISRF}). In addition, given the presence of dust in RCW175, we implement an extinguishing of the radiation field. The final parameterization of the ERF is as follows:

\begin{itemize}
\item $\chi_{ERF}$ $<$ 1 $\Rightarrow$ use the \citet{Mathis:83} radiation field extinguished by dust \\
\item $\chi_{ERF}$ = 1 $\Rightarrow$ use the \citet{Mathis:83} radiation field \\
\item $\chi_{ERF}$ $>$ 1 $\Rightarrow$ use the \citet{Mathis:83} radiation field plus a 40,000~K blackbody
\end{itemize}

In Figure~\ref{Fig:ISRF} it is possible to see the parameterization of the ERF for the full range of values used in this analysis. Also displayed is the~\citet{Mathis:83} radiation field which represents the boundary between values of~$\chi_{ERF}$ which are greater than, or less than, equal to 1. Regarding our parameterization of the ERF, from Figure~\ref{Fig:ISRF} it is apparent that the dust extinction only affects the shorter wavelengths and that for values of $\chi_{ERF}$~$\gtrsim$~2500, the blackbody becomes totally dominant.

\begin{figure*}
\begin{center}
\includegraphics[scale=0.45, viewport=0 0 500 330]{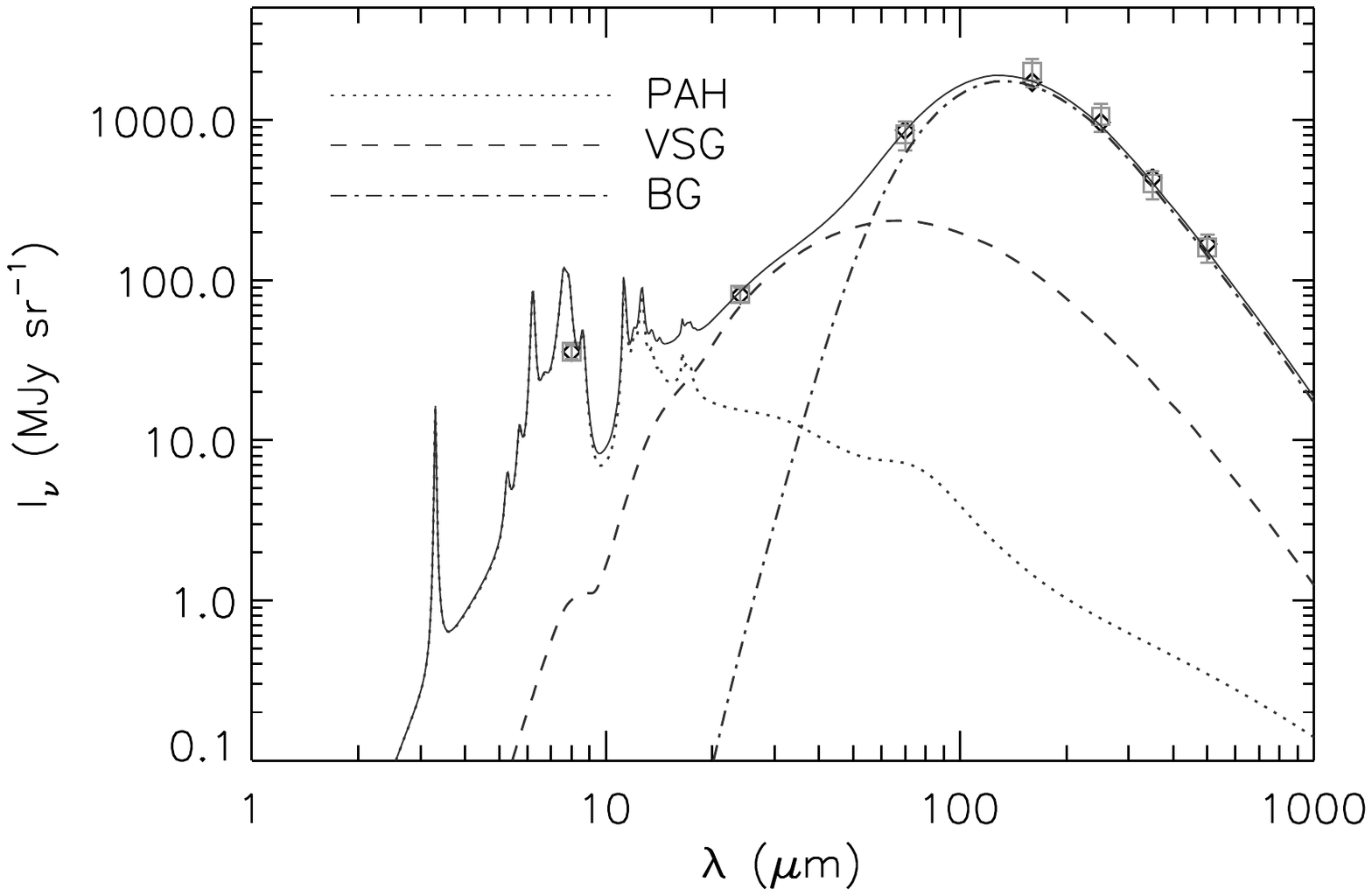} 
\end{center}
\caption{Spectrum of a single pixel located within the centre of the shell of G29.1-0.7 showing the original IR data points~(\textit{squares}) and the modeled best fitting values~(\textit{diamonds}). Also displayed are the three dust components required to reproduce the observed emission and the total IR emission.}
\label{Fig:DUSTEM_Spectra}
\end{figure*}

The result of running our SED-fitting routine and \textsc{DustEm} on the RCW175 data was parameter maps of $Y_{PAH}$, $Y_{VSG}$, $\chi_{ERF}$ and $N_{H}$. An example of one of the fits obtained is displayed in Figure~\ref{Fig:DUSTEM_Spectra}. This is the spectrum for a pixel within the centre of the shell of G29.1-0.7 and it shows the IR data, the best fitting values from our fitting routine and the three dust components required to reproduce the observed emission.

\begin{figure*}
\begin{center}$
\begin{array}{ccc}
\includegraphics[scale=0.320,viewport=150 100 680 700]{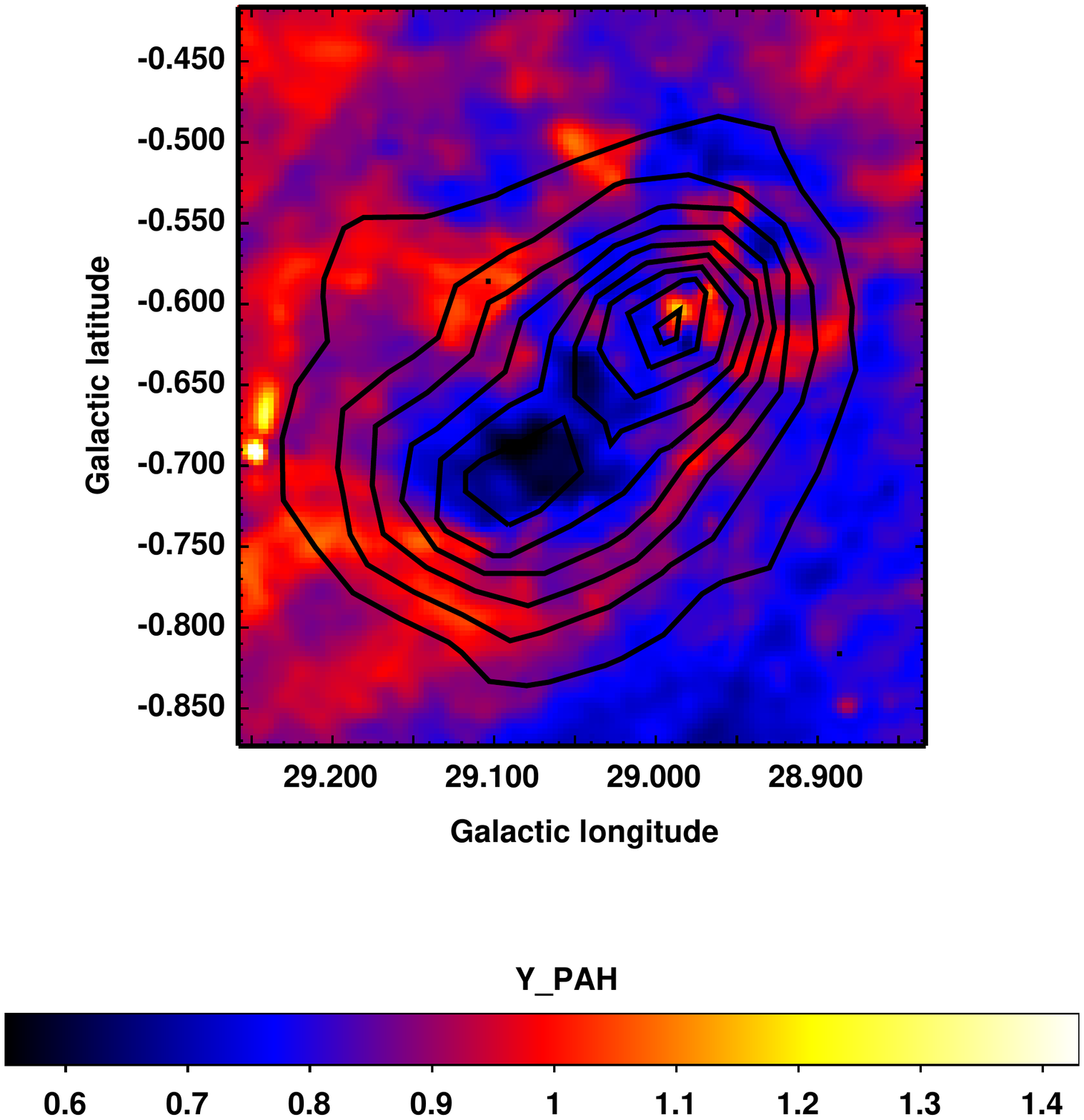} &
\includegraphics[scale=0.320,viewport=150 100 680 700]{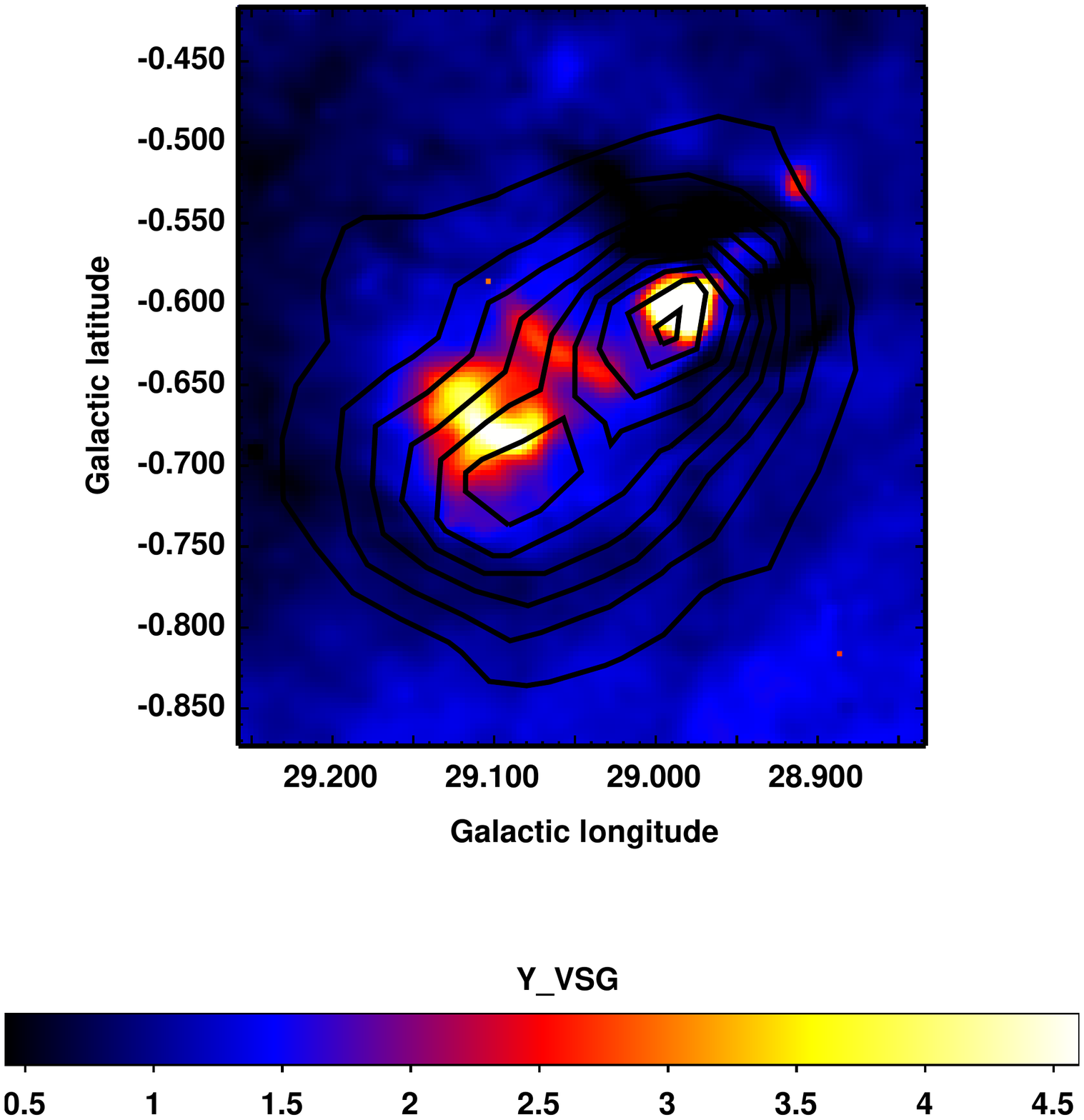} &
\includegraphics[scale=0.320,viewport=150 100 680 700]{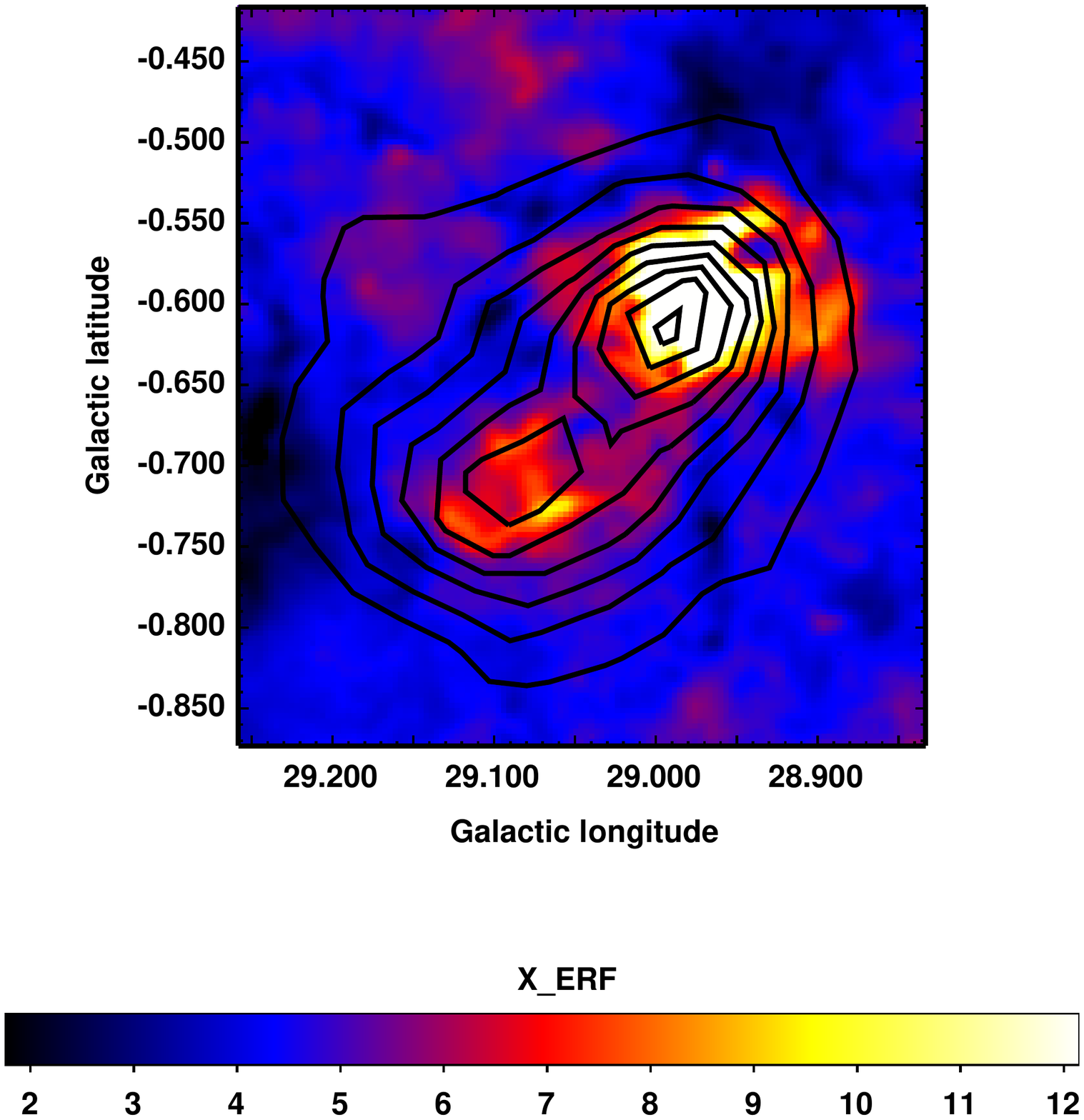} \\
\end{array}$
$
\begin{array}{cc}
\includegraphics[scale=0.320,viewport=0 100 680 700]{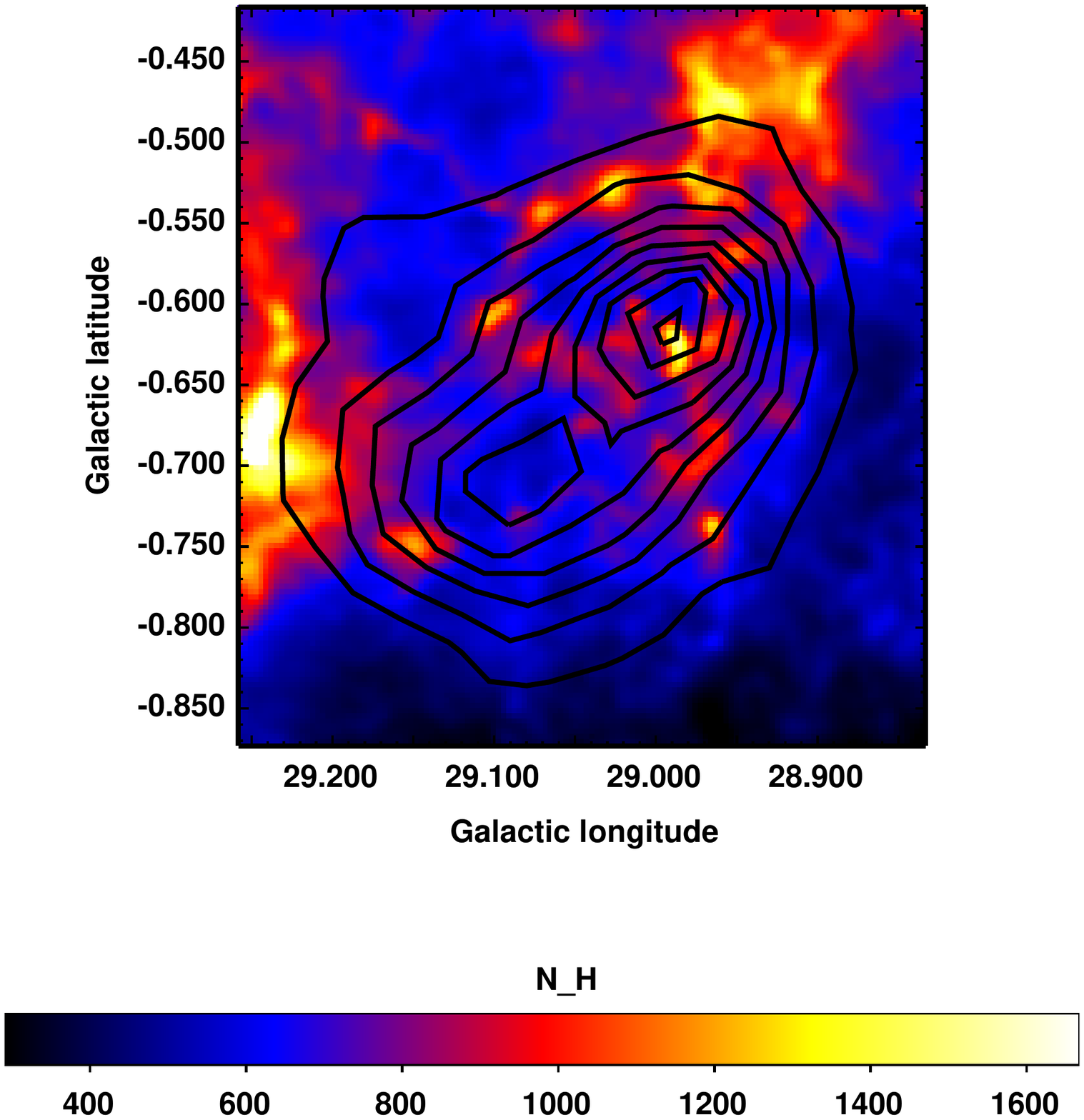} &
\includegraphics[scale=0.320,viewport=100 100 680 700]{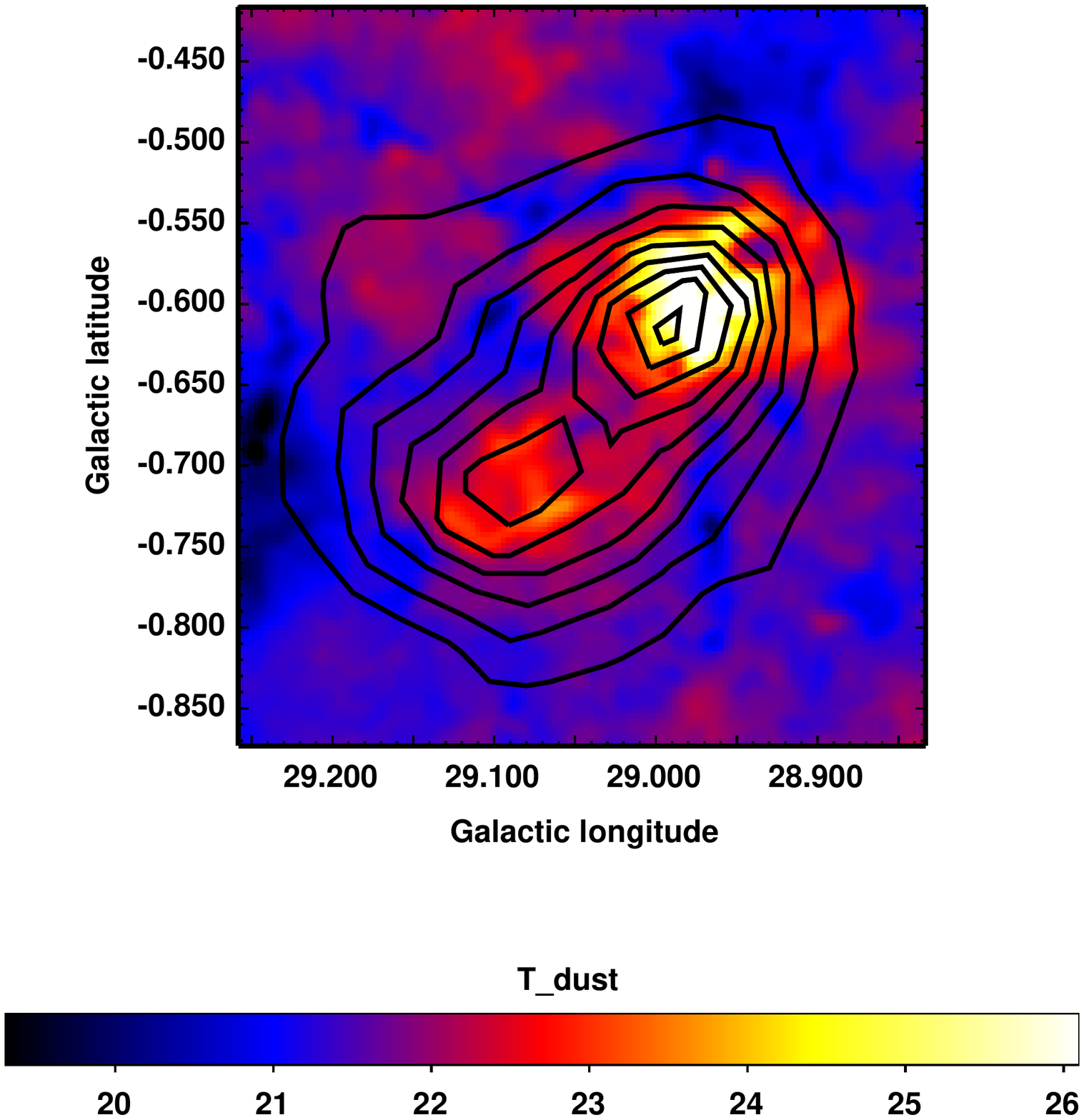} \\
\end{array}$
\end{center}
\caption{Parameter maps ($Y_{PAH}$, $Y_{VSG}$, $\chi_{ERF}$, $N_{H}$ and $T_{dust}$) of RCW175 produced as a result of fitting the IR data using the \textsc{DustEm} model. $Y_{PAH}$ and $Y_{VSG}$ are given relative to the value for the diffuse high Galactic latitude~\citep[see][]{Compiegne:11}, $\chi_{ERF}$ is dimensionless and $N_{H}$ is is units of 10$^{20}$~H~cm$^{-2}$. Also displayed are the CBI contours at 10, 20, 30, 40, 50, 60, 70, 80, 90~$\%$ of the peak flux which is 0.98~Jy~beam$^{-1}$, which demonstrate that the cm-wave emission is correlated with $\chi_{ERF}$ and $T_{dust}$. All maps are at a common angular resolution of 35~arcsec.}
\label{Fig:Param_Maps_Contours}
\end{figure*}

In addition to producing the four parameter maps, \textsc{DustEm} combines the optical and thermal properties of the grains with the given ERF and calculates a temperature distribution for each size bin for each grain species. These temperature distributions are computed taking into account the multi-photon absorption processes, based on the method described by~\citet{Desert:86}. Therefore, to complement the parameter maps, we computed the median temperature of the BG size distribution, and used it as an estimate of the equilibrium temperature of the dust~($T_{dust}$). We emphasize that it is unphysical to compute a median temperature for the other dust components as these are stochastically heated and hence are not in thermal equilibrium with the ERF. This $T_{dust}$ map, along with the four parameter maps, is displayed in Figure~\ref{Fig:Param_Maps_Contours}. 

These maps allow us to better understand the dust properties and environment within RCW175. Looking at Figure~\ref{Fig:Param_Maps_Contours} we find that the abundances of the PAHs and VSGs relative to the BGs vary between~$\sim$~0.5~--~1.5 and~$\sim$~0.5~--~10, respectively. The maps also show that the VSGs are concentrated in the interior of both G29.1-0.7 and G29.0-0.6, while the PAHs are constrained to the edge of both components. There is also a clear deficit of PAHs within G29.1-0.7. As discussed in Section~\ref{Subsec:Morphology}, this is consistent with the current hypothesis of H\textsc{ii} region formation, with the PAHs confined to the PDR and the warmer VSGs present inside the shell.

The strength of the ERF map reveals a range of values from~$\sim$~2~--~30, with the highest values found towards G29.0-0.6, implying that there is a significant source of excitation present. This is again consistent with what we inferred from the radio and IR maps, and suggests that G29.0-0.6 is also an H\textsc{ii} region. In G29.1-0.7, the ERF appears to be offset from the centre of the shell and concentrated toward the dust filament along the edge of the shell. 

The column density map reveals that the highest densities~($\sim$~1.5~$\times$~10$^{23}$~H~cm$^{-2}$) are also found in the vicinity of G29.0-0.6, implying that there are large quantities of dust present which we believe is the reason for a lack of observed OB stars associated with this component. Also observable is that the density is decreased toward the centre of the diffuse shell of G29.1-0.7 with values of~$\sim$~5~$\times$~10$^{22}$~H~cm$^{-2}$.

The dust temperature map closely resembles the ERF map, which is expected given the relationship between these two physical quantities, and shows that the dust temperature in RCW175 is in the range of 20~--~30~K, which is typical for H\textsc{ii} regions. This map also illustrates that the highest temperatures are found towards G29.0-0.6.

Having already identified that these two H\textsc{ii} regions are associated with each other, and based on the fact that G29.0-0.6 is more compact, has a stronger ERF, a higher dust temperature, and a higher column density, we believe that G29.0-0.6 is younger than the more evolved G29.1-0.7, and hypothesize that it may even have formed as a result of the expansion of G29.1-0.7 into the surrounding ISM. Such triggered star formation is ubiquitous throughout the Galaxy~\citep[e.g.][]{Deharveng:05, Zavagno:06} essentially because the gravitational timescale of molecular gas is shorter than the typical lifetime of an O-type star.~\citet{Elmegreen:98} identified three distinct triggering mechanisms: i) direct compression of pre-existing density enhancements in a cloud; ii) accumulation of gas into a dense filament that collapses gravitationally into dense cores~("collect and collapse"); and  iii) cloud collisions. Given the environment of RCW175 and the significant differences in size, and hence age, it is unlikely that this was produced as the result of cloud collisions. Assuming that G29.1-0.7 formed first, then the expansion into the surrounding ISM could have triggered the formation of G29.0-0.6 by either causing an existing density enhancement to collapse or by the collect and collapse method. It is difficult to determine which method is more favorable, however, given that G29.0-0.6 is located at the edge of the shell of G29.1-0.7, it is likely that triggering has occured. This idea of triggered star formation will be developed further in Section~\ref{Subsec:YSOs}.


\subsection{Kinematic Distances}
\label{Subsec:Kinematic_Distances}

Previous estimates of the distance to RCW175 have been presented in the literature. Using the rotation curve for the northern hemisphere given by~\citet{Georgelin:76} with values of $R_{0}$~=~10~kpc and $\Theta_{0}$~=~250~km~s$^{-1}$, \citet{Crampton:78} found a kinematic distance of 3.5~kpc to RCW175. Assuming a value of the ratio of total to selective absorption $R_{V}$~=~3.0, \citet{Forbes:89} computed the reddening and distance to the ionizing B1~II~star in RCW175 using the intrinsic colors given by~\citet{Fitzgerald:70} and the absolute magnitude~--~spectral type calibration of~\citet{Turner:80}, and found a value of 3.6~kpc.

\begin{figure*}
\begin{center}
\includegraphics[scale=0.5,viewport=10 70 750 550]{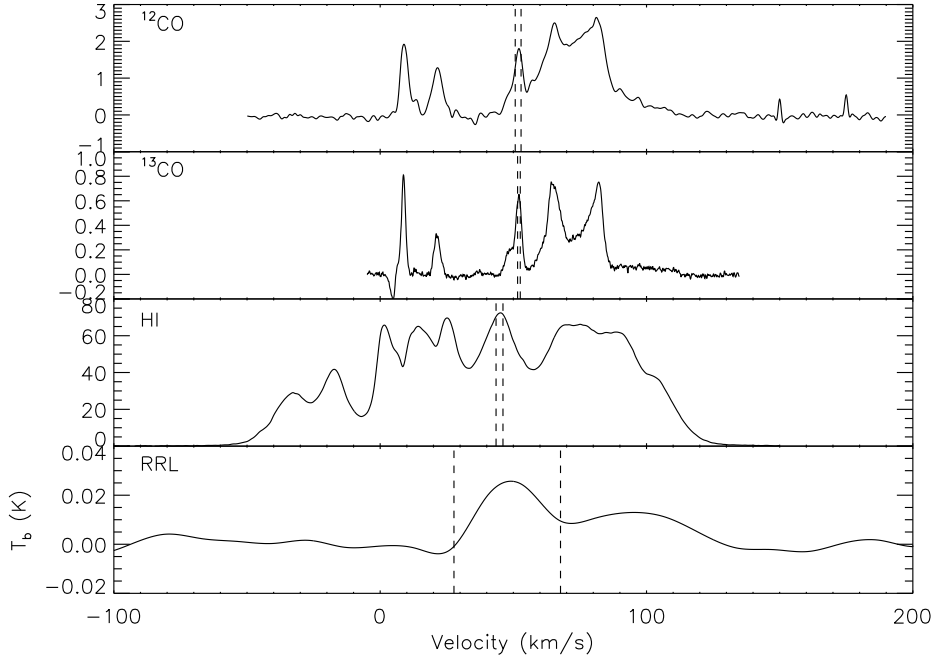} 
\end{center}
\caption{Velocity spectra of the $^{12}$CO, $^{13}$CO, H\textsc{i} and RRL data discussed in Section~\ref{Subsec:Spectral_Data}. It is possible to identify the velocity components associated with RCW175 at~$\sim$~50~km~s$^{-1}$. The dashed lines indicate the 1$\sigma$ values of the corresponding velocity components listed in Table~\ref{Table:Velocity_Components}.}
\label{Fig:Velocity_Components}
\end{figure*}

With the spectral data described in Section~\ref{Subsec:Spectral_Data} now available, it is possible to update the distance estimates to RCW175. The first step in this process is to compute the mean velocity component associated with RCW175 for each of the spectral tracers. To do this, we produced the mean spectrum within a 0.45~$\times$~0.45~square degree region centred on RCW175, and these spectra are displayed in Figure~\ref{Fig:Velocity_Components}. Based on the velocity measurements in the literature and by inspecting the individual data cubes~(as displayed in Figures~\ref{Fig:12CO_Data}, \ref{Fig:13CO_Data} and~\ref{Fig:RRL_Data}), it is clear that the velocity component associated with RCW175 occurs around~$\sim$~50~km~s$^{-1}$. The velocity component at~$\sim$~50~km~s$^{-1}$ in each of the $^{12}$CO, $^{13}$CO, H\textsc{i} and RRL spectra were simultaneously fitted with a Gaussian and a linear baseline, resulting in a measure of the mean velocity component associated with RCW175 for each of the tracers. These mean velocities, along with some measurements from the literature, are listed in Table~\ref{Table:Velocity_Components}.

\begin{deluxetable*}{l c c c c l}
\tabletypesize{\normalsize}
\tablecaption{Velocity components, and corresponding kinematic distances, for RCW175.}
\tablewidth{0pt}
\tablehead{
\colhead{Tracer} & \colhead{Velocity} & \colhead{$D_{near}$} & \colhead{$D_{far}$} & \colhead{$R$} & \colhead{Reference for} \\
\colhead{} & \colhead{(km~s$^{-1}$)} & \colhead{(kcp)} & \colhead{(kpc)} & \colhead{(kpc)} & \colhead{velocity measurement}
}
\startdata
H$\alpha$ & 49.5~$\pm$~5.0 & 3.2~$\pm$~0.5 & 10.8~$\pm$~1.5 & 5.4~$\pm$~0.8 & \citet{Crampton:78} \\
$^{12}$CO & 52.4~$\pm$~1.0 & 3.3~$\pm$~0.3 & 10.7~$\pm$~1.1 & 5.3~$\pm$~0.5 & \citet{Blitz:82} \\ 
$^{12}$CO & 51.8~$\pm$~1.1 & 3.3~$\pm$~0.3 & 10.7~$\pm$~1.1 & 5.4~$\pm$~0.5 & This work \\
$^{13}$CO & 52.1~$\pm$~0.5 & 3.3~$\pm$~0.3 & 10.7~$\pm$~1.1 & 5.4~$\pm$~0.5 & This work \\
H\textsc{i} & 44.8~$\pm$~1.3 & 2.9~$\pm$~0.3 & 11.0~$\pm$~1.1 & 5.6~$\pm$~0.6 & This work \\
RRL & 47.7~$\pm$~20.0 & 3.1~$\pm$~1.3 & 10.9~$\pm$~4.7 & 5.5~$\pm$~2.4 & This work \\
\enddata
\label{Table:Velocity_Components}
\tablecomments{Both the near and far distance from the Sun, $D_{near}$ and $D_{far}$, and the Galactocentric distance, $R$, are reported. References for the velocity measurements are also listed. All velocities are given with respect to the local standard of rest.}
\end{deluxetable*}

\begin{figure*}
\begin{center}$
\begin{array}{ccc}
\includegraphics[scale=0.35, viewport=80 50 750 550]{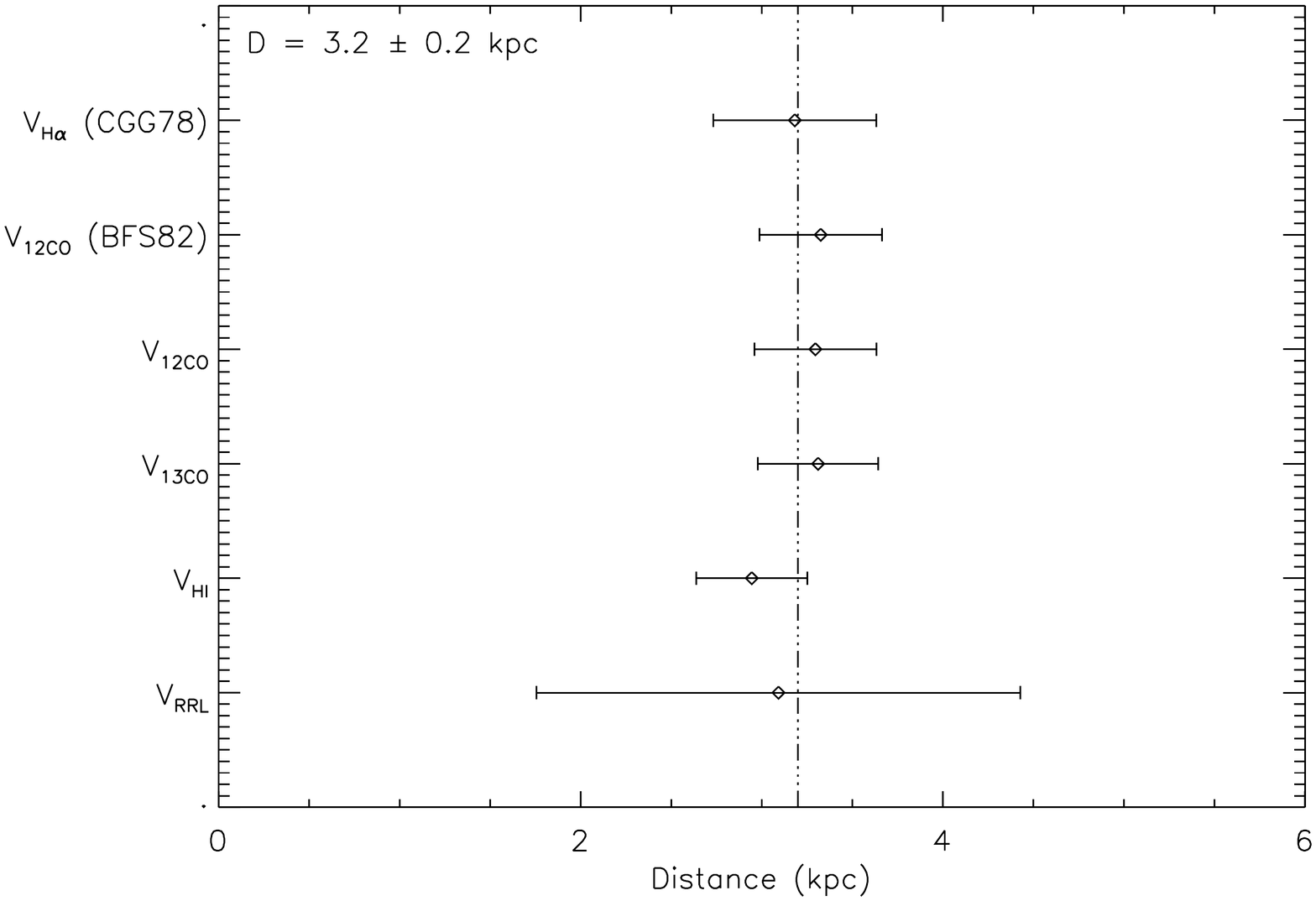} &
\includegraphics[scale=0.35, viewport=80 50 750 550]{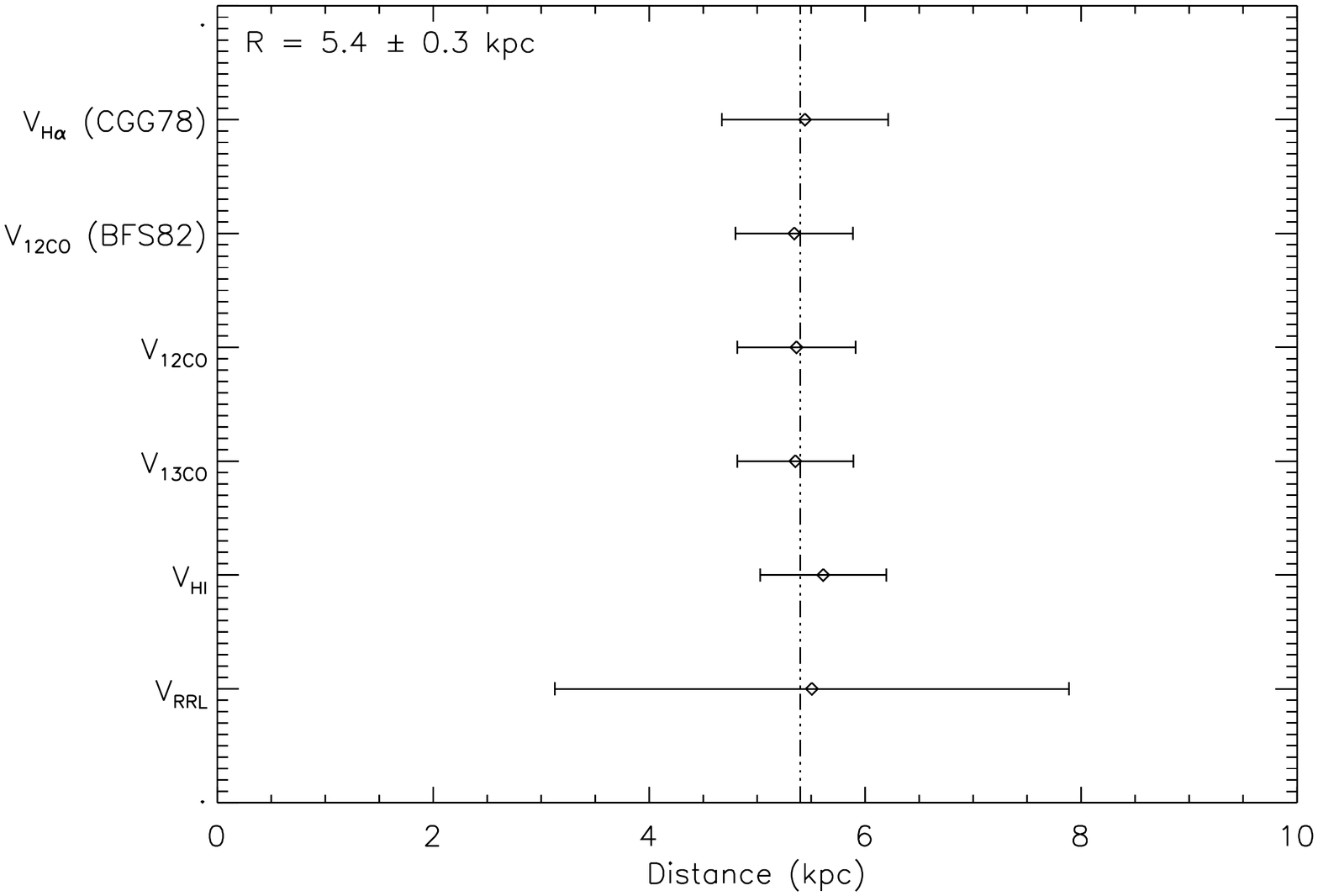}  \\
\end{array}$
\end{center}
\caption{Plots showing the distances and Galactocentric distances to RCW175 obtained from the different tracers listed in Table~\ref{Table:Velocity_Components}. All distances are kinematic distances and have been computed using the~\citet{Fich:89} rotation curve with values of $R_{0}$~=~8.0~kpc and $\Theta_{0}$~=~220~km~s$^{-1}$.}
\label{Fig:Distances}
\end{figure*}

From both the spectra displayed in Figure~\ref{Fig:Velocity_Components} and the measurements listed in Table~\ref{Table:Velocity_Components}, it is possible to see that the velocities of the molecular gas tracers~($^{12}$CO and $^{13}$CO) all have values which are larger than the values for the ionized gas tracers~(H$\alpha$ and RRL). This suggests that the molecular gas and the ionized gas may have become dissociated and that the ionized gas has ruptured and perhaps escaped the molecular gas. This idea is consistent with what we observe when we look at RCW175 in the mid- to far-IR~(see Figure~\ref{Fig:Data}), with only one side of the shell of G29.1-0.7 being visible as a filament like structure at the longer wavelengths. This is the same structure that is visible in the CO data cubes, implying that it consists of colder, denser dust that has been swept up by the radiation pressure within the centre of the shell. The fact that this structure is not fully enclosing the ionized gas implies that the shell has become broken, allowing the ionized gas to leak out into the surrounding ISM. A similar conclusion can be drawn regarding the H\textsc{i} component, which has a lower velocity than both the molecular gas tracers and the ionized gas tracers. A lack of H\textsc{i} associated directly with RCW175 is not surprising since this is an H\textsc{ii} region.

The spectra in Figure~\ref{Fig:Velocity_Components} also highlight the clumpy nature of the CO gas compared to the H\textsc{i} gas. The H\textsc{i} gas is more smoothly distributed throughout the galaxy than the CO, and this is evidenced by the $^{12}$CO and $^{13}$CO spectra containing much more structure than the H\textsc{i} spectra. 

To estimate the kinematic distance to RCW175, we combined the computed velocity components with a Galactic rotation curve. We used the rotation curve models of~\citet{Fich:89}, which holds for Galactocentric distances less than 17~kpc making it adequate for use in this situation. We adopted the~\citet{Fich:89} linear rotation curve of the form

\begin{equation}
\Theta = a_{1}\Theta_{0} + a_{2}\omega_{0}R
\end{equation}

\noindent
where $\Theta$ is the rotation velocity around the Galactic centre of an object at a Galactocentric distance, $R$, $\Theta_{0}$ and $\omega_{0}$ are the rotation velocity and angular velocity of the Sun, respectively, and $a_{1}$ and $a_{2}$ are the fitting coefficients. For the adopted values of $R_{0}$~=~8.0~kpc and $\Theta_{0}$~=~220~km~s$^{-1}$,~\citet{Fich:89} found $a_{1}$~=~0.99334, $a_{2}$~=~0.0030385 and $\omega_{0}$~=~27.5~km~s$^{-1}$. 

We note that the rotation curves of~\citet{Fich:89} are based on both H\textsc{i}~\citep{Burton:78} and CO~\citep{Blitz:82} measurements, with 150 and 104 data points, respectively. The observations by~\citet{Blitz:82} targeted all 313 objects in the~\citet{Sharpless:59} catalog and they found detections for 194 of these objects, including RCW175. This implies that RCW175 was therefore one of the data points used by~\citet{Fich:89} to compute their rotation curve, using a distance of 3.5~$\pm$~1.2~kpc. However, given that there were a total of 254 data points, the weight of any individual region is much less than 1~$\%$, and hence we feel confident in using this rotation curve to compute the kinematic distance to RCW175. 

For sources with Galactic radii less than that of the Sun~(i.e. sources within the solar circle) the estimated solar distance becomes degenerate as there are two solutions for every velocity, corresponding to a near distance, $D_{near}$, and a far distance, $D_{far}$. This is known as the kinematic distance ambiguity. The kinematic distances, $D_{near}$ and $D_{far}$, and Galactocentric distances, $R$, for the data discussed in Section~\ref{Subsec:Spectral_Data} and for the values from the literature are tabulated in Table~\ref{Table:Velocity_Components}. The distances obtained using the velocity measurements from the literature are more accurate then the corresponding previous estimates due to the updated rotation curve and updated values of $R_{0}$ and $\Theta_{0}$. 

The kinematic distance ambiguity makes it difficult to determine whether $D_{near}$ or $D_{far}$ is the correct distance, however, based on the distance estimate of 3.6~kpc to the ionizing source S65-4 located at the centre of G29.1-0.7~\citep{Forbes:89}, we select the near kinematic distance for this component. Since we have seen from the $^{12}$CO and $^{13}$CO data in Figures~\ref{Fig:12CO_Data} and~\ref{Fig:13CO_Data}, that G29.1-0.7 and G29.0-0.6 are interacting, we believe that both components are situated at the near distance.

We note that~\citet{Anderson:09} placed G29.0-0.6 at the far distance. However, they used two separate methods to break the ambiguity, with each method giving a different result: one placing it at the near distance and the other placing it at the far distance. Given this uncertainty, we feel confident choosing the near distance for both components of RCW175. The adopted kinematic distances, $D$, and Galactocentric distances, $R$, are plotted in Figure~\ref{Fig:Distances}. By combining all the distance estimates listed in Table~\ref{Table:Velocity_Components}, we compute the mean distance to RCW175,~$D$~=~3.2~$\pm$~0.2~kpc, and the mean Galactocentric distance, $R$~=~5.4~$\pm$~0.3~kpc.


\subsection{Star Formation Rate}
\label{Subsec:SFR}

Various works have attempted to use the SFR from individual H\textsc{ii} regions~\citep[e.g.][]{Smith:78, Chomiuk:11} and apply them to the entire Galaxy to compute the Galactic SFR. Such an analysis requires investigating multiple H\textsc{ii} regions, which is beyond  the scope of this work, but computing the SFR of RCW175 will be helpful to others estimating the Galactic SFR.

\citet{Calzetti:07} developed a calibration to use monochromatic 24~$\mu$m emission from extragalactic H\textsc{ii} regions as a SFR indicator. Similar analyses have been performed by~\citet{Alonso-Herrero:06},~\citet{Perez-Gonzalez:06} and~\citet{Relano:07}, each derived based on slightly different 24~$\mu$m luminosity ranges. \citet{Calzetti:10} estimated an intrinsic scatter of the order of 2 between these different calibrations. 

Recently, \citet{Chomiuk:11} performed an analysis investigating various SFR determinations, and in doing so they extended the~\citet{Calzetti:07} extragalactic mid-IR diagnostic to the Galactic H\textsc{ii} region M17. Following their example, we use the extragalactic calibrations to compute the SFR in RCW175.

To estimate the 24~$\mu$m luminosity we computed the flux density within RCW175 and converted this to a luminosity using

\begin{equation}
L_{24~\mu m}~=~4\times10^{7} \pi D^{2} S_{24~\mu m} \nu~\mathrm{erg~s^{-1}},
\end{equation}

\noindent
where $D$ is the distance in m, $S_{24~\mu m}$ is the flux density in W~m$^{-2}$~Hz$^{-1}$ and $\nu$ is the frequency in Hz. The flux density was derived using aperture photometry in a circular aperture centred on RCW175 with a radius of 12~arcmin. An estimate of the background was obtained by computing the median value of all the pixels within an annulus, also centred on RCW175, with an inner radius of 14~arcmin and an outer radius of 17.5~arcmin. 

Given that we compute a 24~$\mu$m luminosity of the order of 1~$\times$~10$^{38}$~erg~s$^{-1}$ for RCW175, the calibrations established by~\citet{Perez-Gonzalez:06} and~\citet{Relano:07} are the most applicable. Therefore, using the calibration of~\citet{Relano:07}

\begin{equation}
\mathrm{SFR}_{24~\mu m} = 5.66\times10^{-36} L_{24~\mu m}^{0.826}~\mathrm{M_{\sun}~yr^{-1}},
\label{equ:SFR_24um}
\end{equation}

\noindent
where $L_{24~\mu m}$ is the 24~$\mu$m luminosity in units of erg~s$^{-1}$, results in a SFR of (12.6~$\pm$~1.9)~$\times$~10$^{-5}$~M$_{\sun}$~yr$^{-1}$. This result is consistent with the SFR computed using the \citet{Perez-Gonzalez:06} calibration, which is (12.6~$\pm$~1.8)~$\times$~10$^{-5}$~M$_{\sun}$~yr$^{-1}$.

\citet{Chomiuk:11} explored other methods of SFR determination, including the thermal radio continuum diagnostic. Unlike the 24~$\mu$m diagnostic, which uses dust emission as an indirect measure of the Lyman continuum photon flux, $N_{c}$, this method is based on the ionized gas. The Lyman continuum photon flux is the driving force of H\textsc{ii} regions and the ionized gas can be observed at frequencies of a few GHz, due to the emission of thermal bremsstrahlung or free-free radiation~(as can be seen in Figure~\ref{Fig:Data}). \citet{Chomiuk:11} adopted the following relationship between the SFR and the Lyman continuum rate

\begin{equation}
\mathrm{SFR} = 7.5\times10^{-54} N_{c}~\mathrm{M_{\sun}~yr^{-1}}.
\label{equ:SFR_ff}
\end{equation}

\noindent
Therefore to compute the SFR based on the thermal radio emission, we combined this relationship with the expression for $N_{c}$ derived by~\citet{Mezger:74} 

\begin{eqnarray}
N_{c} = 4.761\times10^{48} a(\nu, T_{e})^{-1} \left(\frac{\nu}{\mathrm{GHz}}\right)^{0.1} \nonumber \\
\times \left(\frac{T_{e}}{\mathrm{K}}\right)^{-0.45} \left(\frac{S_{\nu}}{\mathrm{Jy}}\right) \left(\frac{D}{\mathrm{kpc}}\right)^{2}~\mathrm{photons~s^{-1}},
\label{equ:Nc}
\end{eqnarray} 

\noindent
where 

\begin{eqnarray}
a(\nu, T_{e}) = 0.366 \left( \frac{\nu}{\mathrm{GHz}} \right)^{0.1} \left(\frac{T_{e}}{\mathrm{K}}\right)^{-0.15} \nonumber \\
\times \left\{ \mathrm{ln} \left[ 4.995\times10^{-2} \left(\frac{\nu}{\mathrm{GHz}}\right)^{-1} \right] + 1.5~\mathrm{ln} \left(\frac{T_{e}}{\mathrm{K}}\right)  \right\},
\label{equ:a}
\end{eqnarray}

\noindent
as defined by~\citet{Mezger:67}, and $T_{e}$ is the electron temperature.

Having already obtained the mean Galactocentric distance to RCW175, we used the relationship between the Galactocentric distance of H\textsc{ii} regions and their electron temperature, empirically derived by~\citet{Paladini:04} based on a sample of 404 H\textsc{ii} regions, to compute $T_{e}$. The expression from \citet{Paladini:04} is

\begin{equation}
T_{e} = (4166 \pm 124) + (314 \pm 20)R~\mathrm{K},
\label{equ:Te}
\end{equation} 

\noindent
where $R$ is in units of kpc. This results in $T_{e}$~=~5861.6~$\pm$~189.5~K. Throughout the rest of this analysis we adopt a value of $T_{e}$~=~5800~K.

Incorporating this value of $T_{e}$ into Equation~(\ref{equ:Nc}) and the flux density, $S_{\nu}$, computed by integrating the flux in the Parkes 5~GHz map~(using the same aperture as for the 24~$\mu$m map) we found that $N_{c}$~$\sim$~5~$\times$~10$^{48}$ photons~s$^{-1}$ which, when substituted into Equation~(\ref{equ:SFR_ff}), results in a SFR~$\approx$~3.6~$\times$~10$^{-5}$~M$_{\sun}$~yr$^{-1}$. 

This value is of the order 3~--~4 times smaller than the SFR computed from the mid-IR luminosity indicator. This discrepancy can partially be explained in terms of the intrinsic uncertainties in the different diagnostics, which is of the order~2~--~3, with an additional contribution arising from the fact that $N_{c}$ is an underestimate of the true Lyman continuum rate. Equation~(\ref{equ:Nc}) only computes the Lyman continuum rate of the Lyman photons that are ionizing the gas. This is not the same as computing the Lyman continuum rate of all the Lyman photons emitted by the ionizing star, as some of the Lyman photons are absorbed by dust before having the opportunity to ionize the gas. In light of this effect, we believe that the SFR of RCW175 is most accurately computed based on the 24~$\mu$m emission.


\subsection{Young Stellar Objects}
\label{Subsec:YSOs}

\begin{deluxetable*}{c c c}
\tabletypesize{\normalsize}
\tablecaption{YSO candidate selection criteria adopted from~\citet{Rebull:10}.}
\tablewidth{0pt}
\tablehead{
\colhead{Magnitudes} & \colhead{YSO selection} & \colhead{Number of sources}
}
\startdata
24 & [24] < 7 & 65 \\

K$_{s}$/24 & K$_{s}$ < 14 AND K$_{s}$$-$[24] > 1 & 58 \\

\multirow{2}{*}{8/24} & [8]$-$[24] > 0.5 AND & \multirow{2}{*}{60} \\
 & (([8]$-$[24] < 4 and [8] < 10) or ([8]$-$[24] $\ge$ 4 and [8] < 2.5$\times$([8]$-$[24]))) & \\

\multirow{2}{*}{4.5/8} & [4.5] < 6 AND ([4.5] $\ge$ 6 and [4.5] $\le$ 11.5 and [4.5]$-$[8] > 0.4) & \multirow{2}{*}{54} \\
 & OR ([4.5] > 11.5 and [4.5] < 0.6944$\times$([4.5]$-$[8])+11.22) & \\

3.6/4.5/5.8/8 & [3.6]$-$[4.5] > 0.15 and [5.8]$-$[8] > 0.3 and [3.6] < 13.5 & 12 \\
\enddata
\label{Table:YSOc_Selection}
\end{deluxetable*}

\begin{figure*}
\begin{center}
\includegraphics[scale=0.65, viewport=100 80 700 550]{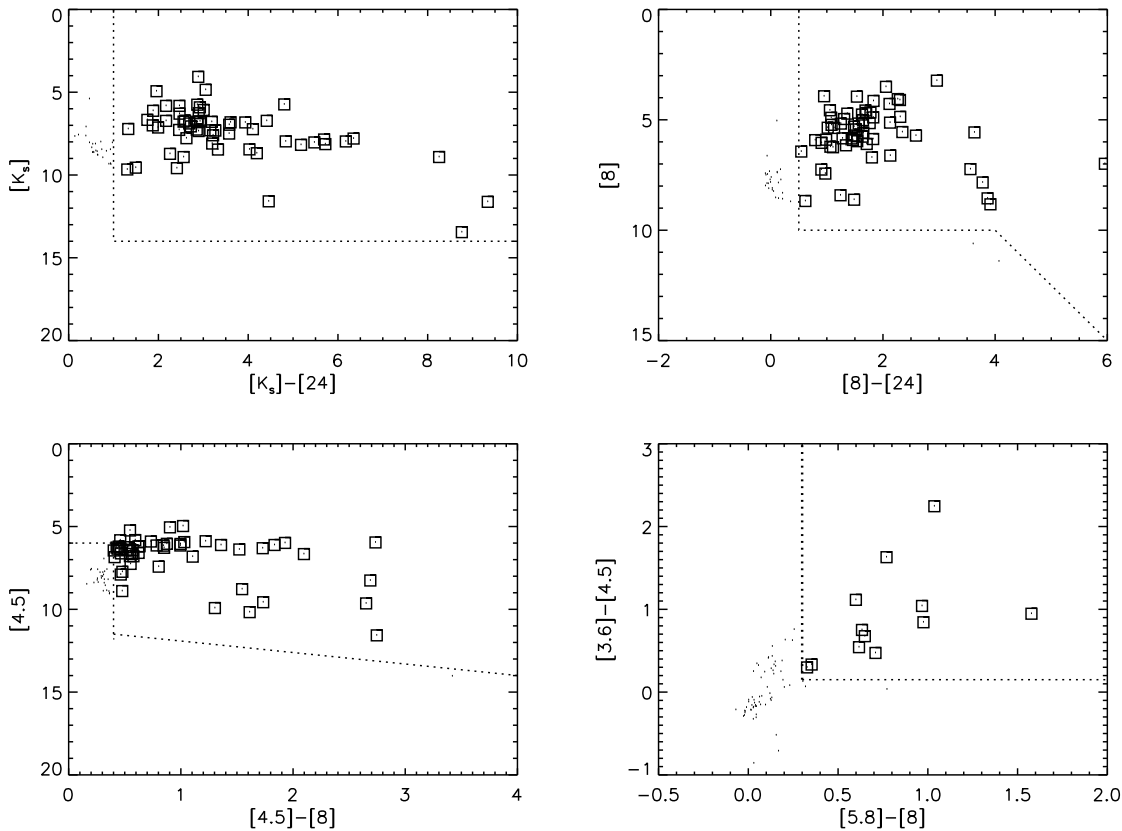} \\
\end{center}
\caption{Color-magnitude and color-color selection criteria implemented to identify YSO candidates. All 95 sources are plotted~(\textit{dots}) along with the selection criteria~(\textit{dotted line}) and the sources that meet these criteria~(\textit{squares}).}
\label{Fig:YSOc_Selection}
\end{figure*}

As discussed by~\citet{Deharveng:05}, and briefly in Section~\ref{Subsec:Dust_Modeling}, H\textsc{ii} regions are known to play an important role in triggering star formation. Having concluded that we believe that G29.0-0.6 was possibly triggered by the expansion of G29.1-0.7 into the surrounding ISM, we would like to investigate if there is any ongoing triggering occurring within RCW175. To do this we looked at the location of candidate young stellar objects~(YSOs) in relation to the general morphology of RCW175. To identify candidate YSOs we used the MIPSGAL point source catalogue~(Shenoy et al. in prep.) which contains all the MIPSGAL sources detected at 24~$\mu$m. Since the coverage of the MIPSGAL survey is similar to that of the GLIMPSE survey, the MIPSGAL catalogue has been band-merged with the GLIMPSE sources. The GLIMPSE point source archive contains sources at all four IRAC bands~(3.6, 4.5, 5.8 and 8~$\mu$m) that have been band-merged with sources in all three 2MASS bands~(J, H and K$_{s}$). The result is a band-merged source catalogue with sources detected at both near- and mid-IR wavelengths that is suitable to identify YSO candidates.

Selecting only 24~$\mu$m sources with a completeness and reliability of $>$~95~$\%$, we find 95 sources within a 0.45~degree square region centred on RCW175. We implemented the color-magnitude and color-color selection criteria of~\citet{Rebull:10} to identify YSO candidates. These criteria are listed in Table~\ref{Table:YSOc_Selection} along with the corresponding number of sources in each selection, and are displayed visually in Figure~\ref{Fig:YSOc_Selection}. Combining these selection criteria resulted in a total of 6 YSO candidates within the vicinity of RCW175, all of which are listed in Table~\ref{Table:YSOc}.

\begin{deluxetable*}{c c c c}
\tabletypesize{\normalsize}
\tablecaption{YSO candidates identified in RCW175.}
\tablewidth{0pt}
\tablehead{
\colhead{Source} & \colhead{RA} & \colhead{DEC} & \colhead{Comments} \\
 & \colhead{(J2000)} & \colhead{(J2000)} & 
}
\startdata
25 & 18:45:59.9 & $-$3:49:12.0 & \\
40 & 18:46:25.2 & $-$3:48:09.1 & \\
53 & 18:46:43.8 & $-$3:44:35.3 & \\
66 & 18:46:51.8 & $-$3:41:59.5 & YSO candidate~\citep{Robitaille:08} \\
67 & 18:47:06.8 & $-$3:43:49.0 & \\
73 & 18:46:31.4 & $-$3:38:27.6 & \\
\enddata
\label{Table:YSOc}
\end{deluxetable*}

A SIMBAD search within 6~arcsec~(the FWHM of the 24~$\mu$m point response function) of the YSO candidate position reveals that 5 of the sources~(sources 25, 40, 53, 67 and 73) have no SIMBAD counterparts and one of the candidates~(source 66) has previously been identified as a possible YSO candidate~\citep{Robitaille:08}. Looking at the location of the 6 candidate YSOs displayed in Figure~\ref{Fig:3_color_RCW175}, two occur on the periphery of RCW175~(sources 25 and 73). Sources 53 and 66 are located within the diffuse shell of G29.1-0.7. The location of these sources occur at both ends of a small filament of warm dust at either side of the ionizing source S65-4. Source 67 and source 40 are found in the denser PDRs surrounding G29.1-0.7 and G29.0-0.6, respectively. Given the lack of any distance estimate to these sources, it is difficult to confirm that these YSO candidates are indeed associated with RCW175, however, given the location is within the dense PDR, sources 40 and 67 are likely associated with RCW175, and show that triggering may be occurring within this region.


\subsection{Dust Mass}
\label{Subsec:Dust_Mass}

To estimate the total dust mass associated with RCW175, we use

\begin{equation}
M_{dust} = \frac{S_{\lambda} D^{2}}{\kappa_{\lambda} B(\lambda, T_{dust})}
\label{equ:mass}
\end{equation}

\noindent
where $S_{\lambda}$ is the flux density, $D$ is the distance to RCW175, $\kappa_{\lambda}$ is the dust opacity and $B(\lambda, T_{dust})$ is the Planck function. The dust opacity was computed using

\begin{equation}
2.92 \times 10^{5} \lambda_{\mu m}^{-2}~\mathrm{cm^{2}/g}
\end{equation}

\noindent
from~\citet{Li:01}, which is accurate for wavelengths between 20~--~700~$\mu$m. Since the cold, BGs contribute most of the dust mass, we evaluate Equation~(\ref{equ:mass}) at a wavelength of 500~$\mu$m. Using the 500~$\mu$m flux density and the equilibrium temperature of the cold dust component in RCW175~(see Section~\ref{Subsec:SED} and Table~\ref{Table:Fluxes}) along with the distance, we compute a total dust mass of 215~$\pm$~53~M$_{\sun}$. We also computed the dust mass using the 250 and 350~$\mu$m data and found that the total dust masses were in agreement~(within the uncertainties) with the one computed using the 500~$\mu$m data.


\subsection{Full SED}
\label{Subsec:SED}

Having discussed the various intricacies of the morphology of RCW175 in the previous sections, we now want to look at the energy budget of the region. To do this we computed the flux density of the entire complex from the radio to the mid-IR, using the same method as was used to compute the 24~$\mu$m flux density described in Section~\ref{Subsec:SFR}.

Unlike the dust modeling analysis performed with \textsc{DustEm} in Section~\ref{Subsec:Dust_Modeling}, we are now looking at the entire region and so no longer restrict ourselves to using only the high resolution \textit{Spitzer} and \textit{Herschel} IR data. Therefore, we also incorporate the IRIS and \textit{WMAP} data, as well as all the radio data discussed in Section~\ref{Subsec:Radio_Data}. As stated previously, we ignore the NVSS data due to a lack of short spacings. All the flux densities and the corresponding wavelengths are listed in Table~\ref{Table:Fluxes}. Both the calibration uncertainty and the flux density uncertainty from the aperture photometry, which is computed based on the r.m.s of the pixel values in the background, have been incorporated into the final uncertainties of the flux densities.

\begin{figure*}
\begin{center}
\includegraphics[scale=0.65, viewport=0 25 500 370]{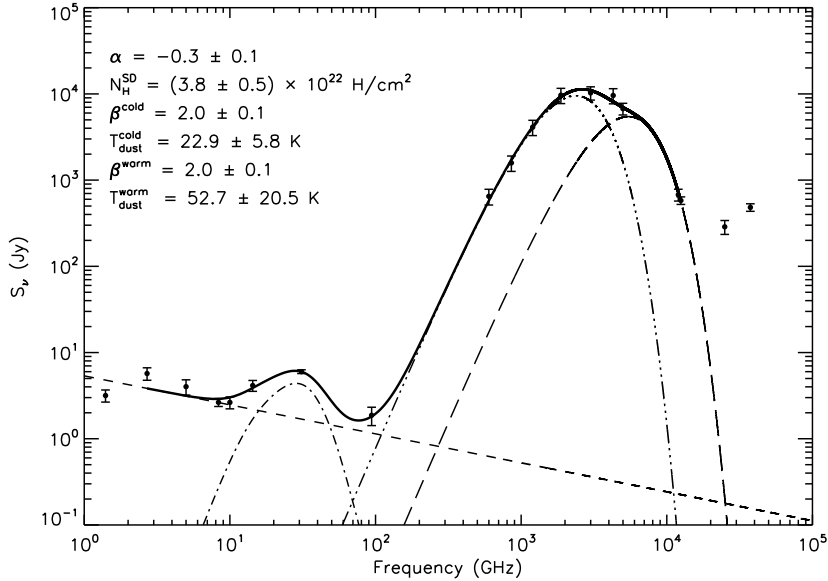} 
\end{center}
\caption{Complete SED for RCW175 from the radio to the mid-IR. The data have been modeled using a power-law representing the free-free emission~(\textit{dashed line}), two modified blackbody curves representing the warm~(\textit{long-dashed line}) and cold~(\textit{dot-dot-dot-dashed line}) thermal dust emission components and a spinning dust curve for the warm ionized medium~(\textit{dot-dashed line}). The best fitting values for $\alpha$, $N_{H}^{SD}$, $\beta^{cold}$ and $T_{dust}^{cold}$, $\beta^{warm}$ and $T_{dust}^{warm}$ are displayed. It is clear that there is an excess over both the free-free and thermal dust emission present in RCW175.}
\label{Fig:Full_SED}
\end{figure*}

The values in Table~\ref{Table:Fluxes} are plotted in Figure~\ref{Fig:Full_SED}. This SED is an updated version of the original SED presented by~\citet{Dickinson:09}. We have used more recent data where available~(e.g. Effelsberg 1.4~GHz) and we have substantially improved the constraint on the Rayleigh-Jeans tail of the thermal dust distribution given the inclusion of the \textit{Herschel} data. It should also be noted that even though~\citet{Dickinson:09} used Gaussian fitting to compute the flux densities, while in this anlaysis we have used aperture photometry, the general consistency between both sets of data points is good.

The data between 2.7 and 12491.4~GHz, or 11~cm and 24~$\mu$m, were simultaneously fitted with a power-law~($S_{\nu}$~$\propto$~$\nu^{\alpha}$), a spinning dust model and two modified blackbody curves~($S_{\nu}$~$\propto$~$\nu^{\beta}B(\nu, T_{dust})$) representing the free-free, AME and thermal dust emission, respectively. These components will be discussed further in Sections~\ref{Subsubsec:Low_Freq},~\ref{Subsubsec:Thermal_Dust} and~\ref{Subsubsec:AME}.

\begin{deluxetable*}{c c c c c}
\tabletypesize{\small}
\tablecaption{Flux densities for RCW175.}
\tablewidth{0pt}
\tablehead{
\colhead{Frequency} & \colhead{Telescope/} & \colhead{Reference for} & \colhead{Angular Resolution} & \colhead{$S_{\nu}$} \\
\colhead{(GHz)} & \colhead{Survey} & \colhead{Data} & \colhead{(arcmin)} & \colhead{(Jy)}
}
\startdata
1.4 & Effelsberg 100~m & \citet{Reich:90a} & 9.4 & 3.2~$\pm$~0.5 \\
2.7 & Effelsberg 100~m & \citet{Reich:90b} & 4.3 & 5.7~$\pm$~0.9 \\

5 & Parkes 64~m & \citet{Haynes:78} & 4.1 & 4.0~$\pm$~0.8 \\

8.35 & Green Bank 13.7~m & \citet{Langston:00} & 9.7 & 2.6~$\pm$~0.3 \\

10 & Nobeyama 45~m & \citet{Handa:87} & 3.0 & 2.6~$\pm$~0.4 \\

14.35 & Green Bank 13.7~m & \citet{Langston:00} & 6.6 & 4.1~$\pm$~0.6 \\

31 & CBI & \citet{Dickinson:09} & 4.3 & 6.0~$\pm$~0.3 \\

94 & \textit{WMAP} & \citet{Jarosik:11} & 13.2 & 1.9~$\pm$~0.4 \\

599.6~(500~$\mu$m) & \textit{Herschel}/SPIRE & \citet{Molinari:10} & 0.6 & 649.0~$\pm$~134.0 \\
856.5~(350~$\mu$m) & \textit{Herschel}/SPIRE & \citet{Molinari:10} & 0.4 & 1580.0~$\pm$~322.0 \\
1199.2~(250~$\mu$m) & \textit{Herschel}/SPIRE & \citet{Molinari:10} & 0.3 & 4110.0~$\pm$~829.0 \\

1873.7~(160~$\mu$m) & \textit{Herschel}/PACS & \citet{Molinari:10} & 0.2 & 9660.0~$\pm$~1940.0 \\

2997.9~(100~$\mu$m) & IRIS & \citet{Miville-Deschenes:05} & 4.3 & 10300.0~$\pm$~1800.0 \\

4282.7~(70~$\mu$m) & \textit{Herschel}/PACS & \citet{Molinari:10} & 0.1 & 9580.0~$\pm$~1920.0 \\

4996.5~(60~$\mu$m) & IRIS & \citet{Miville-Deschenes:05} & 4.0 & 6740.0~$\pm$~1050.0 \\
11991.7~(25~$\mu$m) & IRIS & \citet{Miville-Deschenes:05} & 3.8 & 677.0~$\pm$~106.0 \\

12491.4~(24~$\mu$m) & \textit{Spitzer}/MIPS & \citet{Carey:09} & 0.1 & 581.0~$\pm$~58.1 \\

24982.7~(12~$\mu$m) & IRIS & \citet{Miville-Deschenes:05} & 3.8 & 287.0~$\pm$~52.8 \\

37474.1~(8~$\mu$m) & \textit{Spitzer}/IRAC & \citet{Churchwell:09} & 0.03 & 483.0~$\pm$~48.3 \\
\enddata
\label{Table:Fluxes}
\end{deluxetable*}

\subsubsection{Low-Frequency Emission}
\label{Subsubsec:Low_Freq}

Fitting the radio data in Figure~\ref{Fig:Full_SED} with a single power-law resulted in a spectral index of, $\alpha$~=~$-$0.3~$\pm$~0.1. This is steeper than the typical spectral index of $\sim$~$-$0.1 expected for optically thin free-free emission at these frequencies~\citep{Dickinson:03}, however it is consistent with the value of~$-$0.32~$\pm$~0.15 obtained previously by~\citet{Dickinson:09}. To explain this steeper spectral index we believe that there must be some residual contamination from synchrotron emission in the low frequency radio maps which, given that synchrotron emission generally follows a power-law with a spectral index of~$\sim$~$-$0.7, could be contributing to the steeper spectral index found in RCW175. Given that RCW175 is located on the Galactic plane, there will be a contribution from the diffuse synchrotron emission. Additionally, a search in the~\citet{Green:09} catalogue of Galactic supernova remnants reveals three nearby~($<$ 1~degree away) supernova remnants,~G29.7-0.3~(Kes 75) and G28.6-0.1 both~$\sim$~0.7~degrees away and G29.6+0.1~$\sim$~0.9~degrees away, which may be causing synchrotron relativistic electrons to escape into the surrounding environments.

We therefore conclude that the low frequency emission in RCW~175 is arising from the free-free emission from the ionized gas and some small contribution from background residual synchrotron emission.


\subsubsection{Thermal Dust Emission}
\label{Subsubsec:Thermal_Dust}

To represent the thermal dust emission we used a modified blackbody curve, fitting for both the dust temperature, $T_{dust}$, and the dust emissivity spectral index, $\beta$. Two curves were used to fit both the cold and warm dust components. This is necessary since we are computing the SED for the entire region, and we know that the dust within the entire region is not emitting at a single temperature. In the \textsc{DustEm} analysis, since we were fitting the SED on a pixel-by-pixel basis, we did not have this problem, and hence were able to restrict ourselves to using a single dust temperature.

The best fitting values for the cold dust are $\beta^{cold}$~=~2.0~$\pm$~0.1 and $T_{dust}^{cold}$~=~22.9~$\pm$~5.8~K, while for the warm dust we find best fitting values of $\beta^{warm}$~=~2.0~$\pm$~0.1 and $T_{dust}^{warm}$~=~52.7~$\pm$~20.5~K. These values of the cold and warm dust components are consistent with previous estimates of dust temperatures in H\textsc{ii} regions~\citep[e.g.][]{Paladini:12}. It is also noticeable that by including the \textit{Herschel} data in our SED, the temperatures computed in this analysis are more accurate than the dust temperature estimated by~\citet{Dickinson:09} who fitted a modified blackbody curve to the IRIS 100 and 60~$\mu$m data, and obtained a temperature of 30.1~$\pm$~5.7~K.


\subsubsection{Anomalous Microwave Emission}
\label{Subsubsec:AME}

The excess of emission observed between~$\sim$~10~--~60~GHz in the SED of RCW175 is a clear indication of the presence of AME. The peaked nature of the AME spectrum is consistent with the current theoretical models of spinning dust~\citep{DaL:98, Ali-Haimoud:09, Hoang:10, Ysard:10, Hoang:11, Silsbee:11}, and therefore we modeled the AME as a spinning dust component. To do this we used the spinning dust code, \textsc{spdust}~\citep{Ali-Haimoud:09, Silsbee:11}. For a given set of input parameters, \textsc{spdust} computes the emissivity of the spinning dust emission normalized by the hydrogen column density. We used the updated version of~\textsc{spdust}, which not only accounts for dust grains spinning about their axis of greatest inertia, it also accounts for grains spinning about their non principal axes. 

Since we have only two data points covering the critical frequency range, we decided not to fit for the individual spinning dust parameters within \textsc{spdust}. Instead, we adopted the idealized physical parameters for the various phases of the ISM defined by~\citet{DaL:98}~(warm ionized medium, warm neutral medium, cold neutral medium, molecular cloud, dark cloud, reflection nebula and PDR) within \textsc{spdust} and fitted for the column density. This simply involves scaling the amplitude of the spinning dust curve and has no effect on its shape or peak frequency. We performed the fit using the parameters for each of the idealized environments and found that the warm ionized medium model resulted in the best fit to the data.

The spinning dust curve for the warm ionized medium fits the data extremely well, and from the fit we found that the column density required to produce the observed spinning dust emission is~$N_{H}^{SD}$~=~(3.8~$\pm$~0.5)~$\times$~10$^{22}$~H~cm$^{-2}$, which is consistent with the range of values of the total column density obtained from the \textsc{DustEm} analysis~(Section~\ref{Subsec:Dust_Modeling}). Another method of confirming the strength of the AME is to compare to AME observed in other environments. To do this, the~$\sim$~30~GHz flux density is generally normalized by the 100~$\mu$m flux density. This quantity, often called the dust emissivity, was computed for RCW175 and we found a value of (5.8~$\pm$~1.1)~$\times$~10$^{-4}$, which is consistent with the value computed by~\citet{Davies:06} for the diffuse dust at intermediate latitudes. These two results imply that the parameters we obtain for the AME in RCW175 are both physical and meaningful.

To investigate the origin of the AME component observed in Figure~\ref{Fig:Full_SED} we adopt the CBI 31~GHz map as a tracer of the AME. This is a reasonable assumption given that the free-free and thermal dust components alone only predict a 31~GHz flux density of 1.7~Jy compared to the measured value of 6.0~$\pm$~0.3~Jy. This is an excess of 4.3~Jy implying that~$\approx$~70~$\%$ of the CBI emission is due to AME~(14.3$\sigma$). We note that this is a larger fraction of AME than was found by~\citet{Dickinson:09} as they fixed the spectral index of the low frequency emission at a value of $-$0.12. However, even if we fit a power-law component with a fixed spectral index of~$-$0.12, we find that the fraction of AME is still~$\approx$~60~$\%$~(12.7$\sigma$).

Since we know that there is the possibility of ongoing star formation in RCW175, we investigated the correlation between the AME and the location of the YSO candidates identified in Section~\ref{Subsec:YSOs}. Massive YSOs form in ultra-compact H\textsc{ii}~(UCH\textsc{ii}) regions, which have densities~$\gtrsim$~10$^{4}$ H~cm$^{-3}$ and hence are optically thick. Therefore, the free-free emission originating from the ionized gas is self-absorbed, which results in a turnover in the spectrum from~$\alpha$~=~$-$0.12 to~$\alpha$~$\approx$~2. The exact frequency of this turnover depends on the density of the UCH\textsc{ii} region, but can occur up to tens of GHz. Of the 6 YSO candidates identified within the vicinity of RCW175, only one~(source 40) was located in G29.0-0.6 and this is situated~$\approx$~1~arcmin from the peak of the AME. The pointing accuracy of the CBI is~$\approx$~0.5 arcmin~\citep{Taylor:11}, hence making it unlikely that the source of the excess emission in G29.0-0.6 is due to an UCH\textsc{ii} region. Similarly, none of the 6 YSO candidates are associated with the peak of the AME in G29.1-0.7, again suggesting that the AME in this region is not due to an UCH\textsc{ii} region. 

Having ruled out the possibility of the AME in RCW175 being due to self-absorbed free-free emission, we next want to investigate the connection between the AME and the dust properties. Therefore, we compare the CBI 31~GHz map with the dust parameter maps discussed in Section~\ref{Subsec:Dust_Modeling} to identify which parameters, if any, correlate well with the AME. In Figure~\ref{Fig:Param_Maps_Contours} we display the dust parameter maps with the contours of the 31~GHz emission overlaid. It is apparent that the 31~GHz emission is not strongly spatially correlated with the PAH abundance. The deficit of PAHs within the shell of G29.1-0.7 is coincident with the AME observed in this component, while there appears to be a small collection of PAHs present towards the AME originating from G29.0-0.6. 

Looking at the VSG abundances relative to the 31~GHz emission we find that the VSGs are located in the interior of both G29.0-0.6 and G29.1-0.7. We find that only the VSGs in G29.0-0.6 are correlated with the 31GHz emission. The VSGs present in G29.1-0.7 are centred on the location of the B1 II star at the centre of the shell, which is clearly offset from the AME emission. This suggests that the AME is not being produced by the VSGs, and is not directly associated with the source of ionization.

Although the hydrogen column density is increased towards G29.0-0.6, and decreased towards G29.1-0.7, there is no apparent correlation with the AME. However, the maps of the ERF strength and $T_{dust}$ appear to be very well correlated with the CBI emission. Not only is the correlation present towards G29.0-0.6, but it is also present towards G29.1-0.7. As stated previously, within G29.1-0.7, the ERF is enhanced towards the dust filament along the edge of the shell, implying that the ionization front is not uniform in all directions. The correlation between the AME and the ERF and the warmer dust implies that the AME is dependent on the strength of the ERF, and suggests that it is the excitation of the ERF that is producing the AME. This result is consistent with the result found in the Perseus molecular cloud, where the strength of the ERF appeared to be enhanced in regions of AME~\citep{Tibbs:11}, which therefore could be the result of electric dipole emission from spinning dust grains being spun up by photon--grain interactions. Such photon--grain interactions may also cause the grains to become dehydrogenated and as discussed by~\citet{DaL:98}, this will have a strong effect on the electric dipole moment of the grain, which will also affect the spinning dust emissivity. The observed correlation between the AME and the ERF in both RCW175 and the Perseus molecular cloud is in agreement with the spinning dust theory, where, if all other parameters are fixed, the spinning dust emissivity increases with increasing ERF strength~\citep[e.g.][]{Ali-Haimoud:09, Ysard:11}.

We also note that it is possible that a fraction of the excess emission could be due to magnetic dipole emission produced by thermal fluctuations in the magnetism of the grains~\citep{DaL:99}. According to the theory, the magnitude of the magnetic fluctuations depends on the magnetic properties of the grain material, and any strongly magnetic grain material will almost certainly contain Fe. Indeed, recent results from~\citet{Peimbert:10} have shown that in H\textsc{ii} regions, at least~$\sim$~40~$\%$ of the Fe atoms are embedded in dust grains, with this fraction increasing to~$\sim$~97~$\%$ in high metallicity, high density H\textsc{ii} regions. It is difficult to quantify the exact contribution originating from magnetic dipole emission, but one possible method to help distinguish between the two emission mechanisms would be to observe the polarization of the AME. Electric dipole emission is predicted to be only slightly polarized~(a fractional polarization of <~1~$\%$ above 30~GHz;~\citealp{LaD:00}), while magnetic dipole emission from single domain magnetic grains is expected to be highly polarized~(a fractional polarization of up to~$\sim$~40~$\%$;~\citealp{DaL:99}). No such observations of RCW175 have yet been performed, however, recent observations by~\citet[][]{Harvey-Smith:11} have shown that the magnetic field strength towards Galactic H\textsc{ii} regions is similar to the typical magnetic field strength of the diffuse ISM, implying that there is very little polarized emission in these environments. This leads us to believe that the contribution of magnetic dipole emission from single domain magnetic grains in RCW175 is minimal. However, we note that magnetic dipole emission arising from grains containing magnetic inclusions is expected to be relatively unpolarized~(a fractional polarization of <~1~$\%$;~\citealp{DaL:99}), and as such, we cannot completely rule out a contribution from magnetic dipole emission.

It is interesting that the AME in G29.1-0.7 is not spatially correlated with the surrounding PDR, where the PAHs are concentrated, but is actually originating from a region just on the interior of the PDR. As we have shown in Figure~\ref{Fig:Param_Maps_Contours}, this region has an enhanced ERF compared to the rest of G29.1-0.7. Recent work by~\citet{Bernard-Salas:12} have shown that in the Orion Bar PDR, there is a clear stratification present regarding the major gas ions. They find that although the [C\textsc{ii}] line at 158~$\mu$m traces the dust in the PDR quite well, the [N\textsc{ii}] line at 122~$\mu$m is not spatially correlated with the dust, but actually originates closer to the ionizing source. This is interesting as we appear to be finding a similar result with the AME in G29.1-0.7 not originating from the PDR, but rather originating closer to the ionization source. In fact,~\citet{Hoang:10} investigated the effect of impulsive excitation due to individual ion--grain collisions, and found that these impulses result in a broadening of the spinning dust spectrum to higher frequencies and an increase in the spinning dust emissivity. Furthermore, a recent analysis by~\citet{Ysard:11} has shown that the spinning dust model is sensitive to the abundance of the major gas ions, with both the peak frequency and the spinning dust emissivity increasing with ion abundance. This may be what we are observing here, explaining why the AME is not coincident with the PDR but is  originating from a region closer to the source of ionization due to the stratification of the gas ions in the PDR. Further information regarding the distribution of the major gas ions within RCW175 is required to test this hypothesis but, to date, no such observations exist.


\section{Summary}
\label{Sec:Summary}

In this work we have performed a detailed analysis of the environment of RCW175. For this purpose, we used photometric data from 21~cm~(1.4~GHz) to the 8~$\mu$m~(3.75~THz) in conjunction with HI, CO and RRL spectral data. By looking at the photometric data, we found that RCW175 is not a single H\textsc{ii} region, but consists, in fact, of two separate H\textsc{ii} regions, G29.1-0.7 and G29.0-0.6. G29.1-0.7 is larger, more evolved, and most likely the older of the two regions; while G29.0-0.6 is more compact, more dusty, and likely younger. G29.1-0.7 contains the ionizing B1~II star, S65-4, at its centre, while G29.0-0.6 is characterized by the presence of filamentary pillars. Both components are clearly associated with each other, given the small spread in velocities, and we suggest that G29.0-0.6 may have been produced as a consequence of triggering from the expansion of G29.1-0.7 into the surrounding ISM. 

Using the dust model \textsc{DustEm}, we characterized the dust properties within the region and we found that there is a significant depletion of PAHs within the larger, diffuse shell of G29.1-0.7, and to a lesser extent within G29.0-0.6. We also found that the abundance of the VSGs is increased towards the interior of both G29.1-0.7 and G29.0-0.6. Both these results are consistent with current theories on the formation of H\textsc{ii} regions, with the PAHs concentrated in the surrounding PDRs, while the warmer dust originates from within the centre of the shell. Recent work has hinted that the 24~$\mu$m emission originating from within the H\textsc{ii} region might not be due to VSGs, but to a population of very warm BGs, emitting thermally at 24~$\mu$m~\citep[e.g.][]{Everett:10, Paladini:12, Salgado:12}.

Looking at the available spectral data we hypothesized that the ionized gas has become dissociated from the molecular gas, and we computed a distance of 3.2~$\pm$~0.2~kpc to RCW175 and a corresponding Galactocentric distance of 5.4~$\pm$~0.3~kpc. We used the 24~$\mu$m emission to compute an estimate of the SFR for the region of (12.6~$\pm$ 1.9)~$\times$~10$^{-5}$~M$_{\sun}$~yr$^{-1}$, and also identified 6 YSO candidates in its vicinity. Additionally, we calculated that the total mass of dust in RCW175 is~215~$\pm$~53~M$_{\sun}$.

Finally, we produced a complete SED from the radio to the mid-IR for the entire complex, and found a clear excess of emission at cm wavelengths. This excess was well fitted with a spinning dust model using typical parameters for the warm ionized medium. By comparing the AME with the parameter maps produced by the dust modeling analysis, we found that the AME is not spatially correlated with either the abundance of the PAHs or VSGs, but is much more correlated with the strength of the ERF and hence the dust temperature. This is similar to the result observed for the AME in the Perseus molecular cloud by~\citet{Tibbs:11}, and hence we suggest that the ERF is playing an important role in causing the dust grains to spin. Further investigation of the AME with respect to the location of the YSO candidates rules out the possibility of the AME originating from an UCH\textsc{ii} region, and based on the fact that the AME in G29.1-0.7 originates from the interior of the PDR, we hypothesize that the AME is due to spinning dust, but that there is a significant contribution to the spinning dust model from the major gas ions.

Further studies like the current one need to be performed in order to fully understand if this connection between the AME and the ERF is found throughout the Galaxy or only occurs in certain environments, and further modeling will help us understand the role of the gas ions in the spinning dust models.

\vspace{1cm}
\acknowledgments

We thank the anonymous referee for providing useful comments that improved the content of this paper. We thank Mark Calabretta and Lister Staveley-Smith for help with the RRL data. This work has been performed within the framework of a NASA/ADP ROSES-2009 grant, no. 09-ADP09-0059. CD acknowledges an STFC Advanced Fellowship and EU Marie Curie IRG grant under the FP7. This work is based in part on observations made with the \textit{Spitzer} Space Telescope, which is operated by the Jet Propulsion Laboratory, California Institute of Technology under a contract with NASA. This publication makes use of molecular line data from the Boston University-FCRAO Galactic Ring Survey (GRS). The GRS is a joint project of Boston University and Five College Radio Astronomy Observatory, funded by the National Science Foundation under grants AST-9800334, AST-0098562, AST-0100793, AST-0228993 and AST-0507657. The National Radio Astronomy Observatory is a facility of the National Science Foundation operated under cooperative agreement by Associated Universities, Inc.




\end{document}